\documentclass[final,5p,times,twocolumn]{elsarticle}

\usepackage{amssymb}
\usepackage{amsmath}
\usepackage{amsthm}
\usepackage{lineno}
\usepackage{hyperref}
\usepackage{fleqn}
\usepackage{xcolor}
\usepackage{epstopdf}
\journal{Chaos, Solitons \& Fractals}

\begin{document}

\begin{frontmatter}

\title{Quantum scattering of droplets by wells and barriers in one-dimensional Bose-Bose mixtures}

\author[a,b]{Sherzod R. Otajonov\corref{cor1}}
\cortext[cor1]{Corresponding author.}
\ead{srotajonov@gmail.com}

\author[a,b]{Uktambek R. Eshimbetov}
\author[a]{Bakhram A. Umarov}
\author[a]{Fatkhulla Kh. Abdullaev}

\address[a]{Uzbekistan Academy of Sciences S. A. Azimov Physical-Technical Institute, Chingiz Aytmatov Str. 2-B, 100084, Tashkent, Uzbekistan}

\address[b]{National University of Uzbekistan named after Mirzo Ulugbek,  Department of Theoretical Physics, 100174, Tashkent, Uzbekistan}

\begin{abstract}
We investigate, both analytically and numerically, the scattering of quasi-one-dimensional quantum droplets from P\"oschl-Teller potential wells and barriers.  For attractive wells, we find a sharp transition between complete reflection and transmission at a critical incident velocity for both small and large flat-top droplets. The scattering interactions differ: small, soliton-like droplets form a spatially symmetric trapped mode at the critical velocity, showing their compressibility and coherence characteristics, while large droplets develop a spatially asymmetric trapped state, revealing incompressibility and internal structure. The critical velocity depends non-monotonically on atom number: it rises in the small, compressible-droplet regime, falls in the incompressible, flat-top regime, and turns at the crossover point. We also show that the reflectionless well generates a $\pi$-phase shift, strongly altering droplet-droplet collisions relative to free space. The persistence of a confined mode after collisions between trapped and incident droplets depends sensitively on their relative phase. For the repulsive barrier, we identify regimes of complete reflection, partial return, and full transmission, depending on incident velocity, barrier height, and particle number. Our predictions match direct numerical simulations in all cases.
\end{abstract}


\end{frontmatter}

\section{Introduction}
\label{sec:intro}

Scattering of quantum particles by external potentials is fundamental in quantum mechanics. Recent work on Bose-Einstein condensates (BECs) examines matter-wave scattering from different forms of potentials. They include the cold atom reflection from potential steps and surfaces, which are important for atomic mirrors and atom-optical devices~\cite{Frid,Druzhinina,Cote,Shimizu}. Experiments have observed cold atom reflection from a BEC~\cite{Pasquini}.
These studies matter for both theory and applications, and relate to atomtronics~\cite{atom}, matter-wave interferometry~\cite{QI,QI1}, atomic beam splitting, atomic gravimetry, and coherent control of ultracold wave packets. One key issue is how nonlinear matter waves, especially solitons and quantum droplets, scatter from localised inhomogeneities.

Studies on matter-wave soliton scattering by potential wells have been performed in Refs .~\cite{Goodman,Lee,Cao,Fornish,Marchant,Sakaguchi}. In deep wells, soliton reflection depends on the excitation of nonlinear localised modes. The phase relation between an incoming soliton and the local mode is crucial, creating complex scattering regimes~\cite{Goodman}. Scattering of flat-top solitons with competing nonlinearities appears in Refs.~\cite{Sakkaf,Zeng,Umarov}. Bose-Einstein condensates can produce nonlinear potentials by spatially modulating scattering length~\cite{Kagan,Jap1}. Studies of soliton scattering by such nonlinear defects include Refs.~\cite{Kev,Abd1,Abd2}. A sharp reflection-transmission transition can serve as the filter of atom numbers in a soliton. Studies of more complicated solitonic structures, e.g., solitonic molecule scattering from defects, are performed in Ref.~\cite{AlKhawadja}. Breather splitting by localised potentials was shown in Ref.~\cite{Hulet2}. Related studies of the breather dynamics in arrays with spatially modulated cubic nonlinearity is performed in \cite{Xu2026,Zhao2025}. Both theory and experiment confirm quantum reflection of solitons in BECs~\cite{Marchant,Hulet}.

Another interesting example of localised nonlinear matter waves is the scattering of quantum droplets (QDs). Petrov predicted QDs in Bose-Bose mixtures~\cite{Petr15,Petr16}. In two and three dimensions, binary mixtures with repulsive intra-species and attractive inter-species interactions are unstable in mean-field theory. However, weak mean-field attraction lets quantum fluctuations dynamically stabilise the mixture. The Lee-Huang-Yang (LHY) correction~\cite{LHY} describes these effects and serves as effective repulsion. This yields self-bound QDs. Quantum droplets are liquid-like: in the flat-top regime, density stays constant as atom number grows, while droplet size increases. This contrasts with ordinary matter-wave solitons.
Quantum droplets have been observed in Bose-Bose mixtures~\cite{Cabr18,Seme18} and dipolar BECs~\cite{Ferr16}. Thus, studying QD scattering by potential wells and barriers is of interest. Recent works~\cite{Hu,Debnath} explored QDs scattering from a reflectionless P\"oschl-Teller (PT) well, noting a sharp transition between reflection and transmission for long, flat-top droplets. Droplet bound states with localised defects were studied in Refs.~\cite{Abdullaev,Bighin,Sinha}, and droplet splitting was discussed in Ref.~\cite{Bristy}.

These works focus on the regime of extremely strong transverse confinement, defined by $\xi = gn/\hbar\omega_{\perp} \leq 0.0043$, where $gn$ is the mean-field interaction energy and $\hbar\omega_{\perp}$ the transverse confinement energy~\cite{Zin}. In this regime, the LHY correction acts as an attractive term proportional to $-n^{3/2}$, with $n$ as the condensate density. This regime predicts varied scattering, such as partial or full reflection and droplet fragmentation. As $\xi$ increases and confinement weakens, the LHY correction and resulting scattering change clearly. This motivates studying QD scattering in broader geometric regimes beyond extreme confinement.

Experimentally, extremely strong transverse confinement is challenging. In contrast, BECs in elongated, cigar-shaped traps are more accessible and show a distinct shift in the LHY correction. Here, the LHY term becomes repulsive and proportional to $n^{5/2}$ in the energy density. This gives a quasi-one-dimensional Gross-Pitaevskii equation (GPE) with cubic attraction and quartic repulsion~\cite{Debnaz,Otajonov2019,Otajonov2024,Otajonov2025}. Moving from strong confinement to elongated traps marks a major shift in physics and experimental feasibility. We therefore analyse QD scattering by localised PT defects in these geometries, which is the focus of this work.

The present study should be distinguished from previous related works. Reference~\cite{Khawaja2021} analysed a single-component cubic Gross-Pitaevskii/nonlinear Schr\"{o}dinger equation bright soliton scattered by a reflectionless PT well, where self-localisation arises solely from cubic attraction and the critical velocity is determined by an energy balance with a nonlinear trapped mode. Reference~\cite{Hu} examined strictly one-dimensional QDs with an attractive LHY correction, described by a cubic-quadratic equation that admits an exact free-space droplet solution. In Ref.~\cite{Debnath}, the authors investigated the same strict-1D cubic-quadratic droplet model for narrow localised wells and barriers, primarily represented by delta-function defects. In contrast, this work addresses a distinct elongated one-dimensional regime in which the beyond-mean-field term retains a three-dimensional character after transverse reduction. The resulting effective equation is cubic-quartic, with residual cubic attraction counteracted by quartic LHY repulsion. The presence of this repulsive quartic term alters the droplet energetics, induces a crossover from compressible soliton-like droplets to incompressible flat-top droplets, and modifies the energy landscape of trapped modes. In the absence of a closed analytical profile for free droplets, a super-Gaussian function-based VA is developed and applied to both attractive-well and repulsive-barrier scattering scenarios. This approach accounts for symmetric critical states in small droplets, asymmetric metastable states in large flat-top droplets, transitions between reflection, partial return or splitting, and transmission at barriers, as well as the phase-sensitive interaction between incident droplets and pre-existing trapped modes.

The remainder of the paper is organised as follows. Section~\ref{sec:Model3Dvs1D} describes the three-dimensional origin of the effective one-dimensional model and validates it through comparison with the full three-dimensional formulation.
Section~\ref{sec:VA} introduces the effective quasi-one-dimensional model and develops the super-Gaussian function-based VA. Section~\ref{sec:PTPotwell} analyses the scattering of QDs by an attractive PT well, focusing on critical velocities, trapped modes, effective energy landscapes, and phase-sensitive collision dynamics. Section~\ref{sec:PTPotBar} examines scattering by a repulsive PT barrier and identifies regimes of complete reflection, partial splitting, and complete transmission. Section~\ref{sec:estim} presents the numerical methods and physical estimates for the model parameters, while Section~\ref{sec:conc} summarises the main results. The corresponding linear PT scattering problem, including the reflectionless-well condition, the exact scattering coefficients, transmission phases, and the Born approximation, is presented in~\ref{sec:linear_PT_limit}.

\section{Three-dimensional origin and validation of the effective one-dimensional model}
\label{sec:Model3Dvs1D}

The one-dimensional cubic-quartic GPE used in this study is derived as an effective axial reduction of a three-dimensional parent model with a 3D-like LHY correction. This distinction matters because the beyond-mean-field contribution depends sensitively on dimensionality. In three dimensions, the LHY correction is repulsive and contributes energy density proportional to $n^{5/2}$, whereas in the strict one-dimensional limit it becomes attractive and scales as $n^{3/2}$ \cite{Petr16,Ilg2018,Zin2018,Lavoine2021}. Therefore, the present model should not be interpreted as describing the strict 1D LHY limit. Rather, it applies to an elongated Bose-Bose mixture in which the transverse condensate profile remains close to the harmonic-oscillator ground state, while the beyond-mean-field equation of state is governed by the three-dimensional, or at least 3D-dominated, regime of the dimensional crossover. This interpretation is consistent with Ref.~\cite{Lavoine2021}, which shows that cigar-shaped Bose-Bose mixtures may be geometrically quasi-one-dimensional while retaining an effectively three-dimensional LHY contribution. This describes an elongated, cigar-shaped QD with primarily one-dimensional axial dynamics, while fluctuation-induced repulsion preserves the three-dimensional LHY character. The model should be viewed as an elongated one-dimensional or crossover reduction: the transverse profile is omitted from the dynamical equation, but the beyond-mean-field term remains repulsive and quartic. The reduced cubic-quartic Gross-Pitaevskii equation used here is intended to approximate this regime. Similar reduced cubic-quartic or LHY-corrected GPE models have recently been applied to study analytical droplet profiles, droplet-soliton crossover, collapse suppression, vortical and ring-shaped droplets, impurity dynamics, and self-bound vortex states in one and two dimensions~\cite{Debnath2021,Debnath2022,Shamriz2020,Liu2023,AlMarzoug2026,Huang2026,Arora2021,Chen2024,Pelayo2025}, supporting the effective-model framework adopted here.

The analysis begins with the three-dimensional cubic-quartic Gross-Pitaevskii equation.
\begin{equation}
i\hbar \frac{\partial \Psi}{\partial T}=\left[-\frac{\hbar^2}{2m_0}\nabla^2+U(x,y,z)-\gamma_{3D}^{\rm phys}|\Psi|^2+\delta_{3D}^{\rm phys}|\Psi|^3\right]\Psi .
\label{eq:3D_parent_validation}
\end{equation}
In this context, $\iiint |\Psi|^2 dx dy dz = N_{\rm atoms}$ denotes the total atom number. The trapping potential is assumed to be elongated, $U(x,y,z) = m_0 \omega_\perp^2 (x^2 + y^2)/2 + U_z(z)$. For the stationary validation calculation presented below, $U_z(z)$ is set to zero. This enables a direct comparison of intrinsic self-bound droplet profiles in one and three dimensions without additional axial confinement. The coefficient of the attractive mean-field term is $\gamma_{3D}^{\rm phys} = 4\pi\hbar^2|\delta a|/m_0$, while the repulsive LHY coefficient is $\delta_{3D}^{\rm phys} = 256\sqrt{2\pi}\hbar^2 a^{5/2}/(3m_0)$. Here, $a = \sqrt{a_{11} a_{22}}$ and $\delta a = a_{12} + \sqrt{a_{11} a_{22}}$. According to the sign convention in Eq.~(\ref{eq:3D_parent_validation}), the cubic term is attractive, and the quartic LHY term is repulsive. The interplay between residual mean-field attraction and fluctuation-induced repulsion stabilises the self-bound droplets.

The reduction to an effective one-dimensional equation is achieved by assuming that the transverse mode remains frozen in the Gaussian ground state of the radial harmonic trap. The factorized ansatz $\Psi(x,y,z,T) = R_\perp(x,y) \Phi(z,T)$ is employed.
$$
R_\perp(x,y)=\frac{1}{\sqrt{\pi}l_\perp}\exp\left[-\frac{x^2+y^2}{2l_\perp^2}\right],\qquad l_\perp=\sqrt{\frac{\hbar}{m_0\omega_\perp}} .
$$
This approximation is valid when the transverse confinement energy is sufficiently large to suppress significant radial deformation during axial droplet dynamics.
The dimensional reduction requires the transverse density
mode to remain frozen in the harmonic-oscillator ground
state. Following Refs.~\cite{Ilg2018,Lavoine2021}, we
introduce
$$
\lambda=\frac{a}{l_\perp},
\qquad
\kappa=n_{1D}a,
$$
where $n_{1D}$ is the total axial line density. The regime in
which the axial dynamics are effectively one-dimensional
while the dominant spin fluctuations retain their
three-dimensional character is
$$
\lambda\ll1,
\qquad
1\ll\kappa\ll\lambda^{-2}.
$$
In this parameter, the usual three-dimensional LHY density dependence may be integrated over the frozen transverse Gaussian profile, yielding the quartic
beyond-mean-field term in the effective one-dimensional equation. Strict low-dimensional models are asymptotic limits and may not be quantitatively accurate in the crossover regime $\kappa\sim1$, where a confinement-dependent crossover extended GPE is required \cite{Ilg2018,Lavoine2021,Pelayo2025}. The present model
therefore describes the three-dimensional-dominated quasi-one-dimensional side of the crossover rather than the strict one-dimensional limit.

The transverse zero-point contribution results in a constant energy shift. This shift can be eliminated by a global phase transformation. The nonlinear coefficients are renormalized by the transverse overlap integrals $I_4 \equiv \int |R_\perp|^4 dx dy = 1/(2\pi l_\perp^2)$ and $I_5 \equiv \int |R_\perp|^5 dx dy = 2/(5\pi^{3/2} l_\perp^3)$. Substituting the factorised form into Eq.~(\ref{eq:3D_parent_validation}) yields
\begin{equation}
i\hbar\frac{\partial \Phi}{\partial T}=-\frac{\hbar^2}{2m_0}\Phi_{zz}+U_z(z)\Phi-\gamma_{1D}^{\rm phys}|\Phi|^2\Phi+\delta_{1D}^{\rm phys}|\Phi|^3\Phi .
\label{eq:1D_phys_validation}
\end{equation}
The effective one-dimensional coefficients are $\gamma_{1D}^{\rm phys} = \gamma_{3D}^{\rm phys}/(2\pi l_\perp^2)$ and $\delta_{1D}^{\rm phys} = 2\delta_{3D}^{\rm phys}/(5\pi^{3/2} l_\perp^3)$. Consequently, the cubic-quartic structure of the effective one-dimensional model arises directly from integrating out the frozen transverse degrees of freedom. The cubic term corresponds to the residual attractive mean-field contribution. The quartic term represents the axial remnant of the repulsive 3D-like LHY correction.

We next introduce the dimensionless variables $z_{\rm phys}=x_sx$, $T=t_st$, and $\Phi_{\rm phys}=\psi_s\phi$. The scales are chosen as $\psi_s=\gamma_{1D}^{\rm phys}/\delta_{1D}^{\rm phys}$, $t_s=\hbar(\delta_{1D}^{\rm phys})^2/(\gamma_{1D}^{\rm phys})^3$, and $x_s=(\hbar t_s/m_0)^{1/2}$. With these definitions, Eq.~(\ref{eq:1D_phys_validation}) becomes
\begin{equation}
i\phi_t=-\frac{1}{2}\phi_{xx}+V(x)\phi-|\phi|^2\phi+|\phi|^3\phi .
\label{eq:1D_dimless_validation}
\end{equation}
This normalised cubic-quartic equation is employed throughout this study. The normalisation sets the magnitudes of the attractive and repulsive nonlinearities to unity. The remaining dynamics are determined by the norm, the external potential, and the incident velocity.

For a concrete physical estimate, we consider a symmetric ${}^{39}\mathrm{K}$  mixture with $a_{11}=a_{22}=50a_0$, so that $a=50a_0$, and choose $\delta a=-2a_0$, corresponding to $a_{12}=-52a_0$. The transverse confinement is taken as $\omega_\perp=2\pi\times500\,\mathrm{Hz}$, which gives $l_\perp\simeq0.720\,\mu\mathrm{m}$. For these parameters, the physical nonlinear coefficients are $\gamma_{3D}^{\rm phys}=2.28\times10^{-52}$, $\delta_{3D}^{\rm phys}=1.32\times10^{-62}$, $\gamma_{1D}^{\rm phys}=7.01\times10^{-41}$, and $\delta_{1D}^{\rm phys}=2.55\times10^{-45}$, in the SI units associated with Eqs.~(\ref{eq:3D_parent_validation}) and~(\ref{eq:1D_phys_validation}). The corresponding normalization scales are $\psi_s=2.75\times10^4\,\mathrm{m}^{-1/2}$, $t_s=1.98\times10^{-3}\,\mathrm{s}$, and $x_s=1.80\,\mu\mathrm{m}$. Thus, one dimensionless length unit corresponds to $1.80\,\mu\mathrm{m}$, and one dimensionless time unit corresponds to $1.98\,\mathrm{ms}$. A dimensionless chemical potential $\mu$ is converted according to $\mu_{\rm phys}/h=\mu/(2\pi t_s)$, with $1/(2\pi t_s)=80.28\,\mathrm{Hz}$. The atom-number conversion is $N_{\rm atoms}=\psi_s^2x_sN_{\rm dim}$. Therefore, $N_{\rm atoms}=1.5\times10^4$ corresponds to $N_{\rm dim}=11.01$.

To validate the effective one-dimensional reduction, the corresponding dimensionless three-dimensional parent equation was solved and compared with Eq.~(\ref{eq:1D_dimless_validation}). In normalized units, the transverse frequency is $\Omega_\perp = \omega_\perp t_s = 6.228$, and the dimensionless transverse oscillator length is $l_{\perp,\rm dim} = l_\perp / x_s = 0.4007$, consistent with $1/\sqrt{\Omega_\perp} = 0.4007$. The transverse overlap integrals are $I_4 = 1/(2\pi l_{\perp,\rm dim}^2) = 0.9912$ and $I_5= 2/(5\pi^{3/2} l_{\perp,\rm dim}^3) = 1.1165$. Accordingly, $\gamma_3 = 1/I_4 = 1.0089$ and $\delta_3 = 1/I_5 = 0.8957$ are chosen, yielding $\gamma_3 I_4 = \delta_3 I_5 = 1$. This construction guarantees that, under the Gaussian transverse ansatz, the dimensionless three-dimensional model reduces exactly to Eq.~(\ref{eq:1D_dimless_validation}). It also provides a direct consistency check. Any discrepancy between the one-dimensional and three-dimensional stationary states then results from genuine transverse deformation rather than a mismatch in the nonlinear coefficients.

The stationary validation was conducted using imaginary-time propagation. Initially, Eq.~(\ref{eq:1D_dimless_validation}) was solved at fixed norm. The resulting converged one-dimensional state was then used to initialise the three-dimensional calculation as $\psi_{3D}(x,y,z,0) = R(x,y) \phi_{1D}(z)$. The full three-dimensional equation was subsequently relaxed independently. For the three-dimensional simulations, we use the imaginary-time method~\cite{JYang}. The computational grid is chosen as
$N_x=N_y=128$ and $N_z=1024$, with domain sizes
$L_x=L_y=40$ and $L_z=200$. The time step is set to
$\Delta t=10^{-4}$. From the converged three-dimensional state, the integrated axial density $n_{1D}^{(3D)}(z) = \int |\psi_{3D}(x,y,z)|^2 dx dy$ was computed and compared with the one-dimensional density $n_{1D}^{(1D)}(z) = |\phi_{1D}(z)|^2$. This comparison assesses whether the axial density predicted by the effective one-dimensional model is reproduced by the full three-dimensional parent equation after relaxation without enforcing exact factorisation.

\begin{figure}[t]
\centerline{ \includegraphics[width=8.2cm]{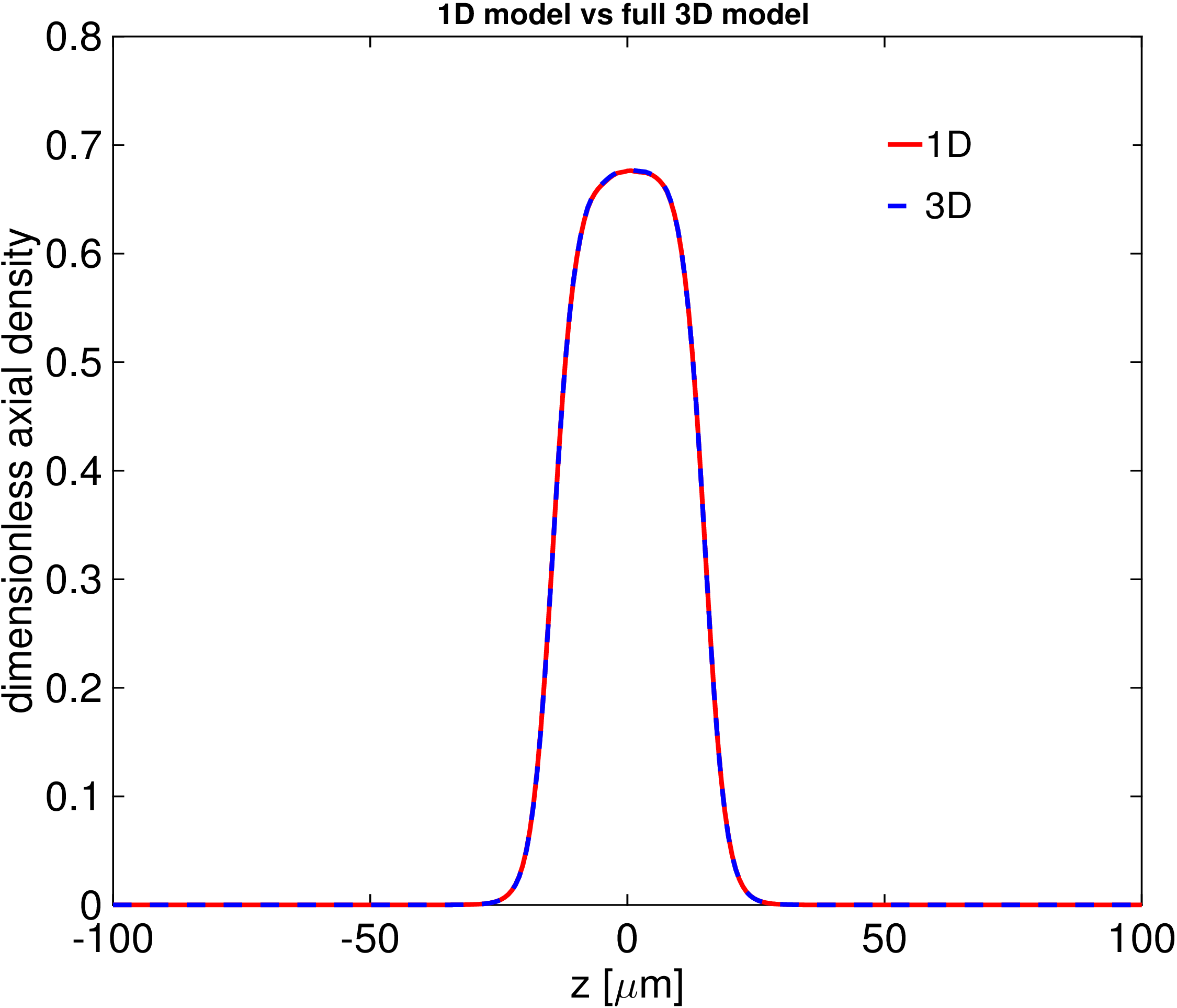}}
\caption{Validation of the effective one-dimensional cubic-quartic model for $a_{11}=a_{22}=50a_0$, $\delta a=-2a_0$, $\omega_\perp=2\pi\times500\,\mathrm{Hz}$, and $N_{\rm atoms}=1.5\times10^4$. The solid curve shows the one-dimensional density $n_{1D}^{(1D)}(z)=|\phi_{1D}(z)|^2$, while the dashed curve shows the integrated three-dimensional density $n_{1D}^{(3D)}(z)=\int |\psi_{3D}(x,y,z)|^2\,dx dy$. The coordinate $z$ is given in micrometres. The two profiles overlap almost completely, the relative axial-density error is $1.32\times10^{-3}$.}
\label{fig:1D_3D_density_validation}
\end{figure}

The comparison shows very close agreement. The one-dimensional stationary state has $E_{1D}=-1.1162$, $\mu_{1D}=-0.1190$, $\mu_{1D}/h=-9.55\,\mathrm{Hz}$, maximum density $\max|\phi|^2=0.6764$, and rms axial size $9.46\,\mu\mathrm{m}$. The full three-dimensional calculation gives $E_{3D}^{\rm axial}=-1.1174$, $\mu_{3D}^{\rm axial}=-0.1192$, $\mu_{3D}^{\rm axial}/h=-9.57\,\mathrm{Hz}$, maximum integrated axial density $0.6772$, and rms axial size $9.45\,\mu\mathrm{m}$. The axial three-dimensional quantities are obtained after subtracting the transverse zero-point contribution. Specifically, $E_{3D}^{\rm axial}=E_{3D}-N\Omega_\perp$ and $\mu_{3D}^{\rm axial}=\mu_{3D}-\Omega_\perp$. This allows direct comparison with the effective one-dimensional energy and chemical potential.
The relative errors are small. The axial-density error is
$1.32 \times 10^{-3}$, the peak-density error is
$1.19 \times 10^{-3}$, the rms-size error is
$1.07 \times 10^{-3}$, the chemical potential error is
$1.36 \times 10^{-3}$, and the axial-energy error is
$1.12 \times 10^{-3}$. The transverse structure also remains close to the assumed Gaussian form. The factorisation error, obtained by comparing $|\psi_{3D}|^2$ with $|R|^2 n_{1D}^{(3D)}(z)$, is
$3.22 \times 10^{-3}$, and the maximum relative deviation of the transverse width from $l_{\perp,\mathrm{dim}}$ is $4.13 \times 10^{-3}$. All these errors are below $0.5\%$, showing that the transverse mode remains effectively frozen and that the axial density, energy, chemical potential, and spatial extent are accurately reproduced by the effective one-dimensional model.
The robustness of the reduction was further tested for several atom numbers at the same interaction parameter $\delta a = -2a_0$. In all cases examined, the integrated three-dimensional axial density remained very close to the corresponding one-dimensional density, and the relative errors remained small. This result indicates that the agreement is not limited to a single norm, but persists across the relevant atom-number range considered in this analysis. The accuracy is highest when the transverse confinement is sufficiently strong to prevent significant radial swelling or compression of the droplet.
As an additional check, the validation was repeated with a stronger residual attraction, $\delta a = -7a_0$. In this scenario, the relative errors increase to approximately $5\%$, which is consistent with physical expectations. A larger value of $|\delta a|$ enhances the mean-field attraction and increases the equilibrium density, thereby promoting nonlinear self-compression and making the transverse profile more sensitive to deformation. As a result, the frozen Gaussian transverse approximation becomes less accurate than for $\delta a = -2a_0$. Nevertheless, even for $\delta a = -7a_0$, the effective one-dimensional reduction remains a reasonable approximation: the axial density profiles are still reproduced with acceptable accuracy, and the dominant droplet properties are captured by the reduced model. The primary calculations in this study are therefore conducted in the parameter regime where the validation errors are below $0.5\%$, while the stronger-attraction test provides an estimate of the range over which the reduction remains physically meaningful.
This validation provides the physical justification for employing Eq.~(\ref{eq:1D_dimless_validation}) in the scattering calculations. The observed agreement between the one-dimensional and three-dimensional stationary states demonstrates that the droplet dynamics under consideration occur in a regime where the axial degree of freedom dominates the deformation, while transverse excitations remain weak. Furthermore, the inclusion of the quartic repulsive term aligns with the intended 3D-like or crossover interpretation of the LHY correction, rather than the strict one-dimensional limit. Therefore, the effective one-dimensional cubic-quartic equation offers a quantitatively reliable and computationally efficient framework for describing the elongated QDs analysed in this study.

\section{The variational approximation}
\label{sec:VA}

The reduced model is employed to investigate the scattering dynamics of QDs after establishing the three-dimensional origin and validating the effective one-dimensional description. The system under consideration is a two-component BEC with equal atomic masses, repulsive intra-component interactions, and attractive inter-component coupling. In this regime, the residual mean-field attraction competes with the repulsive LHY correction. For nearly symmetric components, the two-component system can be reduced to an effective single-field description for the condensate wave function.

The axial dynamics are governed by the dimensionless extended Gross-Pitaevskii equation.
\begin{equation}
i\psi_t = -\frac{1}{2}\psi_{xx} + V(x)\psi - q|\psi|^2\psi + g|\psi|^3\psi .
\label{eq:gpe}
\end{equation}
In this context, the coefficient $q>0$ quantifies the strength of the residual attractive mean-field interaction, while $g>0$ represents the strength of the repulsive beyond-mean-field, or quantum fluctuation, contribution. As demonstrated in the previous section, the physical one-dimensional equation can be scaled to the normalised form with $q=g=1$. However, $q$ and $g$ are retained in Eq.~(\ref{eq:gpe}) to maintain generality and to clarify the distinct roles of the attractive mean-field and repulsive LHY terms. Alternative choices of dimensional scales yield the same dimensionless equation with non-unit values of $q$ and $g$. Conversely, results can be converted to the fully normalised convention $q = g = 1$ using the scaling described above.

Based on this framework, the scattering of a self-bound QD from a modified PT potential is investigated.
We take the modified PT potential in the following form,
\begin{equation}
V(x) = -U_0\,\mathrm{sech}^{2}(\alpha x)
\end{equation}
where $\alpha=\sqrt{|U_0|}$ sets the inverse width of the defect and $U_0$ controls its strength. In the present convention, $U_0>0$ corresponds to an attractive potential well, while $U_0<0$ would describe a repulsive barrier.
For the attractive case, $U_0>0$, the condition
$\alpha^2=U_0$ selects the $\ell=1$ member of the reflectionless PT family. Consequently, the corresponding linear problem has zero reflection at all incident wave numbers, although the transmitted wave acquires a
velocity-dependent phase. For $U_0<0$, the same expression describes the repulsive barrier $V(x)=|U_0|\operatorname{sech}^2(\alpha x)$. In this case,
the relation $\alpha^2=|U_0|$ fixes the barrier width relative to its height but does not make the barrier reflectionless. The exact linear reflection and transmission coefficients, the associated scattering phases, and their Born approximation limits are presented in~\ref{sec:linear_PT_limit}. The variational formulation itself is not restricted to the scaling $\alpha^2=|U_0|$, because the external potential enters through the general overlap integral $\int V(x)|\psi|^2\,dx$. Nevertheless, this scaling is retained throughout the present calculations to provide a common width--strength relation for the attractive and repulsive defects.

Scattering on PT wells has already been investigated for bright solitons in the cubic Gross-Pitaevskii equation~\cite{Khawaja2021}. It has also been studied for QDs in cubic-quadratic mean-field models that include beyond-mean-field effects~\cite{Hu}. In those works, the availability of exact analytical soliton or droplet solutions in free space enabled the variational description of trapped (critical) modes in the attractive PT well. They also allowed identification of the critical velocity separating reflection from transmission. In particular, the variational ansatz was formed by multiplying the known free-space solution by a defect-induced factor. This factor was tailored to capture the pinned (trapped) configuration.

Here, we address a situation where no explicit analytical droplet profile exists without the potential. We treat this by developing a double variational approach. First, we determine the stationary (free-space) droplet profile variationally. Next, we embed this profile into a position-dependent variational ansatz to describe turning-point (zero-speed) and trapped modes in the presence of the PT defect. This two-stage procedure extends the trapped-mode and critical velocity analysis to QD models beyond those that admit analytical solutions.

To approximate the QD density profile, we employ a super-Gaussian variational ansatz, which takes the general form~\cite{Otajonov2019,Otajonov2024,Otajonov2020,Otajonov2022}:
\begin{eqnarray}\label{gaussian}
\psi(x,t) &=& A(t)\,
\exp\Bigg[
-\frac{1}{2}\left(\frac{x-\xi(t)}{a(t)}\right)^{2m}
+ i\,b(t)\,(x-\xi(t))^{2} \nonumber \\
&& \hspace{1.5cm}
+ i\,v(t)\,(x-\xi(t))
+ i\,\varphi(t)
\Bigg].
\end{eqnarray}
Here, $|\psi(x,t)|^2$ is the density profile, $A(t)$ is the time-dependent amplitude, $a(t)$ is the width, $b(t)$ is the chirp, $\xi(t)$ is the center-of-mass position, $v(t)$ is the center-of-mass velocity, and $\varphi(t)$ is the global phase. All of these are variational parameters.
The parameter $m>0$ determines the profile shape: $m=1$ gives a Gaussian form, $m>1$ yields a flat-top distribution. Here, $m$ is a stationary, time-independent variational parameter for the droplet profile.

The super-Gaussian function in Eq.~(\ref{gaussian}) is employed to represent QD profiles far from the external potential. It is justified by its ability to interpolate between two physically distinct regimes. For small particle numbers, the optimized profile is bell shaped and describes a compressible, soliton-like droplet. As the particle number increases, the variational shape parameter $m$ increases, and the profile progressively develops a broad flat-top region. Thus, the same variational form provides a unified approximation for compressible and liquid-like droplets.

The motion of a droplet is introduced through the Galilean phase factor $\exp(\pm i v x)$, where the sign determines propagation in the positive or negative $x$ direction. This phase imprint imparts a centre-of-mass velocity to the droplet while preserving its density profile. For trapped modes localized near the PT well, the super-Gaussian profile requires modification to account for the symmetry and phase structure induced by the defect. Accordingly, the droplet profile is multiplied by a $\tanh$ factor, as shown in Eq.~(\ref{eq:TrapModeTrialFunc}). This factor introduces a node and an associated phase change across the centre of the well, enabling the trial function to describe pinned states formed near the critical scattering regime.

The norm,
\begin{equation}
N \equiv \int\limits_{-\infty}^{+\infty} |\psi(x,t)|^{2}dx,
\end{equation}
is a conserved quantity and is directly proportional to the total number of atoms in the condensate. Substituting the ansatz~(\ref{gaussian}) yields
\begin{equation}
N = 2aA^2\Gamma(1+M).
\label{norm}
\end{equation}
Here, for convenience, we introduce a reduced super-Gaussian parameter $M=\frac{1}{2m}$. We also use $\Gamma(\cdot)$ to denote the Gamma function. This relation allows elimination of amplitude $A$ in favour of the conserved norm $N$ in the variational formulation.

As the next step, we construct the averaged (effective) Lagrangian
\begin{equation}
L=\int\limits_{-\infty}^{+\infty}\mathcal{L}dx,
\end{equation}
by substituting the super-Gaussian ansatz~(\ref{gaussian}) into the Lagrangian density associated with the extended 1D Gross-Pitaevskii equation,
\begin{eqnarray}
\mathcal{L} &=& \frac{i}{2}\left(\psi_t^{*}\psi-\psi^{*}\psi_t\right)
+\frac{1}{2}|\psi_x|^{2}
+V(x)|\psi|^{2} \nonumber\\
&&-\frac{q}{2}|\psi|^{4}
+\frac{2g}{5}|\psi|^{5}\,.
\label{lagrdens}
\end{eqnarray}

Carrying out the spatial integration yields the averaged Lagrangian in the form
\begin{eqnarray}\label{lagrangian}\nonumber
L &=& \varphi_tN - \frac{\xi_{t}^{2}N}{2} + \frac{a^2b_tN}{3}\frac{\Gamma(1+3M)}{\Gamma(1+M)} + \\
\nonumber
&& \frac{N}{8a^2M}\frac{\Gamma(2-M)}{\Gamma(1+M)} + \frac{2a^2b^2N}{3}\frac{\Gamma(1+3M)}{\Gamma(1+M)} - \\
&& \frac{q N^2}{2^{M+2}a \Gamma(1+M)} + \frac{(2/5)^{M+1} g N^{5/2}}{(2a\Gamma(1+M))^{3/2}} + L_{\rm pot}\,,
\end{eqnarray}
where $L_{\rm pot} = \int\limits_{-\infty}^{+\infty}V(x)|\psi|^2dx$.

The Euler-Lagrange equation for phase $\varphi$ yields $\frac{dN}{dt}=0$. This confirms the conservation of the norm and, hence, the number of atoms. The equation from variation with respect to centre coordinate $\xi$ produces the expected relation between centre-of-mass velocity and position: $\xi_t = v$. Thus, $v(t)$ acts as the collective coordinate for translational motion. Variations with respect to width $a$ and chirp $b$ give coupled evolution equations for $a(t)$ and $b(t)$.
\begin{eqnarray}\nonumber
b_t &=& -2b^2 + \frac{3}{8a^4M}\frac{\Gamma(2-M)}{\Gamma(1+3M)} - \frac{3qN}{2^{M+3}a^3\Gamma(1+3M)} + \\
\nonumber
&& \frac{9 (2/5)^{M+1} gN^{3/2}}{(2a)^{7/2}\Gamma^{1/2}(1+M)\Gamma(1+3M)} - \\
\nonumber
&& \frac{3}{2aN}\frac{\Gamma(1+M)}{\Gamma(1+3M)}\frac{\partial L_{\rm pot}}{\partial a} \equiv f_b(a,M,N)\, , \\
a_t &=& 2ab\, , \quad \xi_{tt} = -\frac{1}{N}\frac{\partial L_{\rm pot}}{\partial\xi}\, .
\label{at_bt_xitt}
\end{eqnarray}
To determine stationary QD profiles, we focus on time-independent solutions in the variational approach, following previous studies~\cite{Otajonov2019, Otajonov2024, Otajonov2020, Otajonov2022}. Stationarity requires the absence of breathing and chirp dynamics, i.e., $ a_t=0$ and $ b_t=0$. The equation for $a_t$ immediately gives $b=0$, so stationary droplets have no quadratic phase curvature. The remaining equations of motion for variational parameters lead to these equations:
\begin{eqnarray}
f_b(a,M,N)&=0,
\nonumber\\
f_m(a,M,N) &\equiv& \left.\frac{\partial L}{\partial M}\right|_{b=0}=0\, .
\label{fb_fm}
\end{eqnarray}
Solving these equations for given parameters $(q,g, N)$ yields the stationary width and reduced super-Gaussian parameter $M$. Once the width $a$ and $M$ are found, we directly obtain the stationary amplitude $A$ from the norm~(\ref{norm}).

The coupled first-order equations for $b_t$ and $a_t$ in Eq.~(\ref{at_bt_xitt}) can be reduced to a single second-order differential equation for the QD width. Equation~(\ref{at_bt_xitt}) can be rewritten as
\begin{eqnarray}\nonumber
a_{tt} &=& 4ab^2 + 2ab_t = \\
\nonumber
&& \frac{3}{4a^3M}\frac{\Gamma(2-M)}{\Gamma(1+3M)} - \frac{3q N}{2^{M+2}a^2 \Gamma(1+3M)} + \\
\nonumber
&& \frac{9 (2/5)^{M+1} gN^{3/2}}{(2a)^{5/2} \Gamma^{1/2}(1+M)\Gamma(1+3M)} - \\
&& \frac{3}{N}\frac{\Gamma(1+M)}{\Gamma(1+3M)}\frac{\partial L_{\rm pot}}{\partial a}\, ,  \\
\nonumber
\xi_{tt}& = &-\frac{1}{N}\frac{\partial L_{\mathrm{pot}}}{\partial\xi} \, .
\label{att}
\end{eqnarray}
This form is particularly convenient for deriving the corresponding effective potentials. One can introduce Newton-like equations $a_{tt}=-\frac{\partial W(a)}{\partial a}$ and $\xi_{tt}=-\frac{\partial W(\xi)}{\partial \xi}$. Here, $W(a)$ and $W(\xi)$ are the effective potentials for $a$ and $\xi$, respectively. The effective potentials are then obtained by integrating these Newton-like equations:
\begin{eqnarray}\nonumber
W(a) &=& \frac{3}{8a^2M}\frac{\Gamma(2-M)}{\Gamma(1+3M)} - \frac{3q N}{2^{M+2}a \Gamma(1+3M)} + \\
\nonumber
&& \frac{3 (2/5)^M (2N/a)^{3/2} g}{20 \Gamma^{1/2}(1+M)\Gamma(1+3M)} + \\
\nonumber
&& \frac{3}{N}\frac{\Gamma(1+M)}{\Gamma(1+3M)}\int\frac{\partial L_{\rm pot}}{\partial a}~da\, , \\
W(\xi) &=& \frac{1}{N}\int\frac{\partial L_{\rm pot}}{\partial\xi}~d\xi\, .
\label{effpot}
\end{eqnarray}

After determining the stationary parameters of the QD, we next study its scattering from the external potential. In numerical simulations, the VA-predicted stationary droplet profile is used as the initial condition to solve the governing Eq.~(\ref{eq:gpe}). The appropriate initial velocity is set for the scattering process.
We define reflection, trapping, and transmission coefficients as $C_{\mathrm{ref}}(t)$, $C_{\mathrm{trap}}(t)$, and $C_{\mathrm{trans}}(t)$ respectively:
\begin{eqnarray}\nonumber
&& C_{\rm ref}(t) = \frac{1}{N}\int\limits_{-\infty}^{-h}|\psi(x,t)|^2dx \, , \quad C_{\rm trap}(t) = \frac{1}{N}\int\limits_{-h}^{h}|\psi(x,t)|^2dx \, , \\
&& C_{\rm trans}(t) = \frac{1}{N}\int\limits_{h}^{+\infty}|\psi(x,t)|^2dx \, .
\label{rtt:coeff}
\end{eqnarray}
These coefficients are defined by integrating the droplet density over the spatial regions corresponding to the reflected, trapped, and transmitted parts of the wave packet at the final evolution time $t_f$. Here, $\pm h$ denotes the left and right influence boundaries of the potential. In our simulations, $h$ is set to 5 times the width of the potential well, measured at half-maximum (FWHM). The final time $t_f$ is taken sufficiently large so that the scattered portions of the QD are well separated from the potential region, ensuring an unambiguous evaluation of the three coefficients.
The sensitivity of the time-dependent coefficients to the choice of the partition boundary $h$ was also examined. For $h=5\times\mathrm{FWHM}$ and $h=6\times\mathrm{FWHM}$, the dynamics of $C_{\mathrm{ref}}(t)$, $C_{\mathrm{trap}}(t)$, and $C_{\mathrm{trans}}(t)$ are nearly identical. This result indicates that $h=5\times\mathrm{FWHM}$ is already sufficiently far from the centre of the PT well. At this distance, the potential is negligible, and the trapping region fully contains the density localised near the defect.
In contrast, for $h=3\times\mathrm{FWHM}$, a very small but noticeable difference appears in the time evolution of the coefficients. This discrepancy does not indicate a change in the physical scattering process. Rather, it arises because $h=3\times\mathrm{FWHM}$ places the partition boundary closer to the interaction region. During the collision and the subsequent transient dynamics, the droplet tails and density oscillations near the well can temporarily cross this boundary. Consequently, part of the density may be counted earlier as reflected or transmitted rather than as trapped. This produces small changes in the intermediate-time behaviour of $C_{\mathrm{ref}}(t)$, $C_{\mathrm{trap}}(t)$, and $C_{\mathrm{trans}}(t)$.
The suitability of the choice $h=5\times\mathrm{FWHM}$ is also supported by the density evolutions shown in Figs.~\ref{fig:dynSmallQDandCoefC} and \ref{fig:dynLargeQDandCoefC}. In these figures, the spatial density dynamics and the corresponding time evolution of the coefficients $C_{\mathrm{ref}}(t)$, $C_{\mathrm{trap}}(t)$, and $C_{\mathrm{trans}}(t)$ are consistent with one another. The growth and decay of the coefficients accurately follow the motion of the reflected, trapped, and transmitted density components.
Importantly, this sensitivity concerns only the diagnostic representation of the time-dependent coefficients. It does not affect the underlying field dynamics, the density-evolution plots, the critical velocities, the effective-potential analysis, or the final scattering classification. Once the outgoing droplets are well separated from the defect region, the final reflection, trapping, and transmission coefficients remain unchanged within numerical accuracy. Therefore, $h=5\times\mathrm{FWHM}$ is used throughout the manuscript as an appropriate partition boundary for the present parameters.
In the following sections, we build on this framework to study the scattering of QDts in the PT potential in detail, first considering the potential well case and then the potential barrier case.

\section{Scattering of Quantum Droplets in P\"oschl-Teller potential well}
\label{sec:PTPotwell}

\begin{figure}[t]
\centerline{\includegraphics[width=4.65cm]{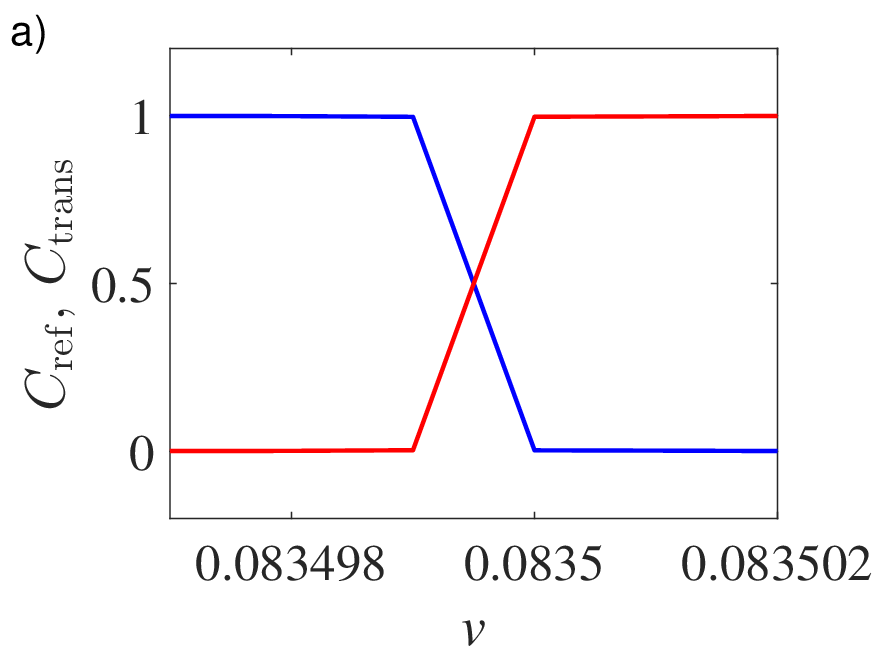} \includegraphics[width=4.65cm]{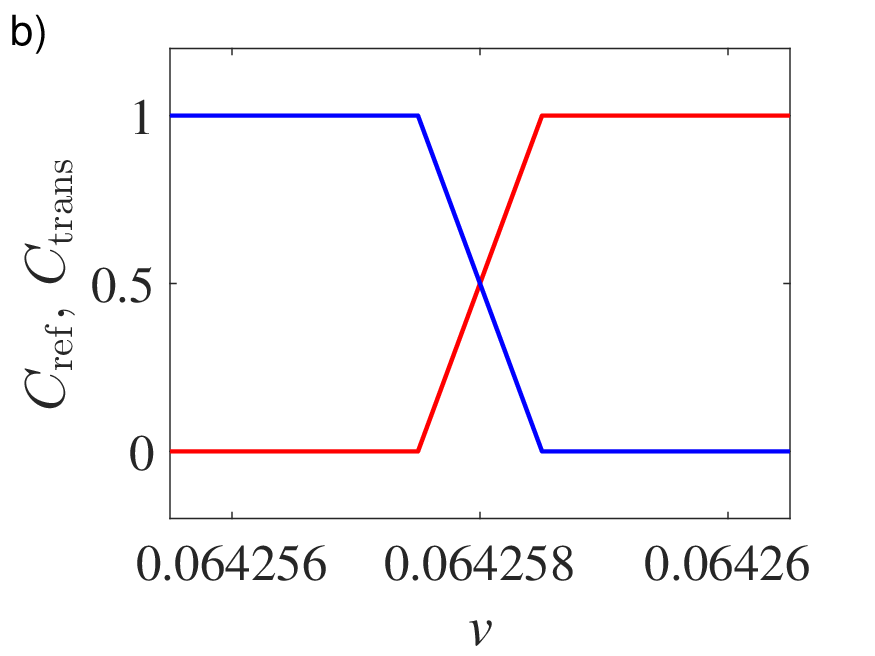}}
\caption{Reflection (blue) and transmission (red) coefficients versus the initial velocity of a QD scattered by a reflectionless PT potential well. Panel (a) corresponds to a small droplet with $N=1$, and panel (b) to a large droplet with $N=10$. Other parameters are $U_0=1$, $q=g=1$, and the initial droplet position is $x_0=-100$.}
\label{fig:RefTrans}
\end{figure}

Figure~\ref{fig:RefTrans} presents the numerically computed reflection (blue) and transmission (red) coefficients as functions of the incident speed for a QD scattered by an attractive, reflectionless PT potential well. For both small and large droplets, a sharp transition is observed between almost complete reflection and complete transmission. This transition defines a well-pronounced critical speed.
The meaning of ``reflectionless'' here refers to the associated
linear $\ell=1$ PT problem. As shown in~\ref{subsec:linear_attractive_PT}, the attractive
potential with $\alpha^2=U_0$ has $\mathcal R_{\rm w}(k)=0$ and
$\mathcal T_{\rm w}(k)=1$ for all incident wave numbers.
Therefore, the reflection observed below is a genuinely
nonlinear droplet effect rather than ordinary linear
backscattering from the defect.

%

\begin{figure}[t]
\centerline{ \includegraphics[width=4.2cm]{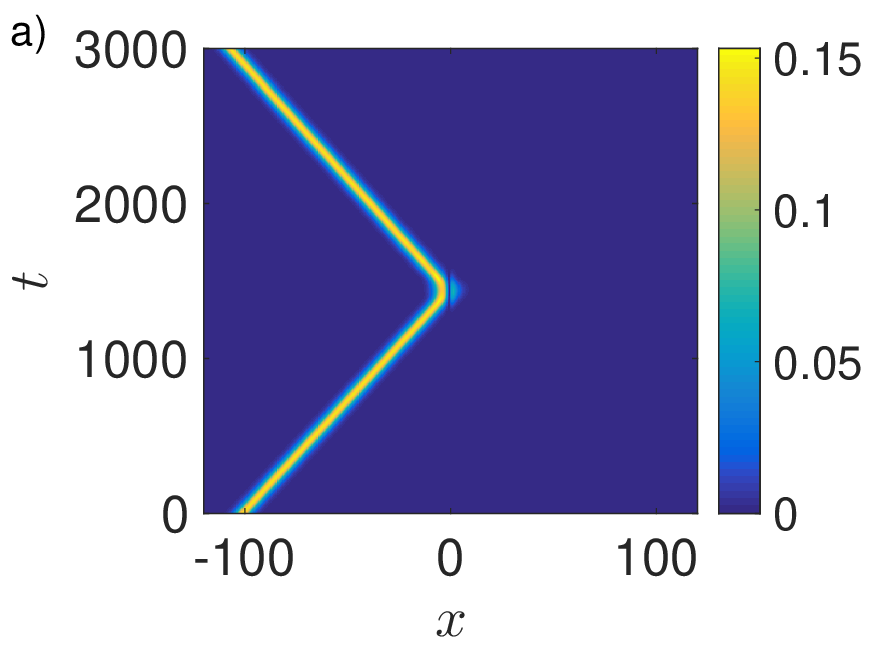} \hskip-0.1cm \includegraphics[width=4.4cm]{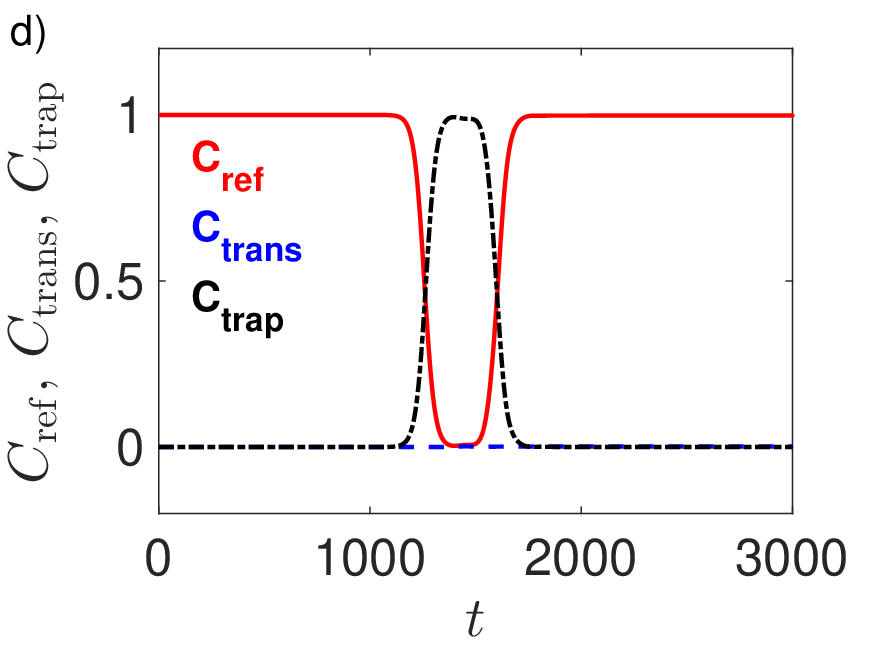}}
\centerline{ \includegraphics[width=4.2cm]{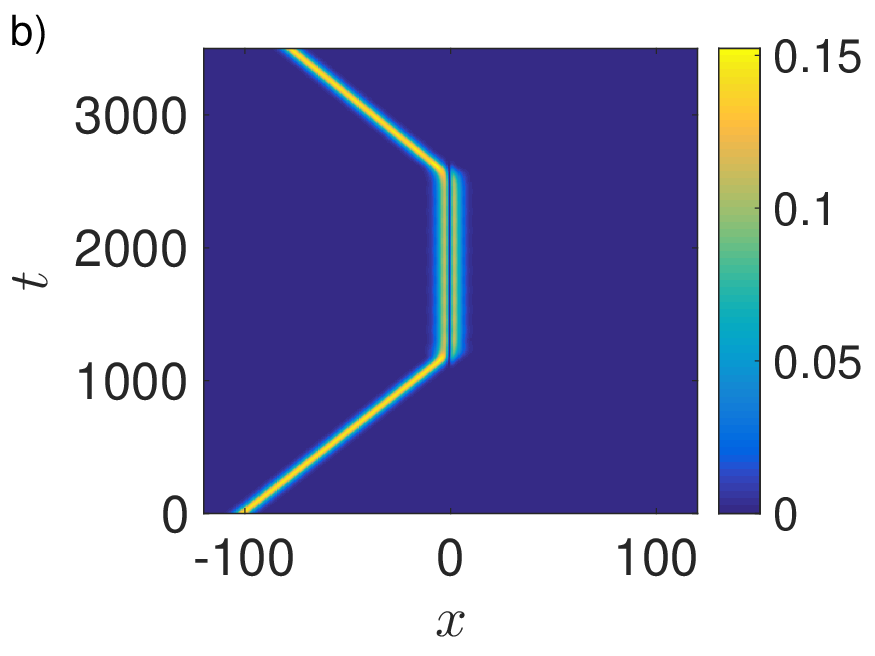} \hskip-0.1cm \includegraphics[width=4.4cm]{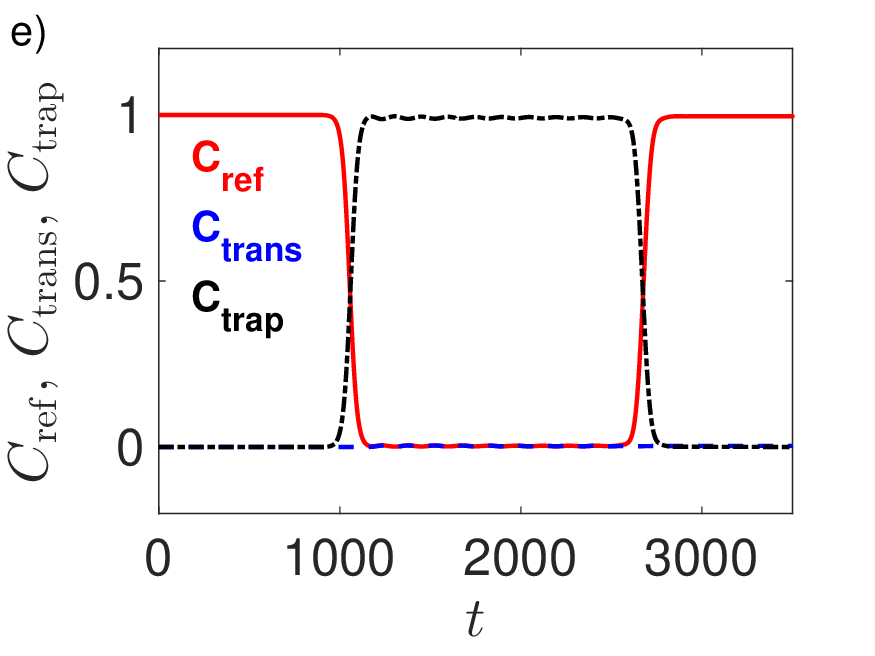}}
\centerline{ \includegraphics[width=4.2cm]{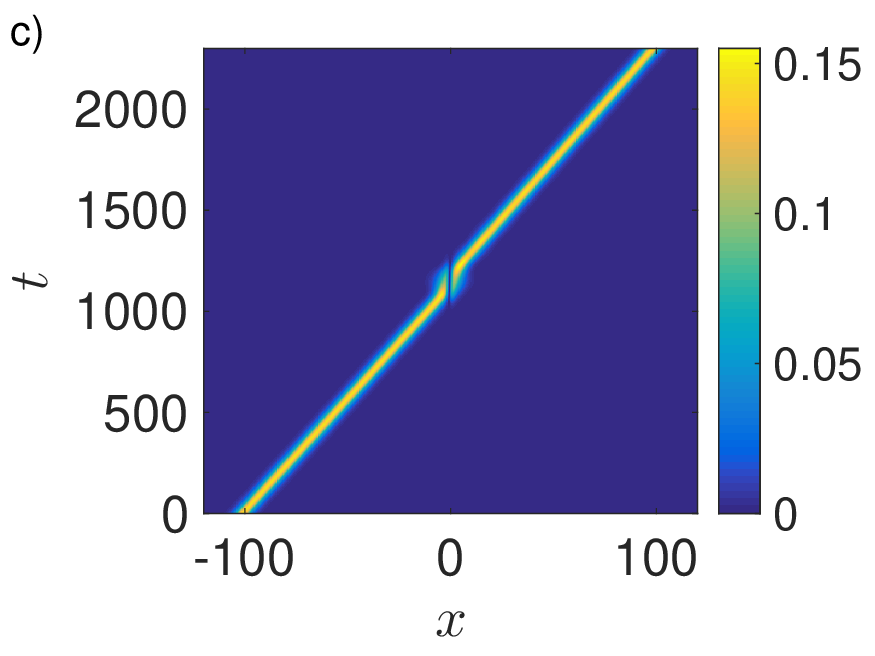} \hskip-0.1cm \includegraphics[width=4.4cm]{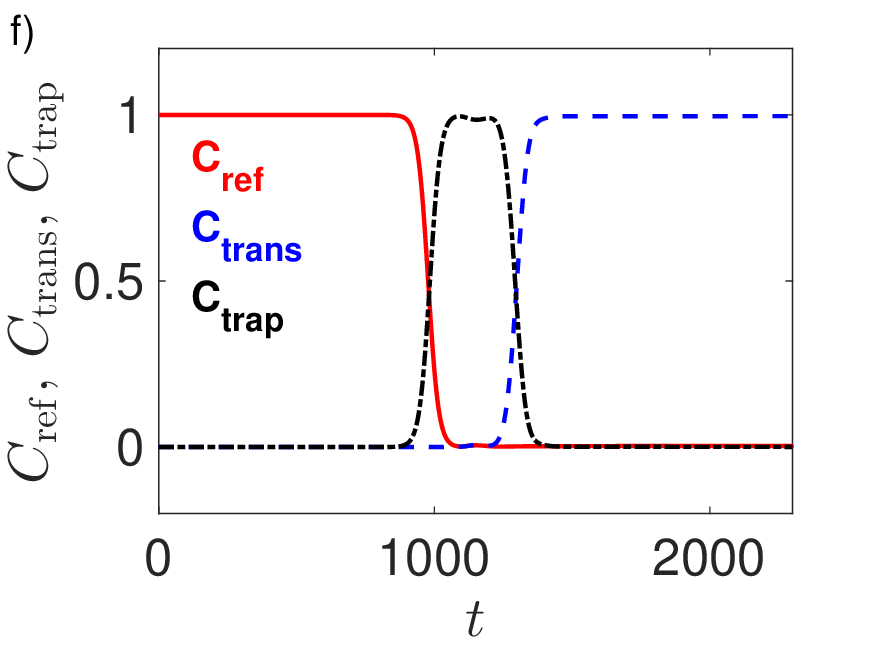}}
\caption{Scattering of a small QD ($N=1$) by a reflectionless PT potential well centred at $x=0$, shown for different incident velocities. The left column presents the spatiotemporal evolution of the droplet density, while the right column shows the corresponding reflection (red solid), transmission (blue dashed), and trapping (black dash-dotted) coefficients. Panels (a,d) correspond to $v=0.07$, panels (b,e) to the critical velocity, $v \simeq v_{\mathrm{cr}}=0.08349974$, and panels (c,f) to $v=0.09$. The colorbar represents the density $|\psi|^2$. The other parameters are $U_0=1$, $q=g=1$, and the initial droplet position is $x_0=-100$.}
\label{fig:dynSmallQDandCoefC}
\end{figure}
%

\begin{figure}[t]
\centerline{ \includegraphics[width=4.2cm]{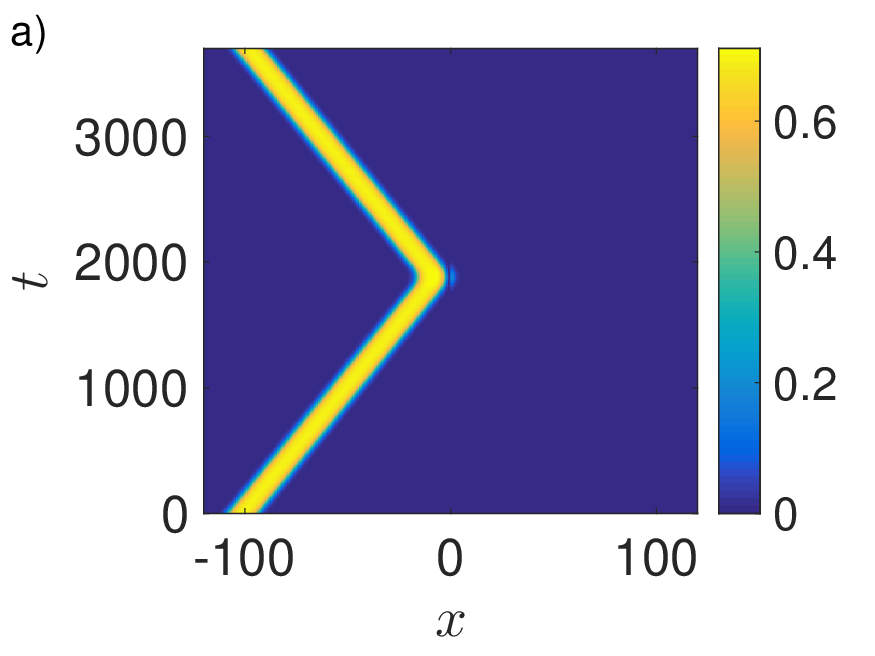} \hskip-0.1cm \includegraphics[width=4.4cm]{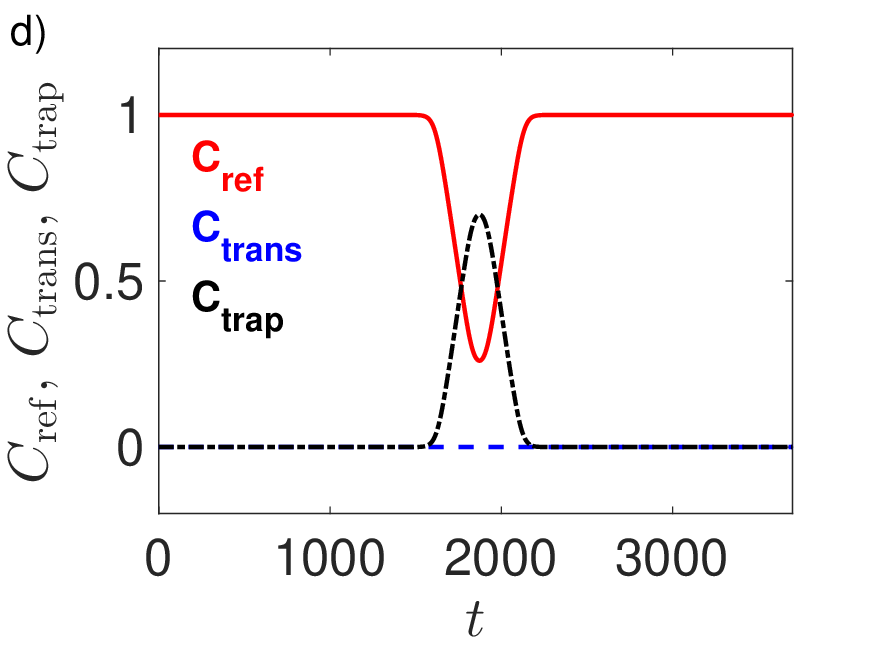}}
\centerline{ \includegraphics[width=4.2cm]{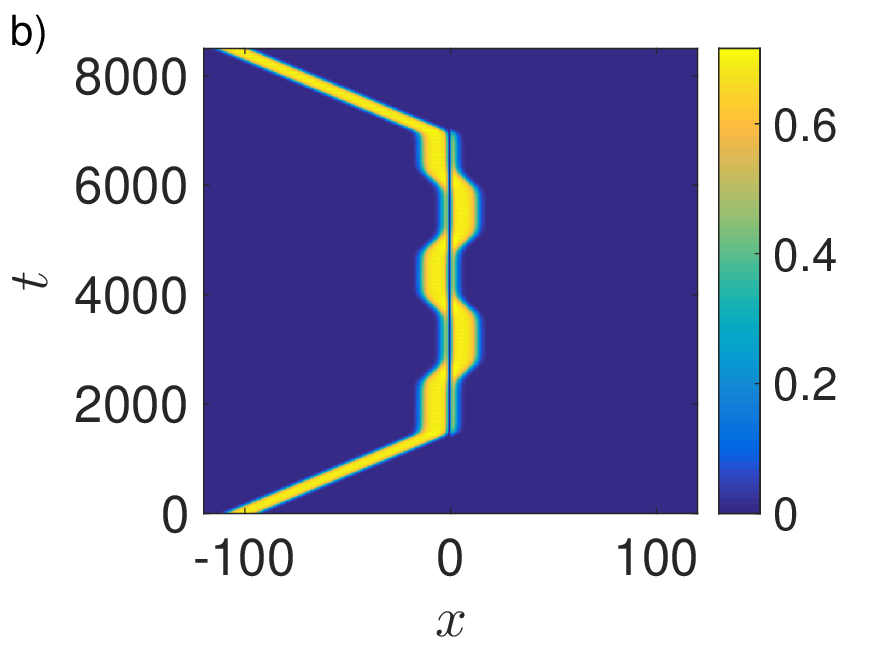} \hskip-0.1cm \includegraphics[width=4.4cm]{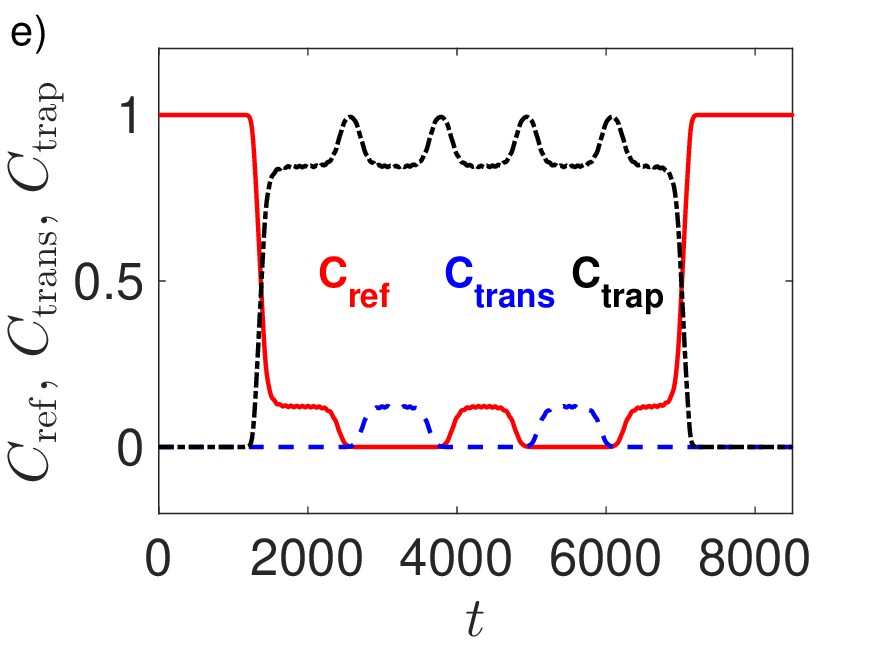}}
\centerline{ \includegraphics[width=4.2cm]{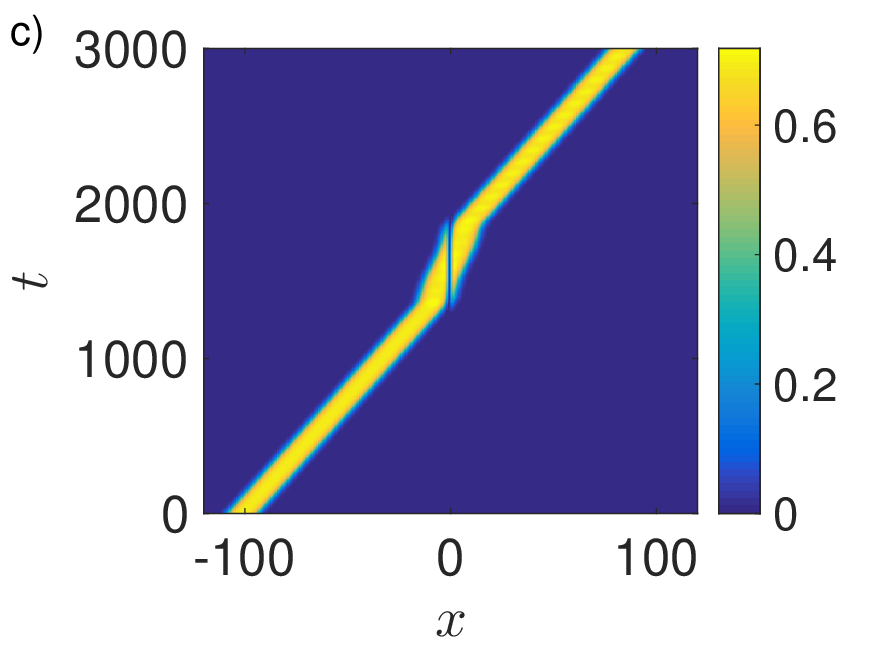} \hskip-0.1cm \includegraphics[width=4.4cm]{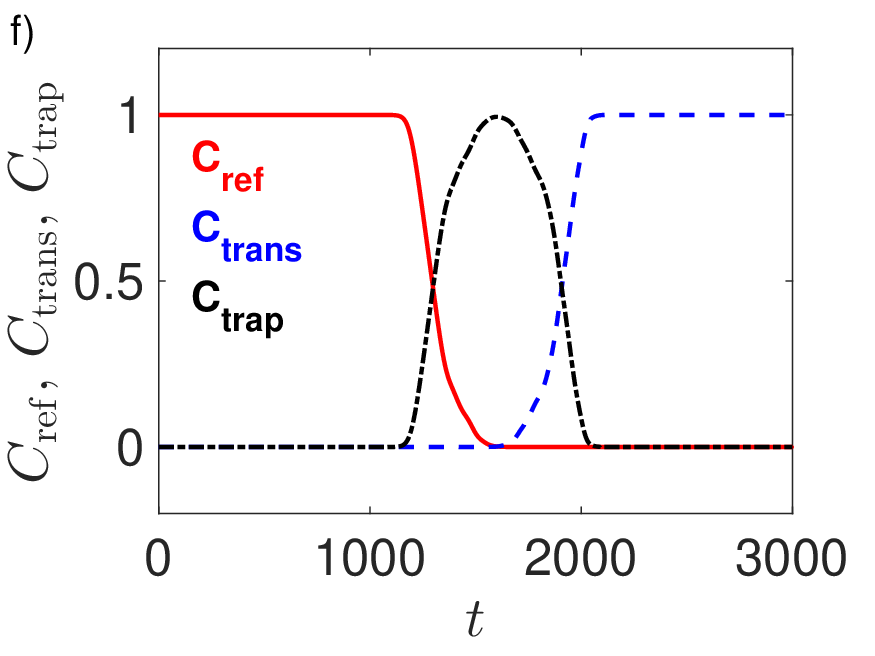}}
\caption{Dynamics of a large QD ($N=10$) scattering from a reflectionless PT potential well centred at $x=0$, shown for three representative incident velocities. The left column presents the spatiotemporal evolution of the droplet density, while the right column shows the corresponding reflection (red solid), transmission (blue dashed), and trapping (black dash-dotted) coefficients. Panels (a,d) correspond to $v=0.05$, panels (b,e) to the critical velocity, $v \simeq v_{\mathrm{cr}}=0.06425806$, and panels (c,f) to $v=0.068$. The colorbar represents the density $|\psi|^2$. In all cases, the remaining parameters are fixed at $U_0=1$ and $q=g=1$, and the droplet is initially located at $x_0=-100$.}
\label{fig:dynLargeQDandCoefC}
\end{figure}

The underlying dynamics are illustrated by the spatiotemporal density plots and by the corresponding time evolution of the reflection, trapping, and transmission coefficients. Two distinct scattering regimes are clearly identified. For small, bell-shaped droplets, representative density plots appear in Figs.~\ref{fig:dynSmallQDandCoefC}(a-c). The associated coefficient dynamics are shown in Figs.~\ref{fig:dynSmallQDandCoefC}(d-f). For large, flat-top droplets, the corresponding results are in Figs.~\ref{fig:dynLargeQDandCoefC}(a-c) and Figs.~\ref{fig:dynLargeQDandCoefC}(d-f), respectively.

For incident speeds below the critical value, the droplet undergoes nonclassical (quantum) reflection. This happens despite the well's attractive, reflectionless character in the linear limit. The critical speed is determined numerically by increasing the incident speed until the droplet spends the maximum time in the vicinity of the well, which signals the formation of a long-lived metastable turning-point state. For speeds above the critical value, the droplet becomes fully transmitted.

\begin{figure}[t]
\centerline{\includegraphics[width=4.47cm]{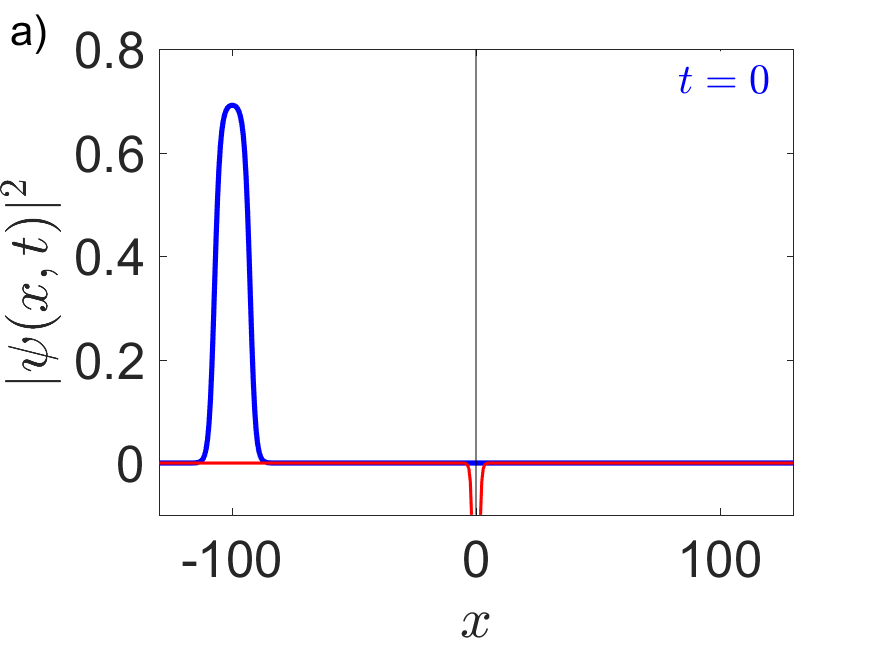} \includegraphics[width=4.47cm]{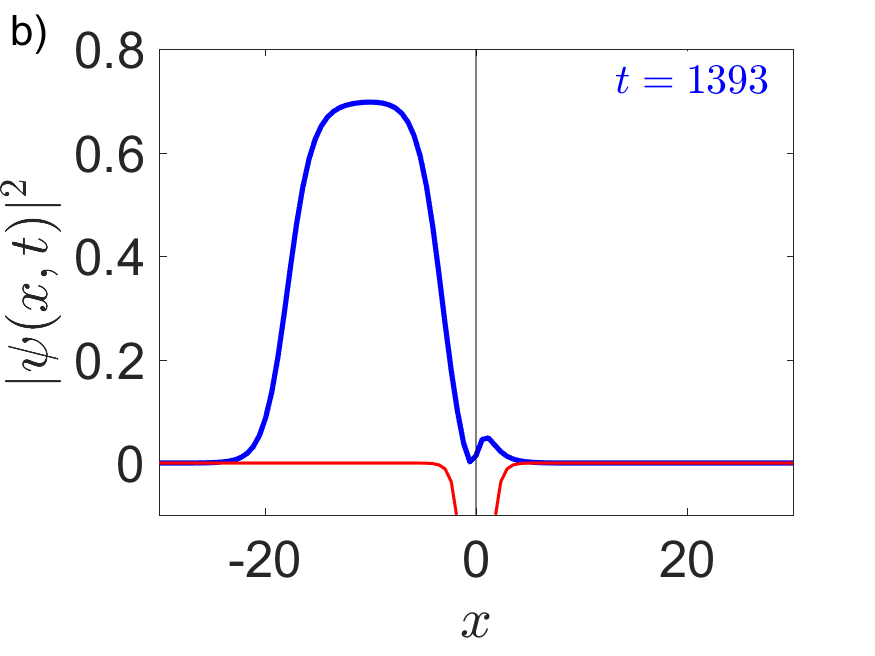}}
\centerline{\includegraphics[width=4.45cm]{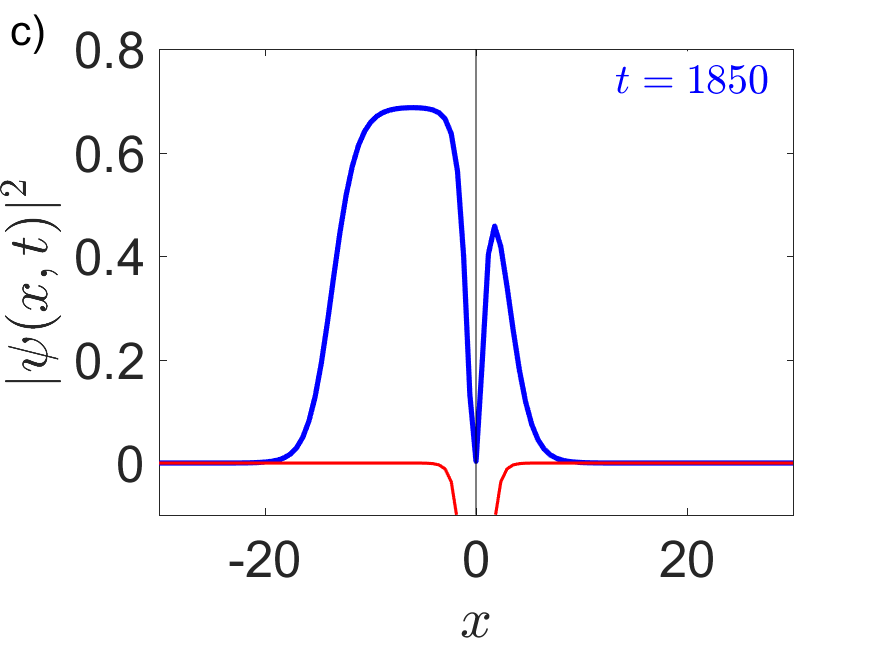} \includegraphics[width=4.47cm]{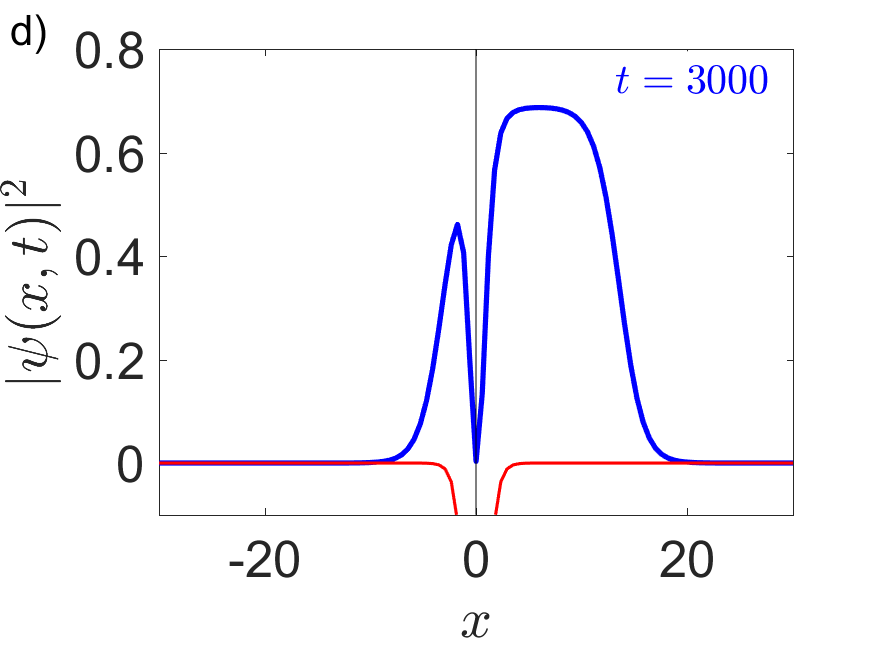}}
\centerline{\includegraphics[width=4.45cm]{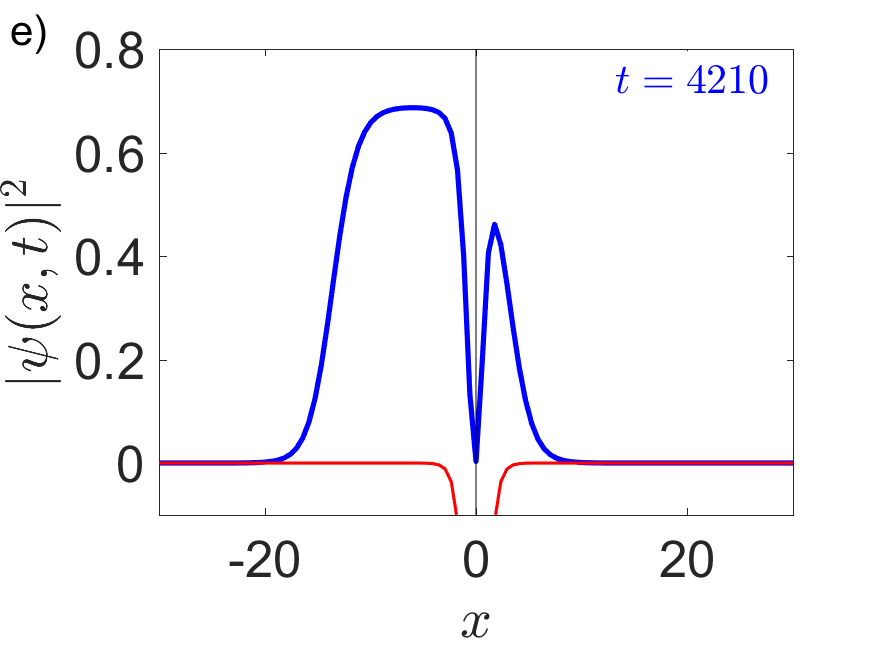} \includegraphics[width=4.47cm]{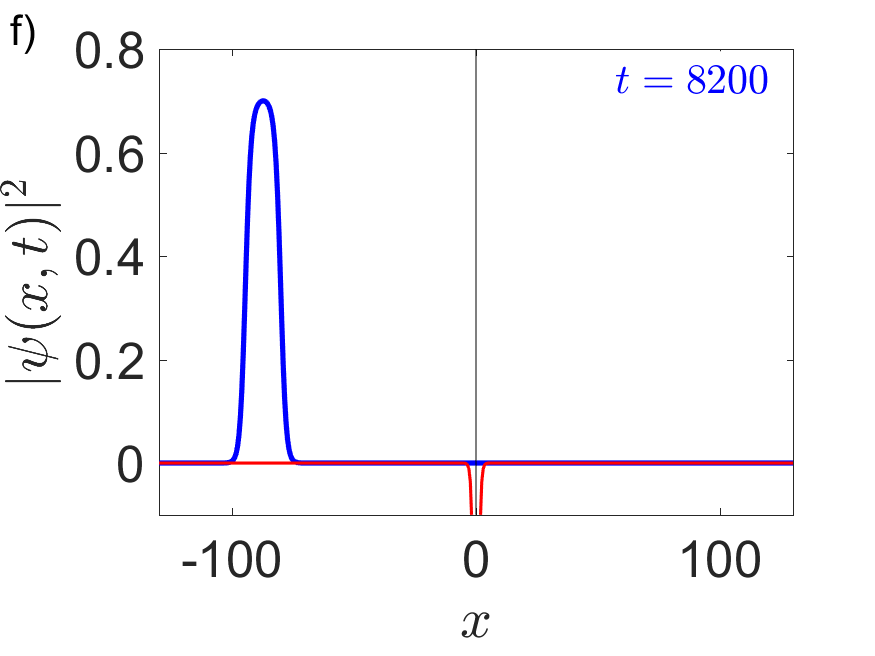}}
\caption{Transmission of the QD through the potential well at an incident velocity close to the critical value, $v \simeq v_{\mathrm{cr}}$. The density profile, $|\psi(x,t)|^2$, is shown at (a) $t=0$, (b) $t=1393$, (c) $t=1850$, (d) $t=3000$, (e) $t=4210$, and (f) $t=8200$. For clarity, only the upper part of the potential well is displayed (red curve). The other parameters are $N=10$, $U_0=1$, $q=g=1$, and $x_0=-100$.}
\label{fig-QD10CrossSec_abcdef}
\end{figure}

A key difference between small and large droplets is the structure of the transient trapped state. For small droplets, the trapped mode stays centred at the well and remains almost symmetric, see Fig.~\ref{fig:dynSmallQDandCoefC} (b). For large droplets, the trapped shape becomes asymmetric: the density peak moves away from $x=0$, and the droplet can switch between the left and right sides of the well before escaping, see Fig.~\ref{fig:dynLargeQDandCoefC}(b) for the density evolution and Fig.~\ref{fig-QD10CrossSec_abcdef} for the corresponding density cross sections at selected times. This asymmetry, not seen in small droplets, indicates that internal deformation and a flat-top density profile play a greater role in larger droplets.

Figure~\ref{fig-QD10CrossSec_abcdef} shows a sequence of density snapshots, $|\psi|^2$, taken at selected times from Fig.~\ref{fig:dynLargeQDandCoefC}(b). These snapshots illustrate the scattering dynamics of a large QD in the potential well. Panel (a) shows the initial droplet profile at $t=0$, when the droplet is still far from the potential and essentially unaffected by it. The first key transition occurs at $t=1393$ in panel (b), where the droplet has approached the well and is already strongly influenced by the potential. At the next major stage, $t=1850$ in panel (c), the droplet becomes trapped inside the well and develops a pronounced spatial asymmetry relative to the well centre: most of the density is concentrated on the left side, with a higher, flatter-top shape, while a smaller fraction remains on the right side in a lower-amplitude, more Gaussian-like form. After remaining in this trapped configuration for a considerable time, the density distribution undergoes repeated left-right rearrangements during subsequent evolution, as seen in panels (d)-(e). These oscillatory exchanges mark another key transition, reflecting the internal dynamics of the trapped state and its positional asymmetry within the well. Eventually, a final transition occurs as the droplet escapes from the potential region and re-emerges in an almost fully restored form, as shown in panel (f).

\subsection{Critical Modes in Droplet Scattering}
\label{subsec:CrModePTwell}

Quantum reflection shows that, during scattering, the droplet can briefly form a zero-velocity (``zero-speed") state. We define this as the droplet's configuration at the velocity turning point, where the centre-of-mass velocity vanishes, but the droplet remains a coherent, finite, localised wave packet with nonzero width and a definite density profile. Let $x_0$ denote the location of the density maximum of the zero-speed state, so the state is $\phi(x-x_0)$. Because the external PT well breaks translational invariance, the zero-speed state's energy depends on its position relative to the well. Shifting $x_0$ changes how the density profile overlaps the localised potential, thereby altering the potential energy and, through induced density deformation, the interaction energy as well.

To model such a state localised near an arbitrary position $x_0$, we introduce the following position-dependent variational ansatz,
\begin{equation}
\phi(x)=A\,\psi_0\!\bigl[\gamma (x-x_0)\bigr]\tanh(\beta x).
\label{eq:TrapModeTrialFunc}
\end{equation}
Here, $\psi_0$ denotes the stationary droplet profile in the absence of the external potential. The $\gamma$ and $\beta$ are variational parameters. The parameter $\gamma$ rescales the internal length scale of the droplet and controls the steepness of the central slope, while $\beta$ sets the spatial scale of the node introduced by the $\tanh(\beta x)$ factor and therefore controls the overall width of the localised mode induced by the defect. The parameter $x_0$ specifies the location of the zero-speed state. The factor $\tanh(\beta x)$ enforces a parity change across $x=0$, introduces a node at the defect centre, and allows the ansatz to capture the reflected or trapped configuration
generated during scattering. This choice is also motivated by the linear threshold structure of the $\ell=1$ reflectionless PT well. As shown in~\ref{subsec:linear_attractive_PT}, the corresponding zero-energy threshold solution is proportional to $\tanh(\alpha x)$. In the nonlinear droplet problem, $\beta$ is treated as a variational parameter because the
nodal length scale is modified by the finite droplet width and
by the cubic and quartic nonlinearities.
The anzatz function~(\ref{eq:TrapModeTrialFunc}) is designed for the lowest single-node trapped mode associated with the $\ell=1$ reflectionless PT well considered in this study. Multinode trapped states, which can arise in broader PT wells and have been examined in the context of nonlinear soliton scattering~\cite{Khawaja2021}, necessitate a more flexible ansatz incorporating additional nodal factors. A comprehensive analysis of these states will be addressed in future research. Similar variational forms have been employed to predict with high accuracy the critical speed separating reflection and transmission for bright solitons and droplets scattered from reflectionless PT potential wells, both for the cubic nonlinear GPE~\cite{Khawaja2021} and for the cubic-quadratic nonlinear GPE model~\cite{Hu}.

Using the normalization condition
$
N = \int\limits_{-\infty}^{+\infty}|\phi(x)|^2\,dx
$
with the trial function~(\ref{eq:TrapModeTrialFunc}) one can find the amplitude
\begin{equation}
A(\gamma,\beta,x_0) = \left(\frac{N}{I(\gamma,\beta,x_0)}\right)^{\!1/2},
\label{eq:A}
\end{equation}
where $I(\gamma,\beta,x_0) = \int\limits_{-\infty}^{+\infty}\left|\psi_0\!\bigl[\gamma (x-x_0)\bigr]\tanh(\beta x)\right|^2\,dx$.

The energy $E_z$ of the zero-speed (turning-point) state can be evaluated from the corresponding energy functional,
\begin{equation}
E_{z}(x_0) = \int\limits_{-\infty}^{+\infty}\left[\frac{1}{2}\,|\phi_x|^{2} - \frac{q}{2}\,|\phi|^{4} + \frac{2g}{5}\,|\phi|^{5} + V(x)\,|\phi|^{2}\right]\,dx ,
\label{eq:Ez}
\end{equation}
where $\phi(x)$ is the zero-speed state centered at $x_0$, $V(x)$ is the external PT well, and the nonlinear terms represent the attractive mean-field contribution and the repulsive beyond-mean-field correction that stabilizes the droplet.

Initially, the droplet is prepared far from the potential center, so that its overlap with $V(x)$ is negligible and the potential-energy contribution can be ignored. The initial droplet energy, computed from the same functional using the initial profile $\psi(x,0)$, is therefore
\begin{eqnarray}
E_{\rm d} &=& \int\limits_{-\infty}^{+\infty}\left[\frac{1}{2}|\psi_x|^{2}-\frac{q}{2}\,|\psi|^{4}
+\frac{2g}{5}\,|\psi|^{5}\right]\,dx \nonumber \\
&=& \frac{1}{2}Nv^{2} + E_{\rm sd}\,,
\label{eq:Ed}
\end{eqnarray}
where we have decomposed the initial state as a Galilean boost of the stationary droplet, $\psi(x,0)=\psi_0(x-x_0)\,e^{ivx}$ (up to an overall phase), so that the kinetic energy splits into the center-of-mass part $\frac{1}{2}Nv^2$ plus the internal energy of the stationary droplet $E_{\rm sd}$. The stationary droplet profile $\psi_0$ and its energy $E_{\rm sd}$ can be obtained from the variational approach developed in Ref.~\cite{Otajonov2024} using a super-Gaussian ansatz,
\begin{eqnarray}
E_{\rm sd} &=& \int\limits_{-\infty}^{+\infty}\left[\frac{1}{2}|(\psi_0)_x|^{2}
-\frac{q}{2}\,|\psi_0|^{4}+\frac{2g}{5}\,|\psi_0|^{5}\right]\,dx = \nonumber\\
&& \frac{N}{8a^2M}\frac{\Gamma(2-M)}{\Gamma(1+M)} - \frac{q N^2}{2^{M+2}a \Gamma(1+M)} + \nonumber \\
&& \frac{(2/5)^{M+1} g N^{5/2}}{(2a\Gamma(1+M))^{3/2}} \, .
\label{eq:Esd}
\end{eqnarray}

Assuming energy conservation during the scattering process,
the total initial energy, consisting of the stationary-droplet
energy and the centre-of-mass kinetic energy, equals the
turning-point energy. This allows one to express the incident velocity that produces a given turning-point energy as
$
v=[2\left(E_z-E_{\rm sd}\right)/N]^{1/2}.
$
As the incident velocity increases, the turning point generally shifts and the associated zero-speed configuration changes, leading to a larger turning-point energy $E_z(x_0)$. Quantum reflection persists up to the largest turning-point energy that can be supported, beyond this value the droplet is transmitted. Consequently, the critical velocity can be determined numerically by scanning $v$ and identifying the maximum attainable turning-point energy $(E_z)_{\max}$:
\begin{equation}
v_{\mathrm{cr}}=\sqrt{\frac{2\left[(E_z)_{\max}-E_{\rm sd}\right]}{N}}.
\label{eq:VelCR}
\end{equation}
Equation~(\ref{eq:VelCR}) therefore imply that droplets incident with different velocities generate distinct zero-speed states localized at different turning-point positions $x_0$ during the reflection process.
The zero-speed states introduced in this work should not be conflated with the linear bound states of the PT potential. The linear bound states are eigenstates of the single-particle linear Schr\"{o}dinger operator and are independent of the droplet norm or nonlinear interaction parameters. In contrast, $E_z(x_0)$ denotes the constrained nonlinear energy of a finite-norm droplet configuration, incorporating kinetic, attractive mean-field, repulsive beyond-mean-field, and external potential contributions. Thus, the zero-speed state characterises a nonlinear droplet configuration that forms at the velocity turning point during scattering.
The maximum of $E_z(x_0)$ corresponds to the highest accessible nonlinear turning-point state. This point marks the threshold at which the droplet transitions from reflection to transmission, thereby determining the critical incident velocity through energy conservation. This interpretation parallels the nonlinear trapped modes described in previous studies of soliton and droplet scattering from reflectionless PT wells~\cite{Hu,Khawaja2021}. However, it should not be regarded as the least-bound eigenstate of the linear PT spectrum. Lower-energy pinned configurations, when present, can be considered nonlinear bound states of the droplet in the well, but these do not determine the reflection-transmission threshold.

The energy of the zero-speed state is obtained by inserting the normalized variational ansatz~(\ref{eq:TrapModeTrialFunc}) into the energy functional~(\ref{eq:Ez}). Owing to the presence of the PT potential and the non-polynomial structure introduced by the $\tanh(\beta x)$ factor, the resulting integrals cannot, in general, be evaluated in the analytical form. Following the numerical variational procedure employed in Refs.~\cite{Hu,Khawaja2021}, we therefore compute both the normalization integral in Eq.~(\ref{eq:A}) and the corresponding integrals entering $E_z(x_0)$ numerically for given parameter sets.

In practice, we treat $\gamma$ and $\beta$ as variational parameters and evaluate the energy functional
$
E_z(\gamma,\beta;x_0)\,,
$
for fixed $x_0$. The numerical results show that, for the relevant parameter ranges, $E_z(\gamma,\beta;x_0)$ develops a well-defined local minimum at $(\gamma,\beta)=(\gamma^\ast,\beta^\ast)$. We interpret the minimized value
$
E_z(x_0)\equiv E_z(\gamma^\ast,\beta^\ast;x_0)
$
as the energy of the turning-point (zero-speed) configuration localized at position $x_0$.
Conservation of the energy then relates this turning-point energy to the incident velocity. Substituting $E_z(x_0)$ into
$
v=[2\left(E_z-E_{\rm sd}\right)/N]^{1/2},
$
yields the initial speed that produces a zero-speed state at the prescribed turning-point position $x_0$ during quantum reflection. Finally, by scanning $x_0$ and identifying the largest attainable turning-point energy,
$
E_T\equiv (E_z)_{\max}=\max_{x_0} E_z(x_0),
$
we obtain the critical velocity from Eq.~(\ref{eq:VelCR}) above which the turning-point state ceases to exist and transmission sets in.

\begin{figure}[t]
\centerline{\includegraphics[width=4.41cm]{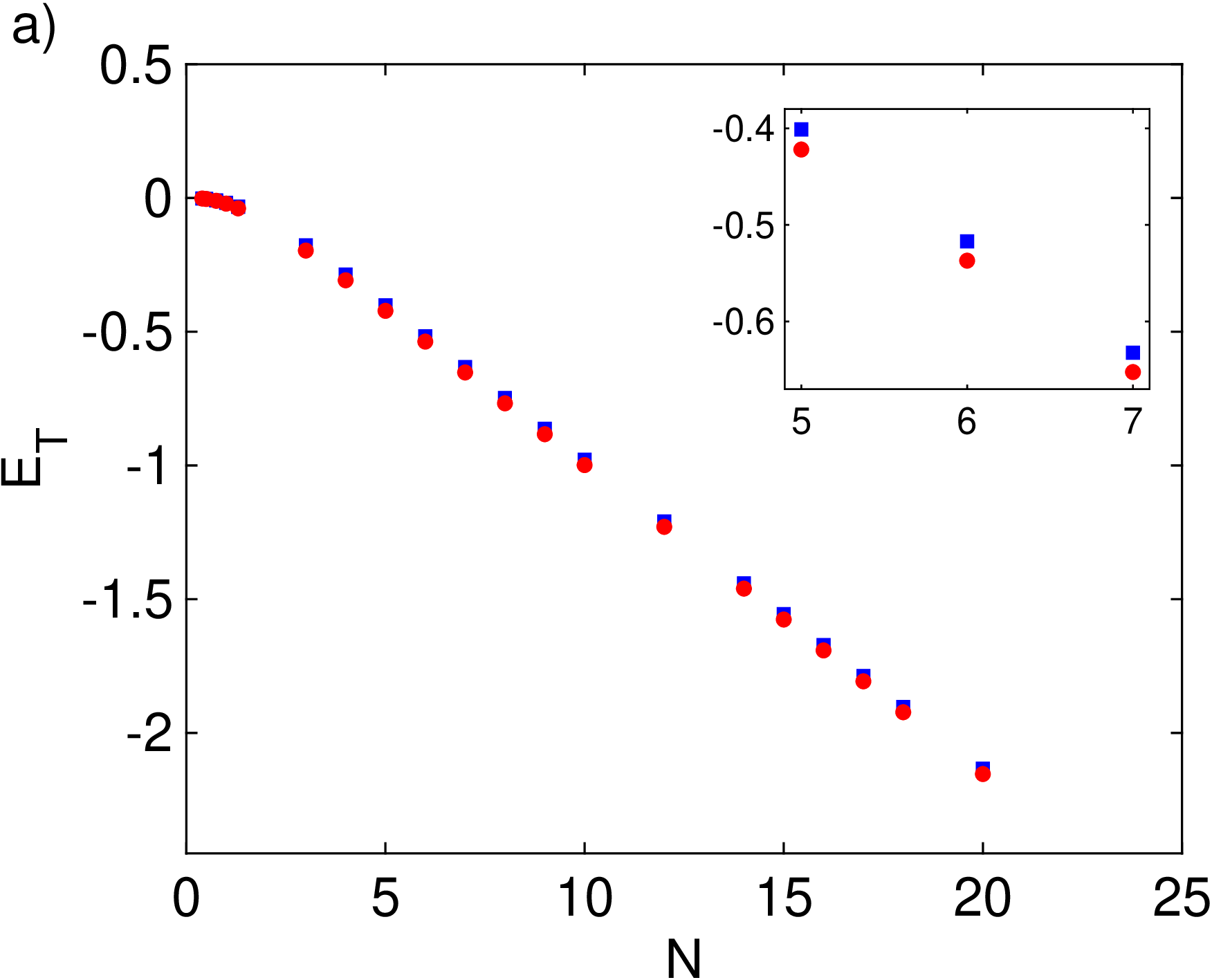} \hskip-0.1cm \includegraphics[width=4.41cm]{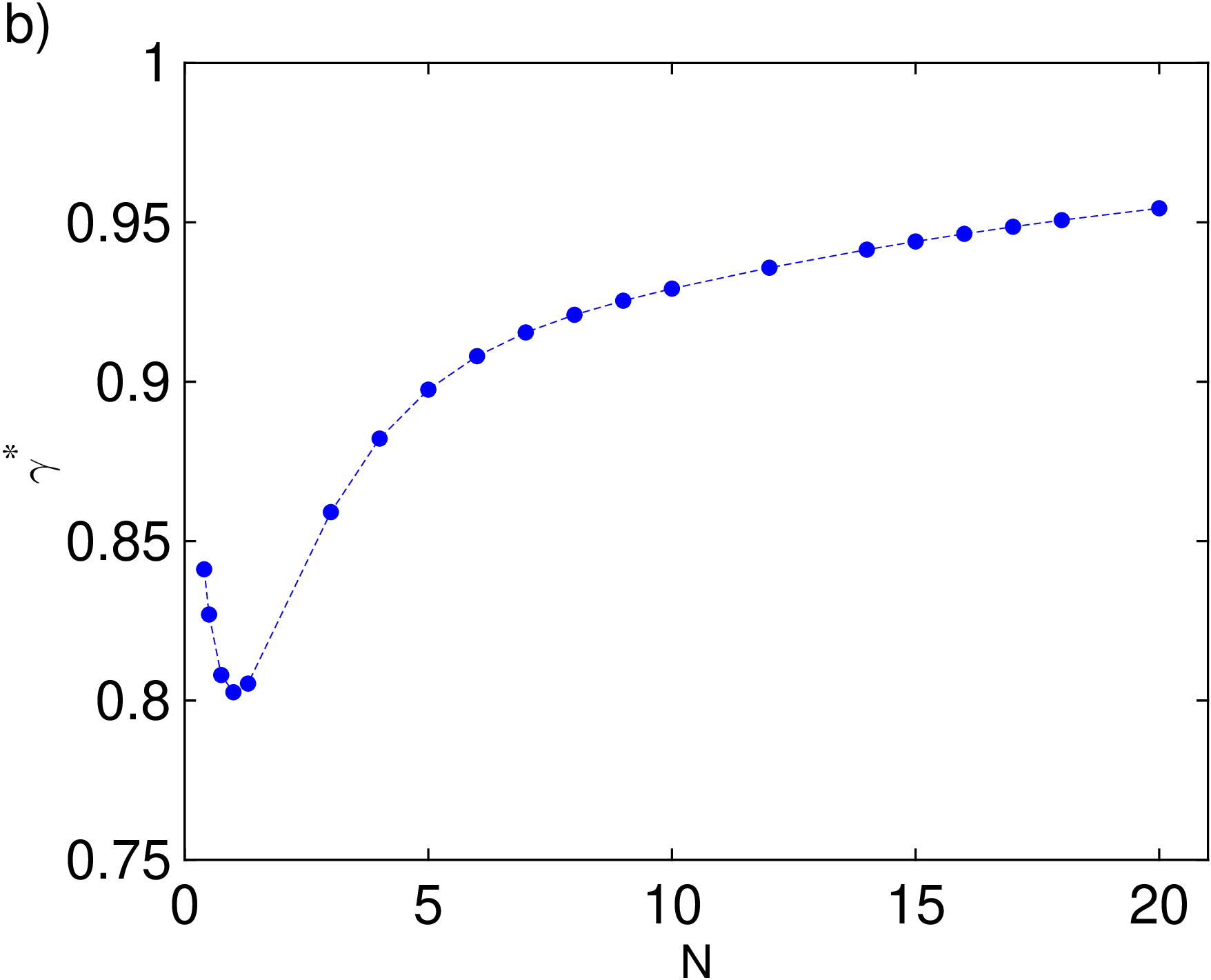}}
\centerline{\includegraphics[width=4.41cm]{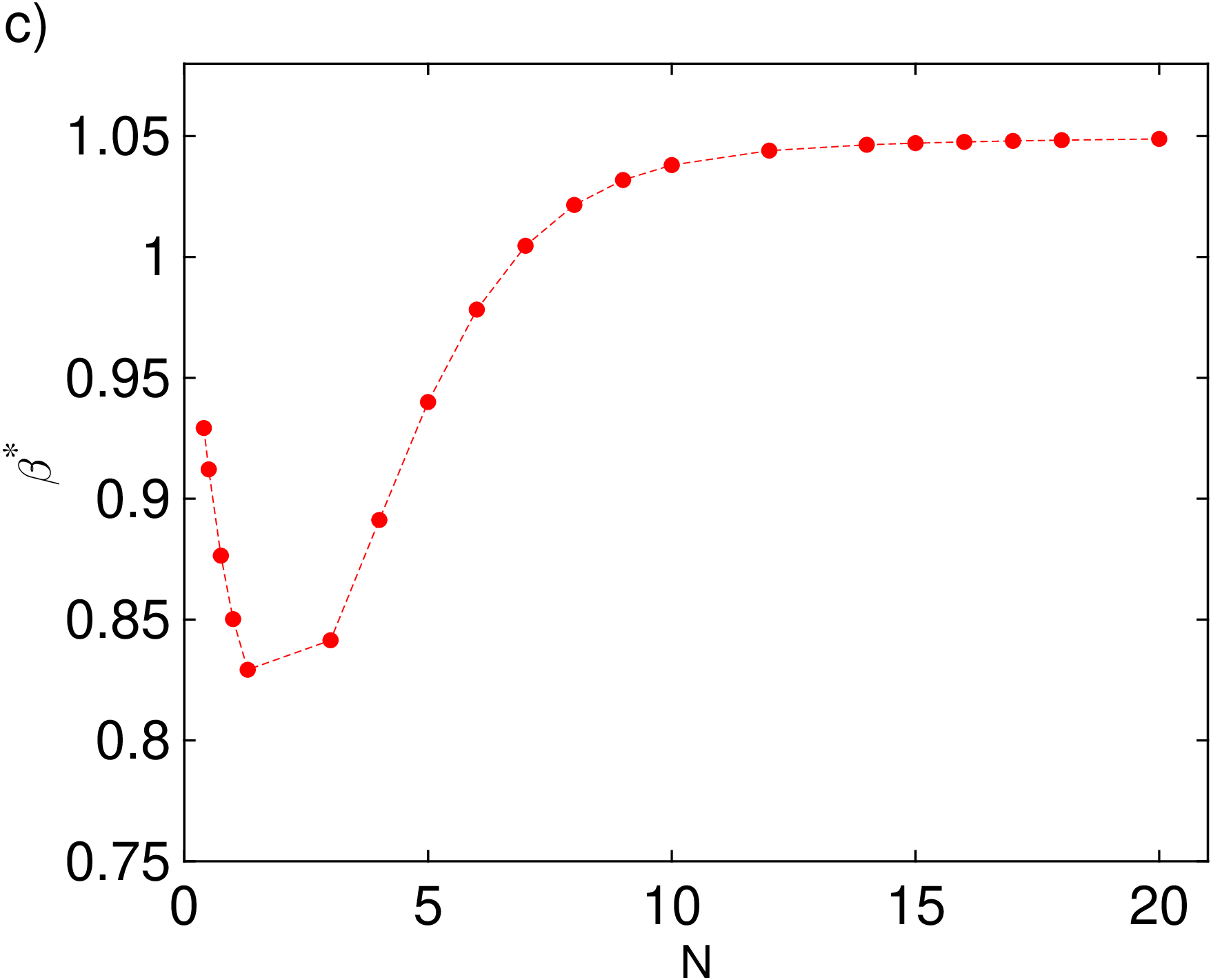} \hskip-0.1cm \includegraphics[width=4.65cm]{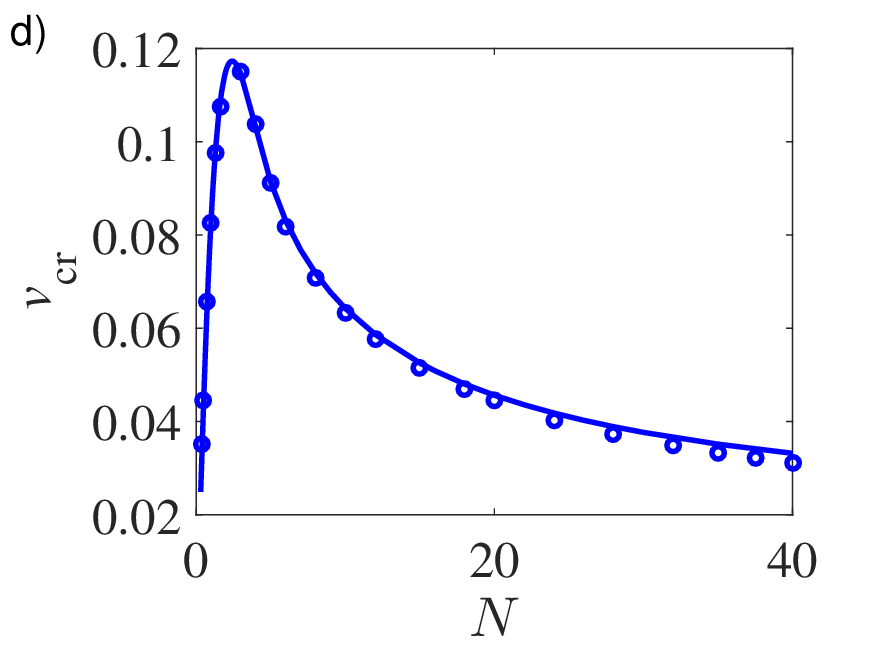}}
\caption{(a) Variational trapped mode energy $E_T$ (circle markers) together with the stationary energy of the incident QD, $E_{\mathrm{sd}}$ (square markers), as functions of the norm $N$. The inset shows a zoomed view of the main panel. (b,c) Optimized variational parameters $\gamma^\ast$ and $\beta^\ast$ corresponding to the minimum of $E_T(\gamma,\beta;x_0)$. (d) Critical velocity $v_{\mathrm{cr}}$ versus norm $N$: the solid curve denotes the variational prediction, while the points show the results of direct numerical simulations of Eq.~(\ref{eq:gpe}).}
\label{fig:EtVcrGamDelVSnorm}
\end{figure}
In the variational approach, the key mode is identified by finding the peak of the minimized turning-point energy $E_z(\gamma^*,\beta^*;x_0)$ as a function of $x_0$. This peak corresponds to the maximum energy at zero velocity in scattering and marks the boundary between reflection and transmission, thereby determining the critical incident velocity.
Building on this approach, Figure~\ref{fig:EtVcrGamDelVSnorm}(a) shows the maximum trapped-mode energy $\max_{x_0}E_z(\gamma^*,\beta^*;x_0)$ (circles) and the stationary droplet energy $E_{\rm sd}$ (squares) as functions of particle number $N$. Both energies are plotted against $N$. In both the main panel and inset, the maximum trapped-mode energy remains above the stationary droplet energy for all values considered. Therefore, $\max_{x_0}E_z(\gamma^*,\beta^*;x_0) - E_{\rm sd}>0$, so a real-valued critical velocity follows from Eq.~(\ref{eq:VelCR}).

Figures~\ref{fig:EtVcrGamDelVSnorm}(b,c) show the optimized variational parameters $\gamma^\ast$ and $\beta^\ast$ versus $N$. These curves indicate a characteristic threshold around $N_{\mathrm{th}}\simeq 2.3$. This threshold marks the critical point where the droplet changes from a compressible, bell-shaped regime (below $N_{\mathrm{th}}$) to an incompressible, flat-top regime (above $N_{\mathrm{th}}$). Below $N_{\mathrm{th}}$, increasing $N$ compresses and shrinks the droplet. Above $N_{\mathrm{th}}$, increasing $N$ grows the droplet's size but not its density.
This behavior is consistent with the crossover reported in Ref.~\cite{Otajonov2024} for the same system in the absence of an external potential. In that work, the stationary droplet parameters as functions of the norm were identified. A clear minimum in droplet width occurs at approximately $N_{\mathrm{th}} \simeq 2.3$. For small norms $0<N\leq N_{\mathrm{th}}$, the width decreases with increasing $N$, reflecting the compressible nature of the droplet. When the norm lies in the interval $N_{\mathrm{th}}<N\leq N_{\mathrm{s}}\simeq 8$, the width begins to increase, while amplitude and peak density rapidly approach their saturation values. The relation between the saturation amplitude and the saturation density is given by $A_{\mathrm{s}} = \sqrt{n_{\mathrm{s}}}$. This interval can be considered a crossover region. Here, the initially bell-shaped profile gradually transitions to a flat-top profile, although the density is not yet fully saturated. For $N\gtrsim N_{\mathrm{s}}$, the peak density becomes practically saturated, indicating the formation of an incompressible, liquid-like flat-top droplet.
This classification is adopted in the present study to distinguish between small and large droplets. Small droplets correspond to the compressible bell-shaped regime, defined by $N<N_{\mathrm{th}}$. Large droplets correspond to the saturated flat-top regime, defined by $N\gtrsim N_{\mathrm{s}}\simeq 8$. Droplets within the intermediate crossover region display mixed characteristics. When the norm is near $N_{\mathrm{th}}$, their dynamics are similar to those of small compressible droplets. Conversely, when the norm approaches $N_{\mathrm{s}}$, their dynamics resemble those of large flat-top droplets. In that case, the density is nearly saturated, and further increases in $N$ primarily affect the droplet width.

The same threshold also appears in the small-amplitude dynamics around the stationary droplet state. As shown in Ref.~\cite{Otajonov2024}, the dependence of the oscillation frequency, or equivalently the oscillation period, on $N$ contains turning points close to $N_{\mathrm{th}}\simeq 2.3$, indicating a change in the dynamical response of the droplet. In the present study, where the scattering of a QD by a PT potential well is considered, the turning point in the critical velocity as a function of $N$ can therefore be interpreted as another manifestation of the same compressible-to-incompressible crossover. Related crossover behavior has also been observed in higher-dimensional QDs stabilized by quantum fluctuations~\cite{Otajonov2020, Otajonov2022}.

The dependence of the critical velocity on particle number appears in Fig.~\ref{fig:EtVcrGamDelVSnorm}(d). The line shows variational results, while the points display the simulation data. The curve is distinctly nonmonotonic, with a maximum critical speed near $N\simeq N_{\mathrm{th}}\simeq 2.3$. For small, compressible droplets ($N\lesssim N_{\mathrm{th}}$), $v_{\rm cr}$ increases with $N$. This behaviour is similar to bright solitons. However, once the droplet enters the flat-top, incompressible regime ($N\gtrsim N_{\mathrm{th}}$), the trend reverses, $v_{\rm cr}$ decreases as $N$ increases.
The post-peak behavior is explained by Eq.~(\ref{eq:VelCR}). Within the flat-top regime, the energy difference $\Delta E=(E_z)_{\max}-E_{\rm sd}$ exhibits only a weak dependence on $N$ and remains approximately $\Delta E\simeq 0.02$ across the parameter range considered. Substituting this value into Eq.~(\ref{eq:VelCR}) yields
$$
v_{\rm cr}\simeq \left(\frac{2\Delta E}{N}\right)^{1/2}
\simeq 0.2N^{-1/2}.
$$
Therefore, the observed decrease in critical velocity within the incompressible regime is primarily attributed to the increase in the droplets effective inertial mass, which is proportional to $N$. The energy excess required to form the critical trapped state remains nearly constant. Fitting the post-peak segment of the $v_{\rm cr}(N)$ curve yields the same inverse-square-root dependence. Consequently, $v_{\rm cr}\simeq 0.2N^{-1/2}$ serves as a practical empirical approximation for the flat-top regime. This reversal characterises QDs rather than 1D solitons. It arises from a change in the stationary size scaling: small droplets become narrower as $N$ increases, whereas large droplets grow. Overall, Fig.~\ref{fig:EtVcrGamDelVSnorm}(d) demonstrates that the variational approach reliably captures the critical velocity well over large intervals of $N$.

\begin{figure}[t]
\centerline{\includegraphics[width=4.51cm]{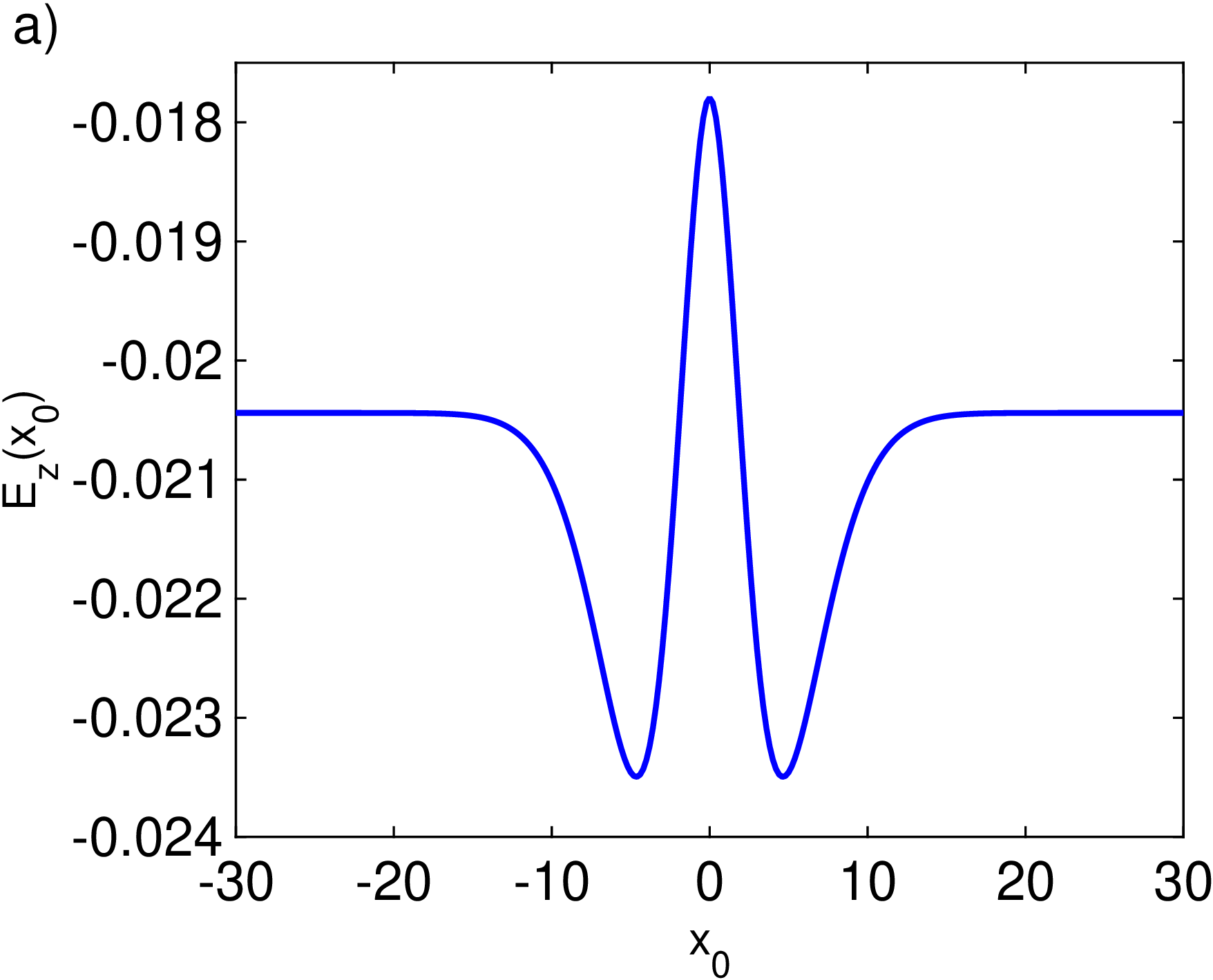} \hskip-0.1cm \includegraphics[width=4.51cm]{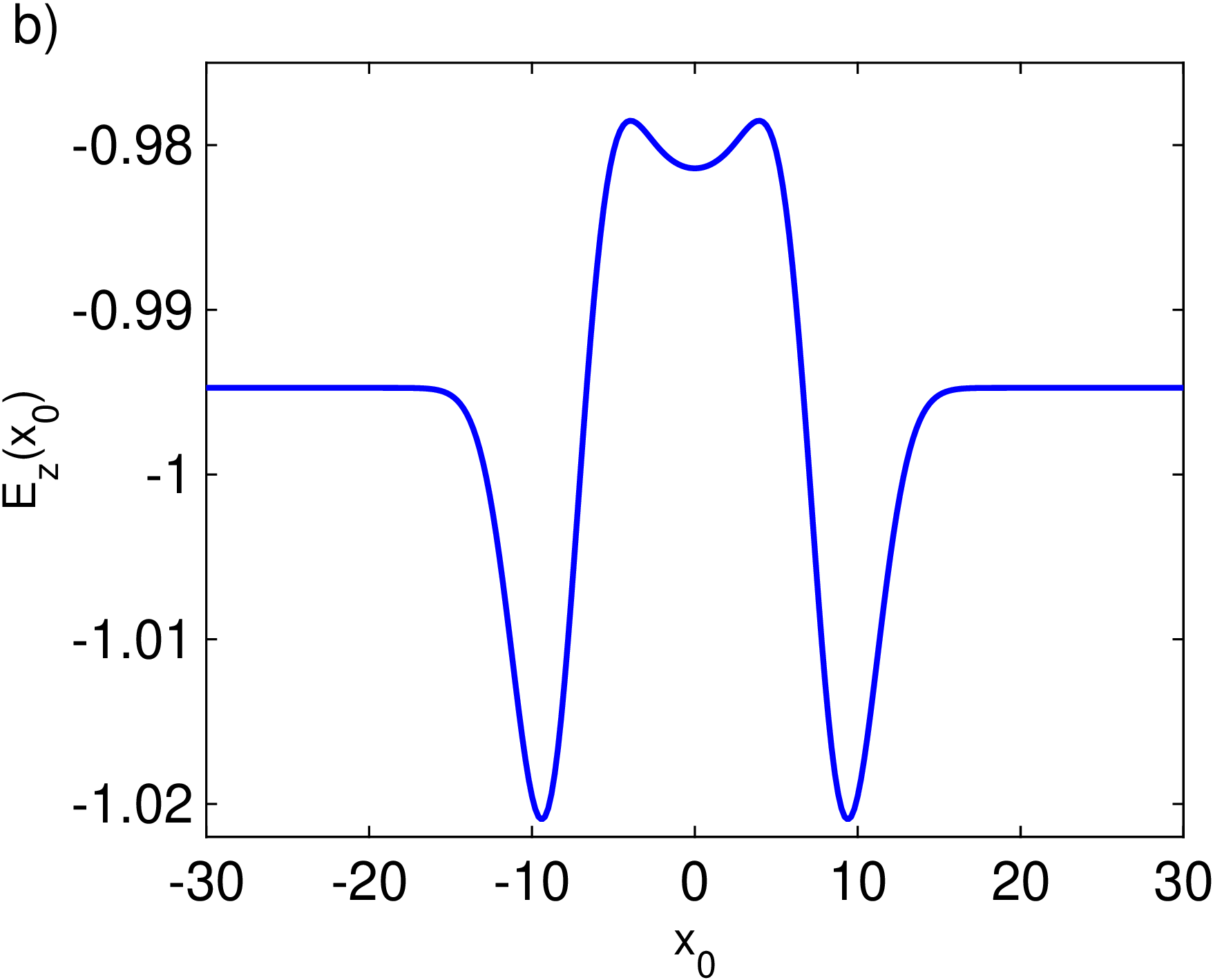}}
\caption{Variationally obtained zero-speed-state energy $E_z(x_0)$ versus position $x_0$ using the position-dependent ansatz~(\ref{eq:TrapModeTrialFunc}). Panel (a) shows a small droplet ($N=1$) and panel (b) a large droplet ($N=10$), with $U_0=1$ and $q=g=1$. The maxima of $E_z(x_0)$ mark metastable trapped (critical) modes in the scattering dynamics: for $N=1$ the peak at $x_0=0$ corresponds to a balanced trapped state, whereas for $N=10$ the curve is left-right symmetric and exhibits two symmetric peaks associated with imbalanced trapped states. Other parameters are fixed as $U_0=1$, $q=g=1$. }
\label{fig:EzVSx0}
\end{figure}
In Fig.~\ref{fig:EzVSx0} we present the variationally obtained turning point (zero-speed) energy as a function of the turning point position $x_0$. For each fixed $x_0$, we minimize the energy functional~(\ref{eq:Ez}) with respect to the variational parameters $\gamma$ and $\beta$. This uses the position-dependent ansatz~(\ref{eq:TrapModeTrialFunc}). The result is the minimized energy $E_z(x_0)\equiv E_z(\gamma^\ast,\beta^\ast;x_0)$. The resulting curves exhibit pronounced extrema. In particular, $E_z(x_0)$ attains a local maximum at specific values of $x_0$. This means that the corresponding zero-speed configuration is unstable with respect to small shifts of its position. The potential energy has negative curvature along the $x_0$ direction at the maximum.

This observation explains near-critical scattering dynamics and different dynamical regimes. When the incident droplet velocity is near the critical value ($v \simeq v_{\mathrm{cr}}$), the droplet's behaviour shifts. Instead of direct transmission or reflection, it forms a transient zero-speed state at the critical extremum of $E_z(x_0)$. In this regime, where the droplet energy is almost equal to the transmission threshold, the dynamics of $x_0$ slow sharply. The droplet stays near the potential for a long time before eventually reflecting or transmitting. As $v \to v_{\mathrm{cr}}$, the trapping time increases. This signals a transition to a metastable (critical) state, as discussed in soliton scattering studies~\cite{Hu,Khawaja2021}. In the absence of external perturbations, these trapped configurations persist for extended periods and may be considered nonlinear quasi-eigenstates of the droplet-well system. To evaluate the robustness of the critical dynamics, simulations were repeated with the addition of a small random perturbation to the initial droplet profile, with a relative amplitude in the range $10^{-4}$ to $10^{-3}$. The system's response is highly dependent on the droplet regime. For small, compressible droplets, the near-critical dynamics remain robust, and perturbed droplets predominantly reflect at the critical velocity. In contrast, for larger droplets, the critical state exhibits greater sensitivity to perturbations. In this regime, small random fluctuations can induce either reflection or transmission and reduce the residence time near the trapped configuration. This increased sensitivity aligns with the metastable character of the nonlinear zero-speed state at the reflection-transmission threshold.
The metastable states observed at $v\simeq v_{\rm cr}$ can be interpreted as nonlinear trapped-mode resonances. These states are distinct from linear PT resonances, as they correspond to finite-norm nonlinear droplet configurations whose energy is determined by the constrained functional $E_z(x_0)$. Dynamically, such resonances are characterized by a pronounced increase in the residence time near the well and a sustained plateau in the trapped fraction. This behavior is analogous to resonant trapping mechanisms previously reported for soliton and droplet scattering from attractive PT wells~\cite{Hu,Khawaja2021,Brand2010, Hansen}.

A key difference between small and large droplets emerges from the location of the maxima of the turning-point energy $E_z(x_0)$. For a small droplet, see Fig.~\ref{fig:EzVSx0}(a), the maximum occurs at the centre of the potential, $x_0=0$, identifying a spatially symmetric critical (metastable trapped) mode. In contrast, for a large droplet, see Fig.~\ref{fig:EzVSx0}(b), the global maxima are displaced from the origin, implying that near the critical velocity the scattering dynamics can transiently populate a metastable trapped configuration with an asymmetric density profile ($x_0\neq 0$). Because the PT well is even, $V(x)=V(-x)$, the variational curve $E_z(x_0)$ is left-right symmetric and the displaced maxima appear as a pair at $x_0=\pm x_m$, consistent with the fact that the droplet may be incident from either side of the well. This two-maximum structure also predicts the behaviour of the effective centre-of-mass potential discussed in the subsequent analysis. For a larger droplet, interaction with the well is mainly determined by the overlap between the localised potential and the droplet's two edges. As a result, the two off-centre maxima of $E_z(x_0)$ correspond to the static variational analogue of the two shoulders that later emerge in the dynamically extracted effective potential $V_{\mathrm{eff}}.$ This qualitative change from a single symmetric critical state for small droplets to asymmetric critical states for large droplets contrasts with the usual 1D bright-soliton scenario, where the critical trapped mode is typically centred at $x_0=0$.

The variational approximation includes turning point states with energies below the stationary droplet energy $E_{sd}$, visible as dips in Fig.~\ref{fig:EzVSx0}. These low-energy configurations are not accessed during our scattering dynamics. The incoming droplet carries more energy than these states allow, so the evolution does not settle into them. For the large droplet, we find a local minimum at $x_0=0$, representing a stable, symmetric trapped mode. Its energy is lower than that of the asymmetric metastable state at the global maxima. Thus, at the critical velocity, scattering dynamics favour excitation of the higher-energy asymmetric metastable configuration (the critical trapped mode) rather than the lower-energy symmetric trapped mode. For velocities above $v_{\mathrm{cr}}$, the incident energy enables transmission through the well, so the droplet is not trapped in the symmetric minimum either.

\begin{figure}[t]
\centerline{\includegraphics[height=3.21cm]{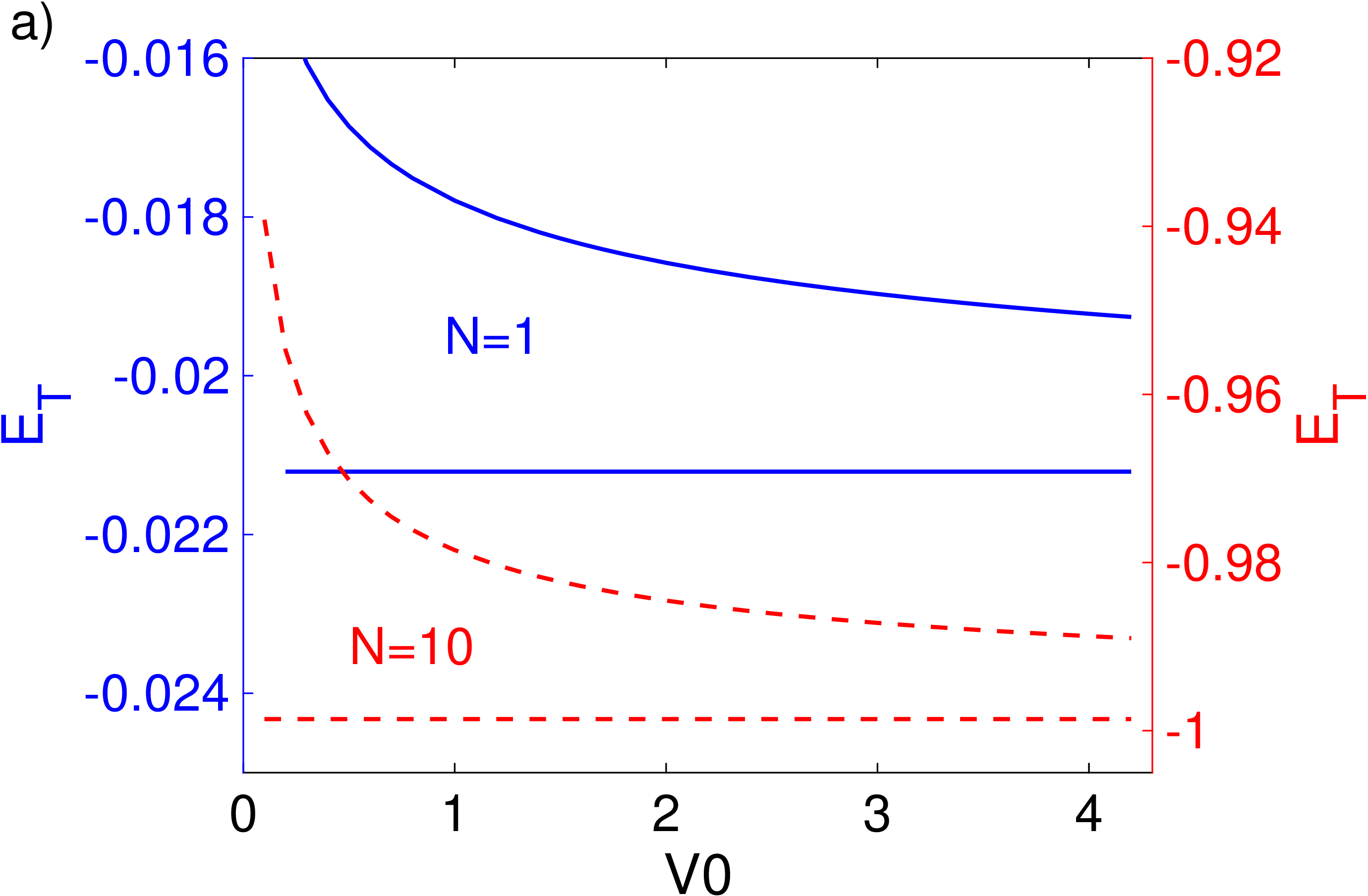} \hskip-0.1cm \includegraphics[width=4.1cm]{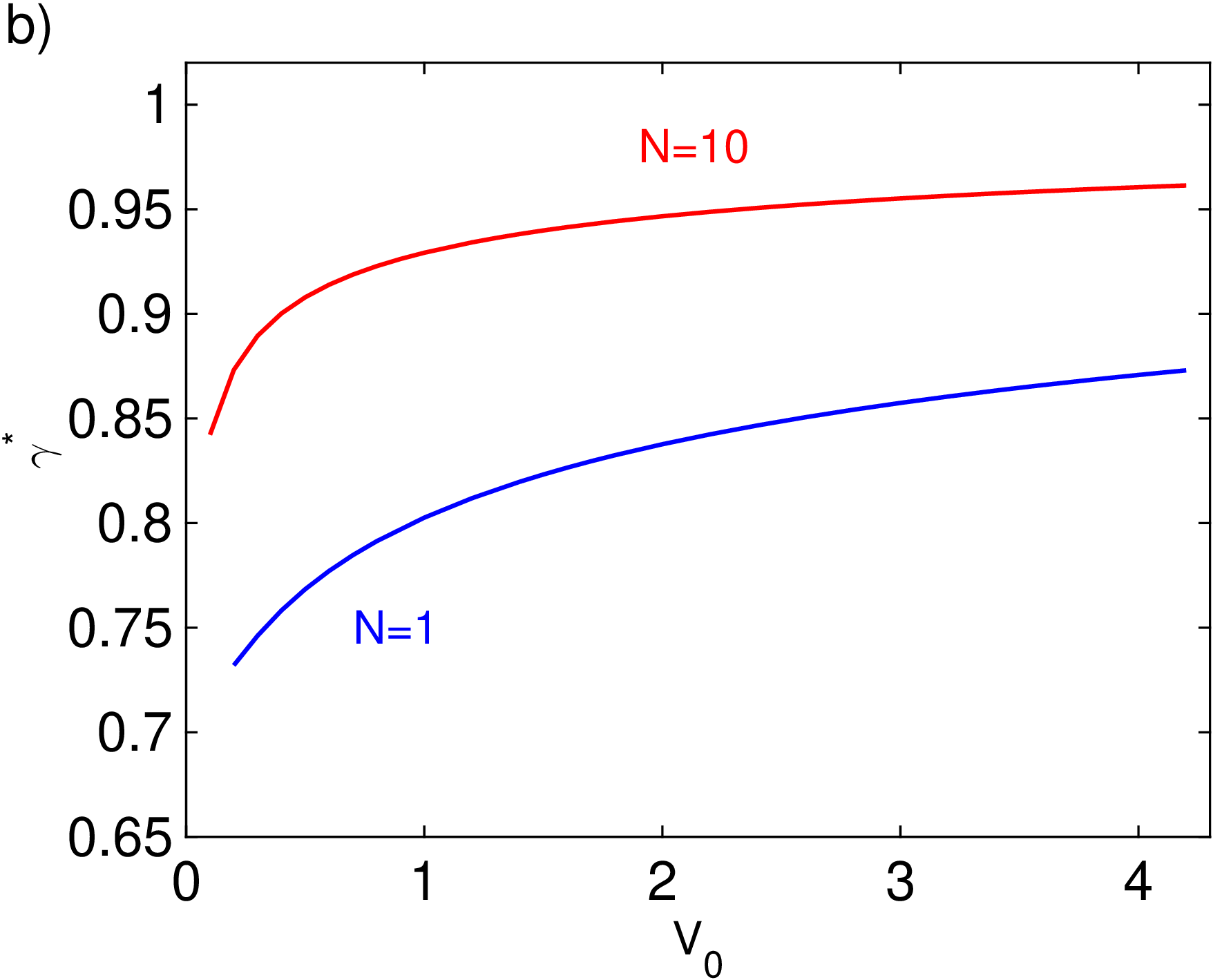}}
\centerline{\includegraphics[width=4.15cm]{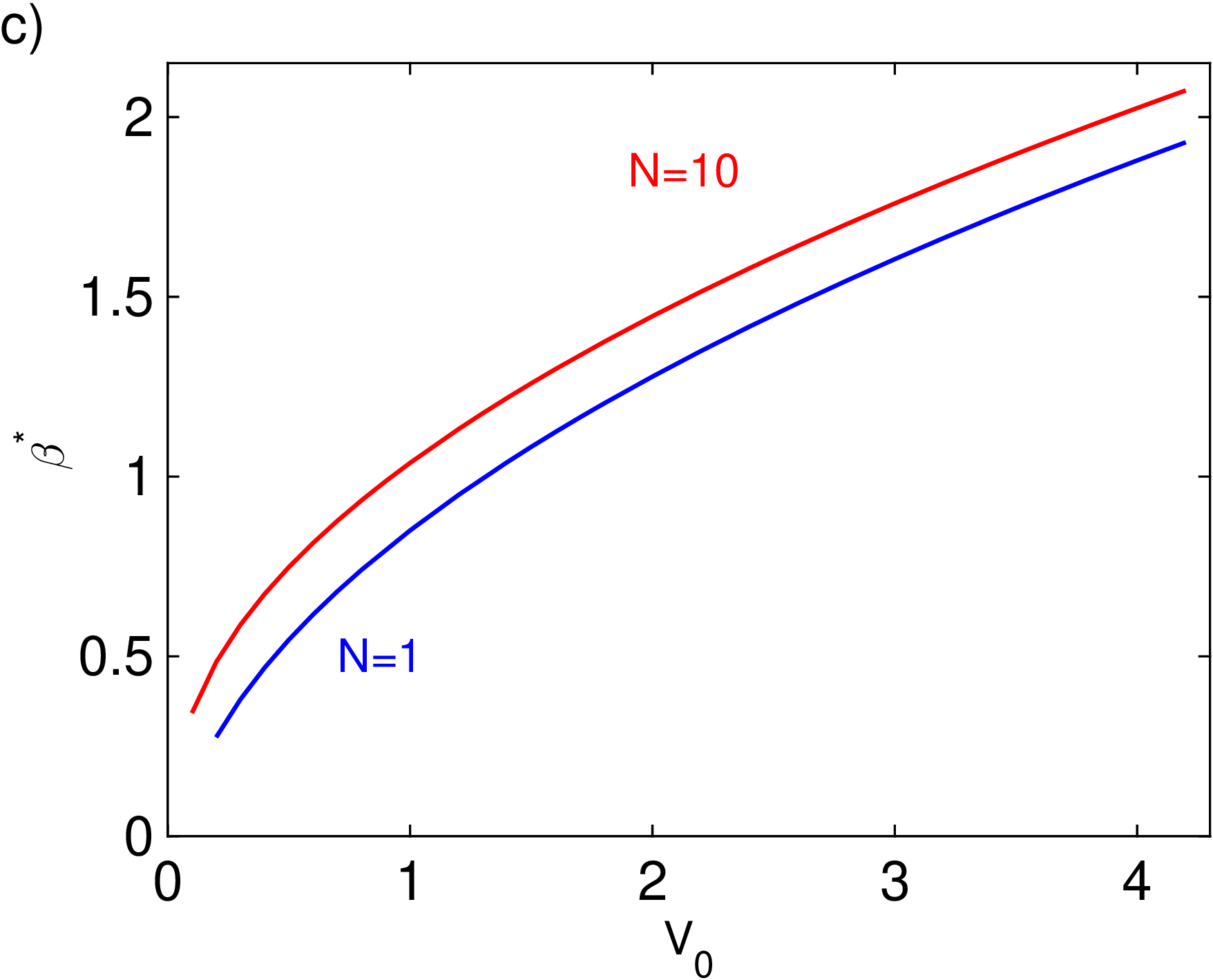} \hskip+0.15cm \includegraphics[width=4.15cm]{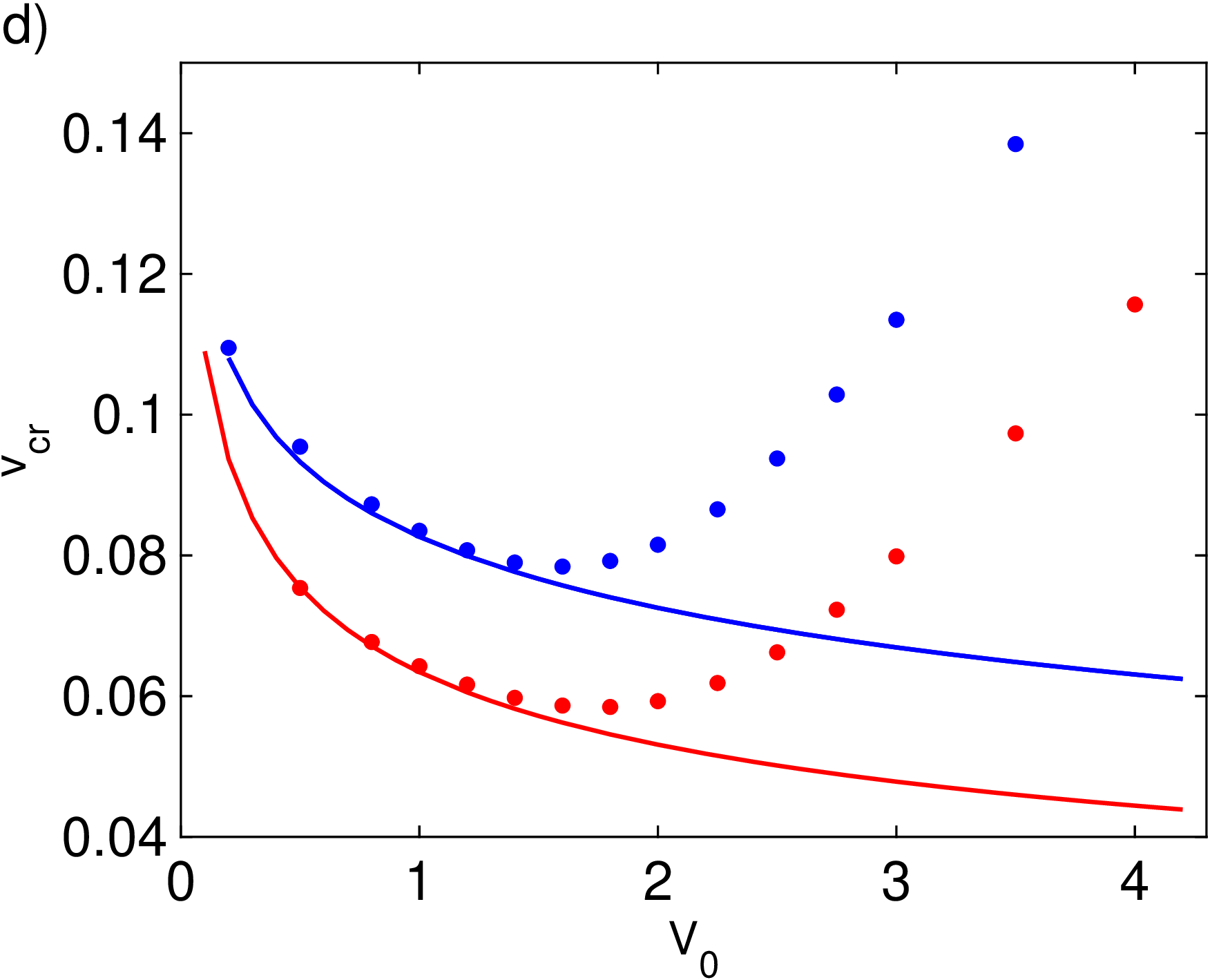}}
\caption{(a) Variational trapped mode energy $E_T$ and stationary energy of the incident QD, $E_{\mathrm{sd}}$, shown as functions of the potential well depth $U_0$ for two different atom numbers $N$. The upper blue solid curve gives $E_T$ for the small, bell-shaped droplet ($N=1$), while the horizontal blue line just below it shows the corresponding $E_{\mathrm{sd}}$. The red dashed curve shows $E_T$ for the larger flat-top droplet ($N=10$), and the lower red horizontal dashed line gives the corresponding $E_{\mathrm{sd}}$.
(b, c) Optimized variational parameters $\gamma^\ast$ and $\beta^\ast$ versus $U_0$ for $N=1$ and $N=10$ (the lower curves correspond to $N=1$, the upper curves to $N=10$).
(d) Critical velocity $v_{\mathrm{cr}}$ as a function of $U_0$. Solid lines denote the variational prediction, while points show direct numerical simulations of Eq.~(\ref{eq:gpe}). The upper blue data set corresponds to the $N=1$ droplet and the lower red data set to the $N=10$ droplet. Other parameters are fixed as $U_0=1$, $q=g=1$, with the droplet initially located at $x_0=-100$.}
\label{fig:EtVcrGamDelVSV0}
\end{figure}
Figure~\ref{fig:EtVcrGamDelVSV0}(a) shows $E_T$ as a function of well depth $U_0$ for two atom numbers. The upper curves are for a small droplet ($N=1$, left axis), and the lower curves are for a larger droplet ($N=10$, right axis). For each, the stationary energy $E_{\mathrm{sd}}$ appears as a horizontal line (blue solid for $N=1$, red dashed for $N=10$), which remains below $E_T(U_0)$. The difference between $E_T$ and $E_{\mathrm{sd}}$ represents the additional kinetic energy required for the droplet to reach the trapped state at the turning point.
Transitioning from energy considerations in panel (a), panels (b) and (c) illustrate how the variational parameters $\gamma^\ast$ and $\beta^\ast$ evolve with $U_0$, comparing $N=1$ (lower curves) to $N=10$ (upper curves). As the well deepens, both the slope and width of the turning-point mode shift together in each case, demonstrating the relationship between the parameters and the well depth.
Building on previous parameter analysis, Figure~\ref{fig:EtVcrGamDelVSV0}(d) shows the critical speed $v_\mathrm{cr}$ needed for the droplets to pass over the well, plotted against $U_0$. It compares values predicted by the variational method (solid lines) with those directly simulating Eq.~(\ref{eq:gpe}) (points), for both small ($N=1$) and large ($N=10$) droplets. The two approaches agree well for shallow and medium-depth wells, but they differ for larger $U_0$. For deeper wells, the simpler two-parameter method can't fully capture what happens when the well causes the droplet to become strongly localised, change shape significantly, or become excited in ways the simplified model doesn't account for.

\begin{figure*}[t]
\begin{center}
\includegraphics[width=6.cm]{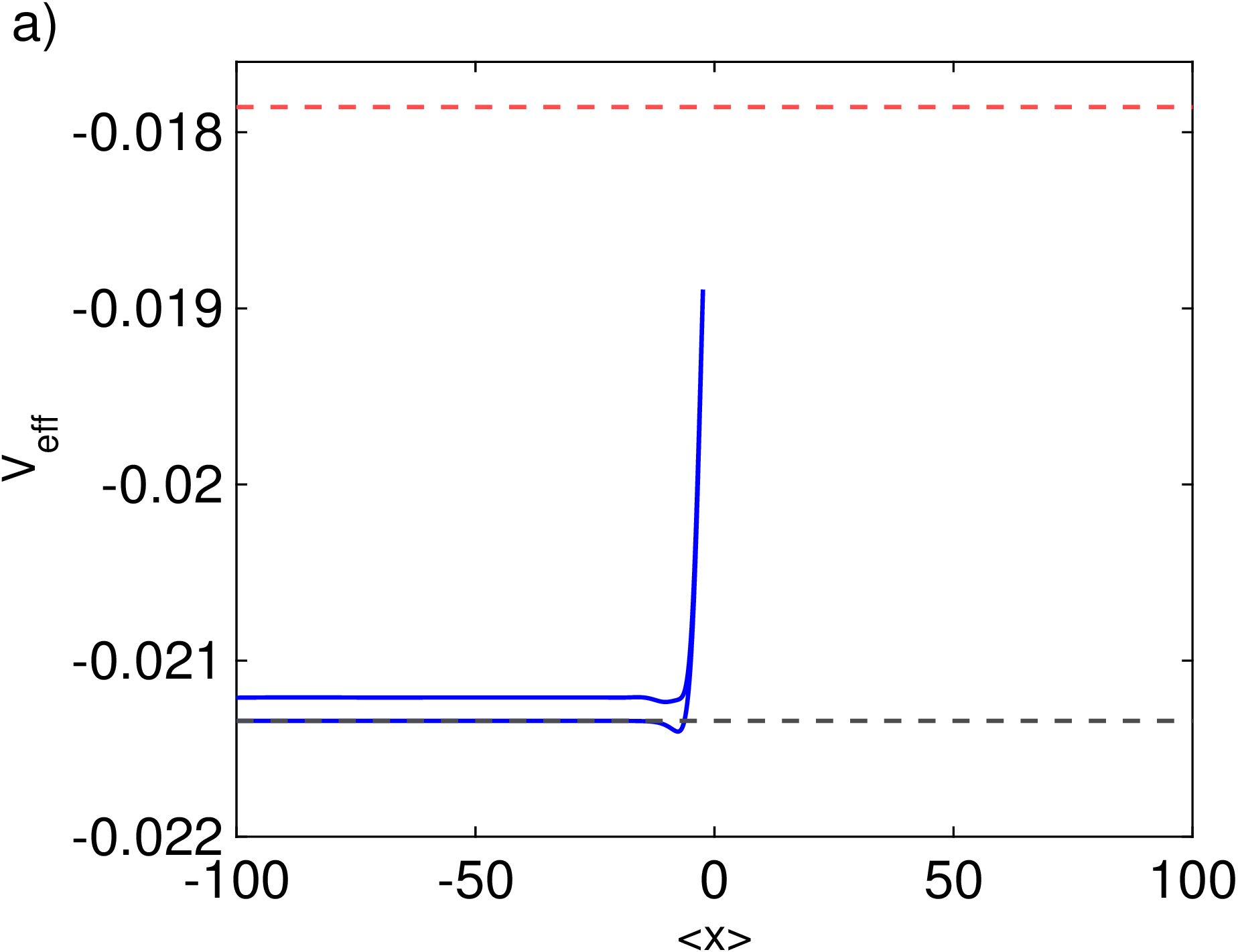}
\includegraphics[width=6.cm]{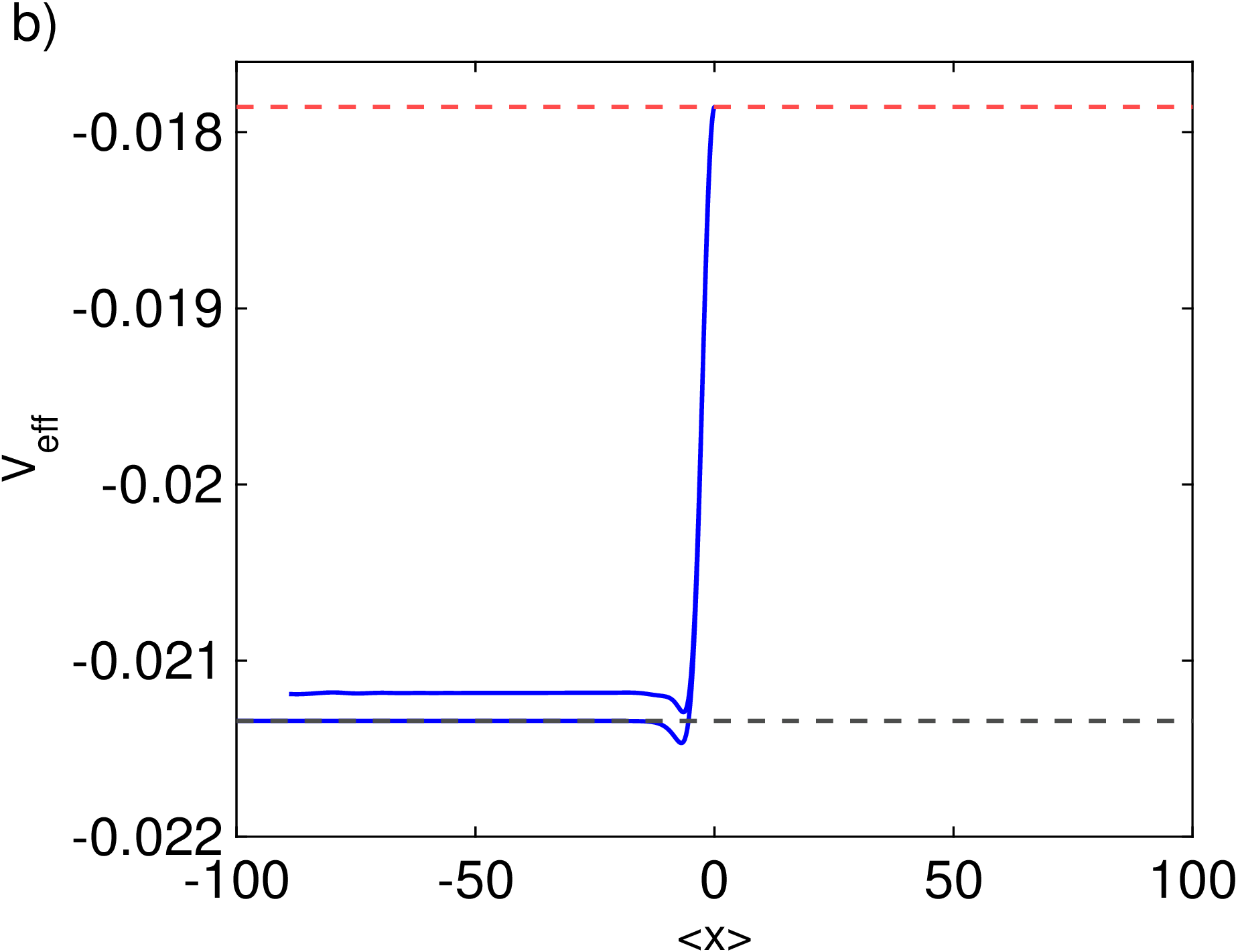}
\includegraphics[width=6.cm]{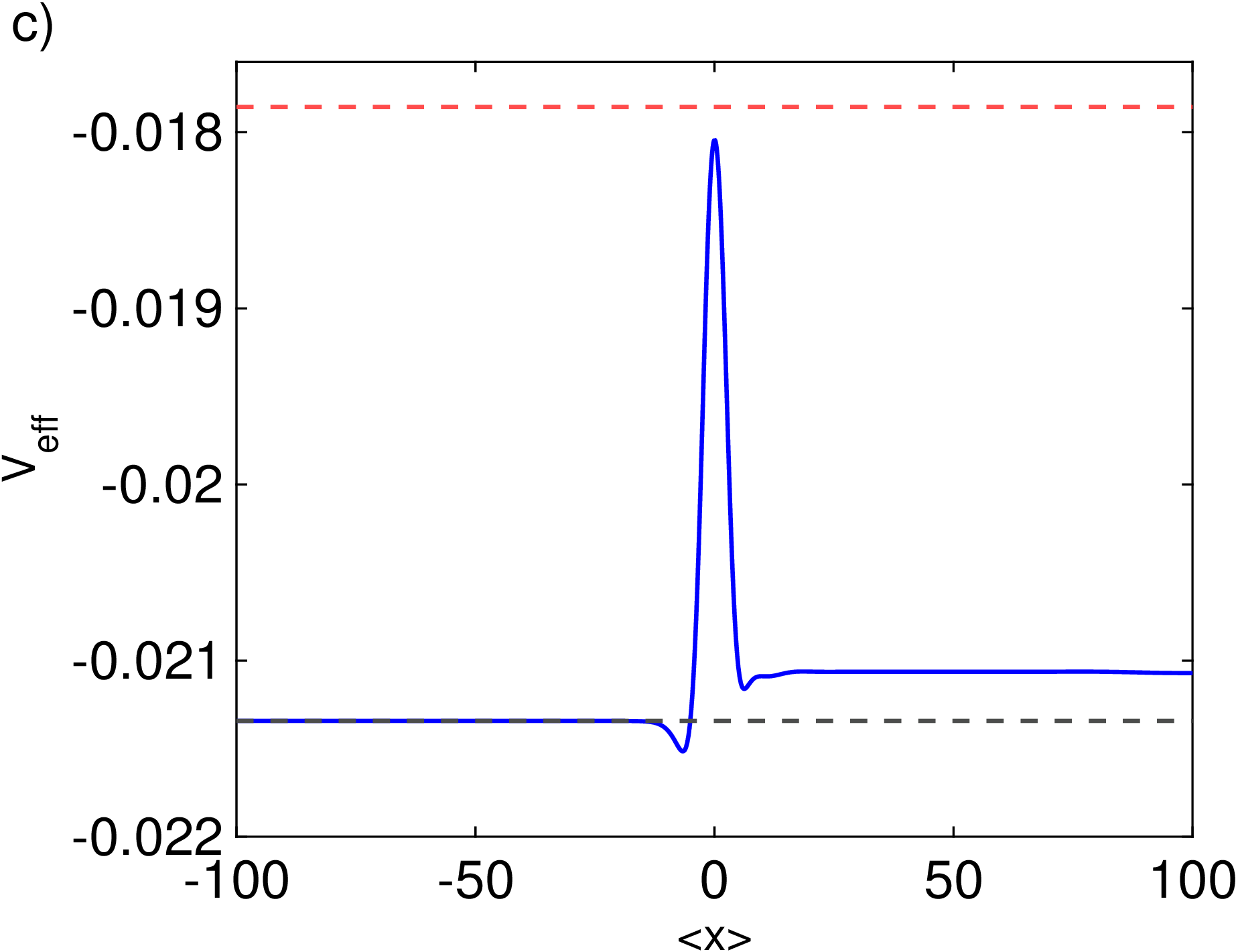}
\includegraphics[width=6.cm]{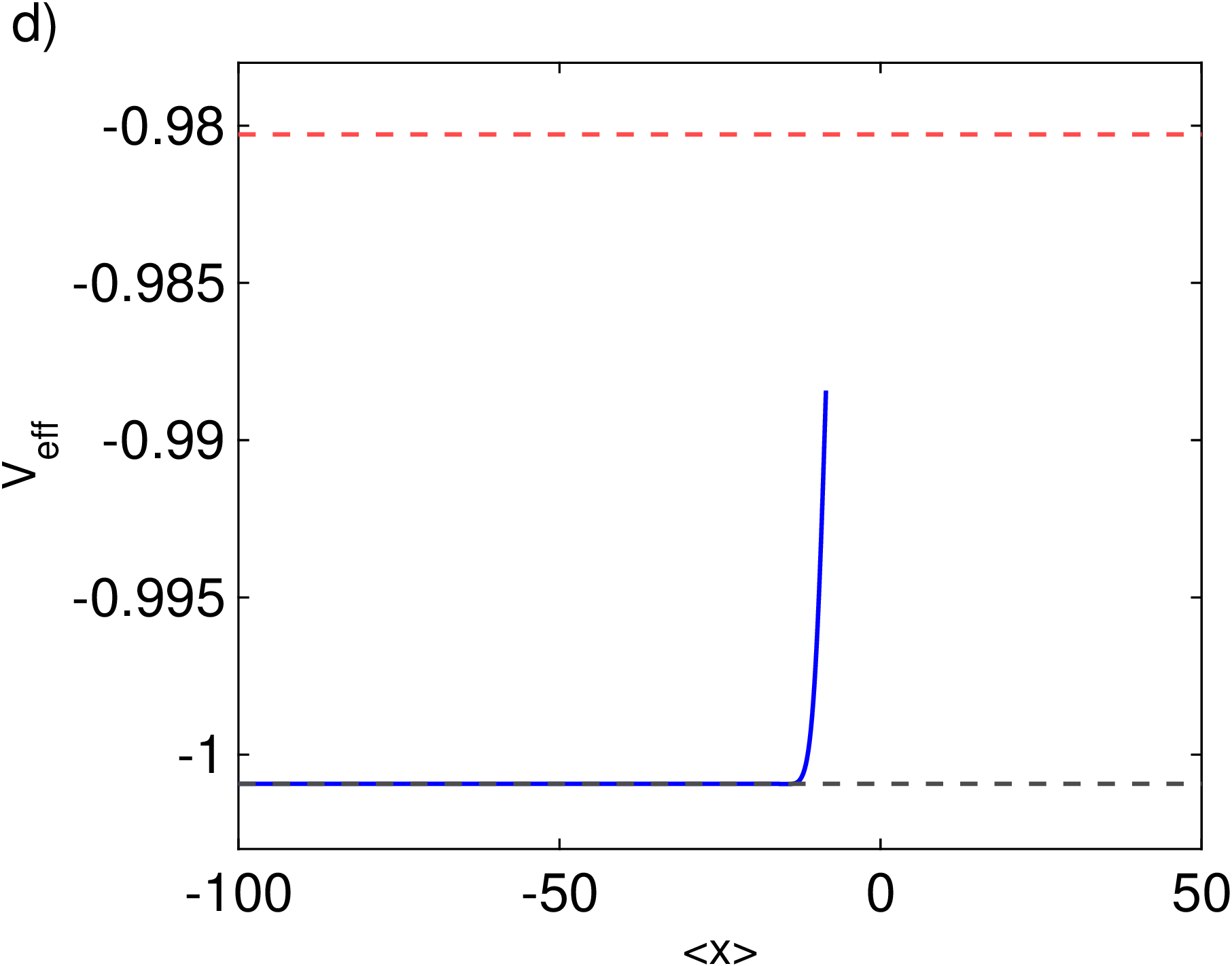}
\includegraphics[width=6.cm]{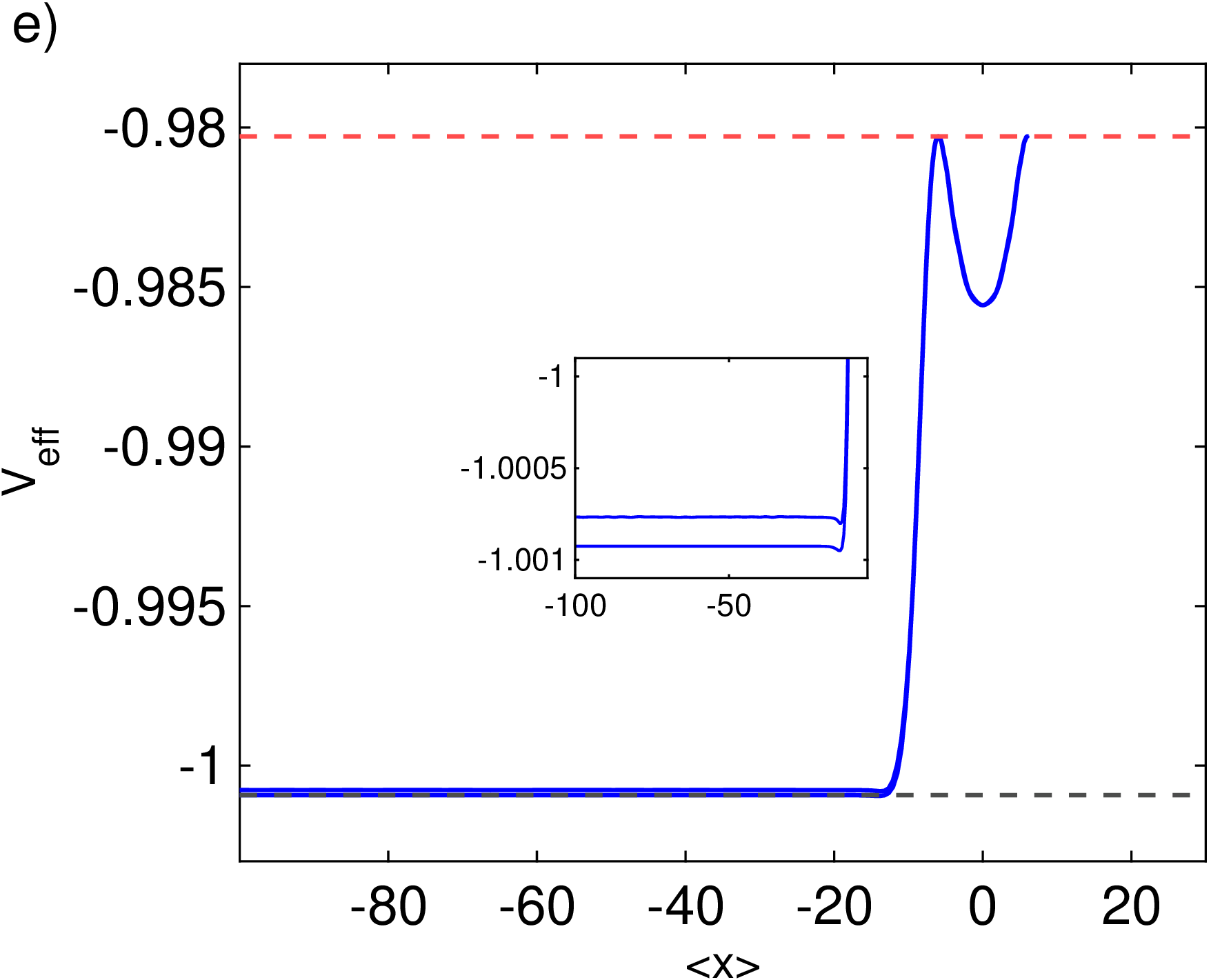}
\includegraphics[width=6.cm]{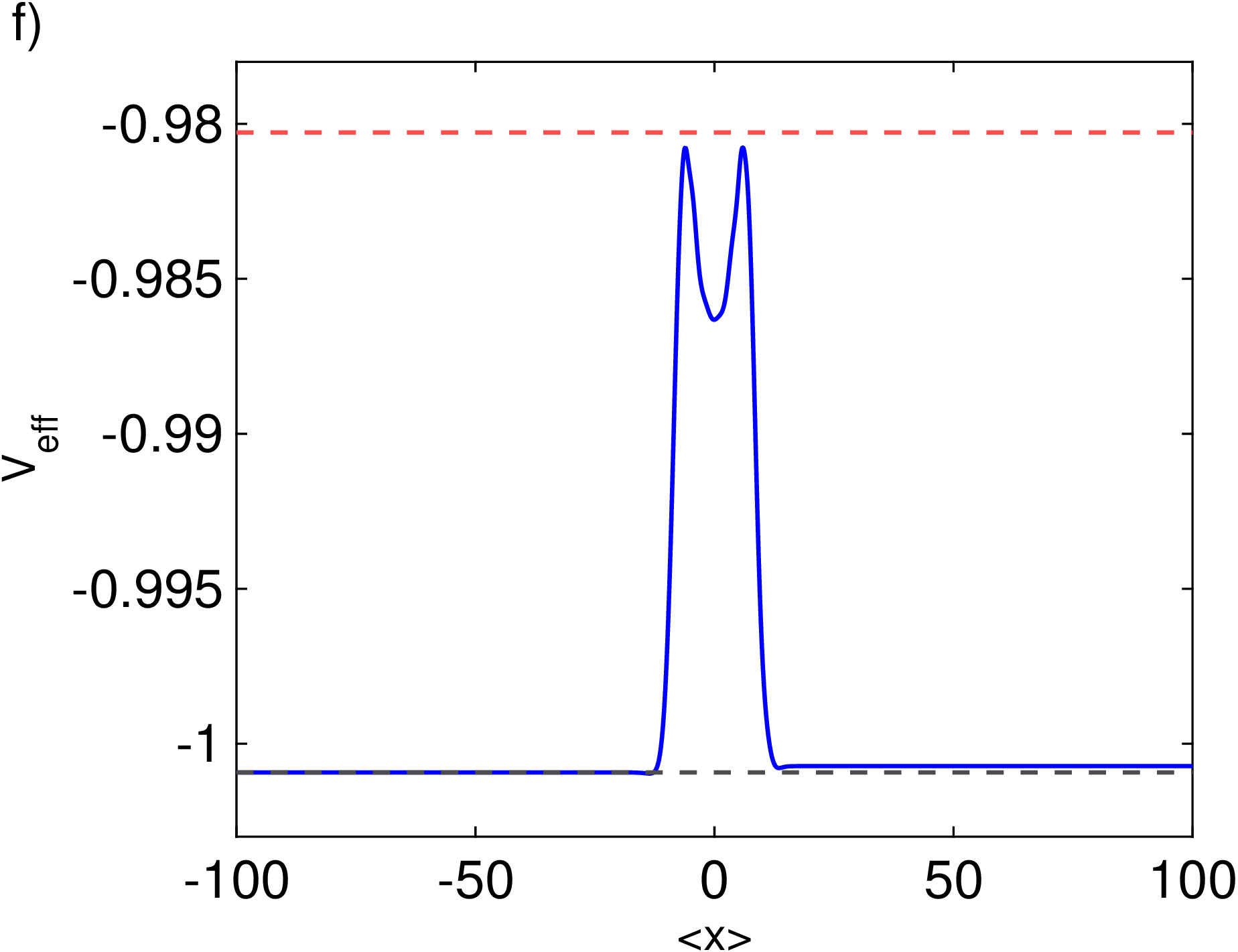}
\end{center}
\caption{Effective potential is plotted as a function of the droplet centre-of-mass position, $\langle x(t)\rangle$, for small ($N=1$, top row) and large ($N=10$, bottom row) QDs scattering from a reflectionless PT potential well. Panels (a) $v=0.07$ and (d) $v=0.05$ show incident speeds below the critical speed, panels (b) $v_{\mathrm{cr}}=0.08349974$ and (e) $v_{\mathrm{cr}}=0.06425806$ correspond to the critical speed, and panels (c) $v=0.09$ and (f) $v=0.068$ show speeds above it. The red dashed horizontal line denotes the trapped-mode energy, while the black dashed horizontal line indicates the stationary initial droplet energy, $E_{\mathrm{sd}}$. All other parameters are the same as in Fig.~\ref{fig:dynSmallQDandCoefC}.}
\label{fig:Veff}
\end{figure*}

To better understand the scattering of QDs, we view the dynamics as those of an effective classical particle in a potential landscape. A time-dependent variational method can construct this potential, but it often yields complicated integrals without exact analytic solutions, making it less practical. Instead, as in Refs.~\cite{Hu,Khawaja2021}, we extract the effective potential directly from the full time-dependent evolution by numerically integrating Eq.~(\ref{eq:gpe}).

To do so, we compute the droplet centre-of-mass coordinate,
$$
\langle x\rangle=\frac{\int\limits_{-\infty}^{+\infty} x\,|\psi(x,t)|^2\,dx}{\int\limits_{-\infty}^{+\infty} |\psi(x,t)|^2\,dx}\, ,
$$
where $\psi(x,t)$ is the numerically evolved wavefunction. Treating the droplet as a collective degree of freedom, we assume that $\langle x(t)\rangle$ obeys an effective Newtonian equation of motion,
$$
N\,\langle \ddot{x}(t)\rangle =-\frac{d}{d\langle x\rangle}\,V_{\mathrm{eff}}\![\langle x(t)\rangle]\, ,
$$
so that the droplet behaves like a classical particle of effective mass $N$ moving in an effective potential $V_{\mathrm{eff}}$. Integrating with respect to $\langle x\rangle$ gives
\begin{equation}
V_{\mathrm{eff}}[\langle x(t)\rangle]
= -N\int \langle\ddot{x}(t)\rangle\,d\langle x\rangle
+V_{\mathrm{eff}}[\langle x(0)\rangle]\, ,
\label{eq:Veff}
\end{equation}
where the integration constant is fixed by $V_{\mathrm{eff}}[\langle x(0)\rangle]=E_{\mathrm{sd}}$, the stationary energy of the initial droplet. This construction compresses the full field dynamics into an intuitive mechanical description in which the centre-of-mass plays the role of a particle coordinate and $V_{\mathrm{eff}}$ encapsulates the combined influence of the external potential and the droplets internal response.

Figure~\ref{fig:Veff} plots the effective potential $V_{\mathrm{eff}}$ against centre-of-mass position $\langle x\rangle$ for small (top) and large (bottom) droplets. Each curve shows incident speeds below, at, and above the critical value $v_{\mathrm{cr}}$. These dynamics mirror a classical object launched from the left, ascending a potential ramp. The transition between regimes occurs at $v_{\mathrm{cr}}$: for $v<v_{\mathrm{cr}}$, the object climbs to a turning point where its kinetic energy vanishes and reverses, mirroring droplet reflection. At the threshold $v=v_{\mathrm{cr}}$, the path just reaches the ramp's top illustrating the trapped-mode energy where the droplet lingers in a critical, trapped state. For $v>v_{\mathrm{cr}}$, the object surmounts the barrier and continues right, paralleling droplet transmission.
For the larger droplet, the effective potential displays two off-centre shoulders, as illustrated in Figs.~\ref{fig:Veff}(e) and~\ref{fig:Veff}(f). This feature corresponds directly to the two off-centre maxima of $E_z(x_0)$ in Fig.~\ref{fig:EzVSx0}(b). Although $E_z(x_0)$ is derived from the static variational zero-speed ansatz and $V_{\mathrm{eff}}$ from the time-dependent centre-of-mass motion, both quantities reveal the same underlying mechanism. In a spatially extended flat-top droplet, centre-of-mass acceleration is determined by a density-weighted average of the external force. The overlap with the localised PT well is maximised when either the left or right edge of the droplet interacts with the defect, producing the symmetric maxima in $E_z(x_0)$ and, dynamically, the two shoulders in $V_{\mathrm{eff}}$. Therefore, the two-shoulder structure in the effective potential represents the centre-of-mass manifestation of the asymmetric critical configurations predicted by the variational turning-point analysis. In the supercritical regime, the corresponding classical analogue traverses the effective barrier through this two-shoulder landscape, as depicted in Fig.~\ref{fig:Veff}(f).
\begin{figure}[t]
\begin{center}
\includegraphics[width=4.3cm]{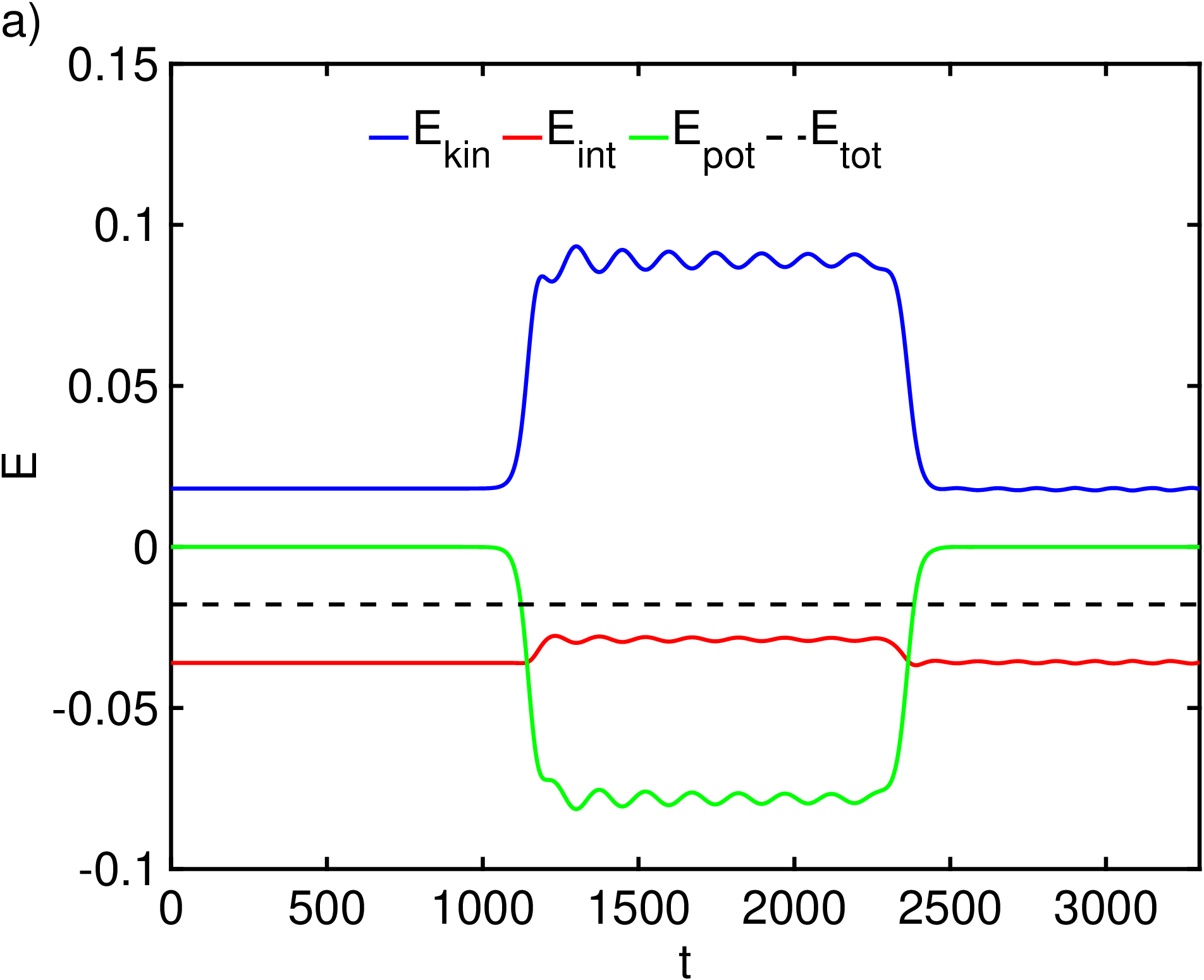}
\includegraphics[width=4.3cm]{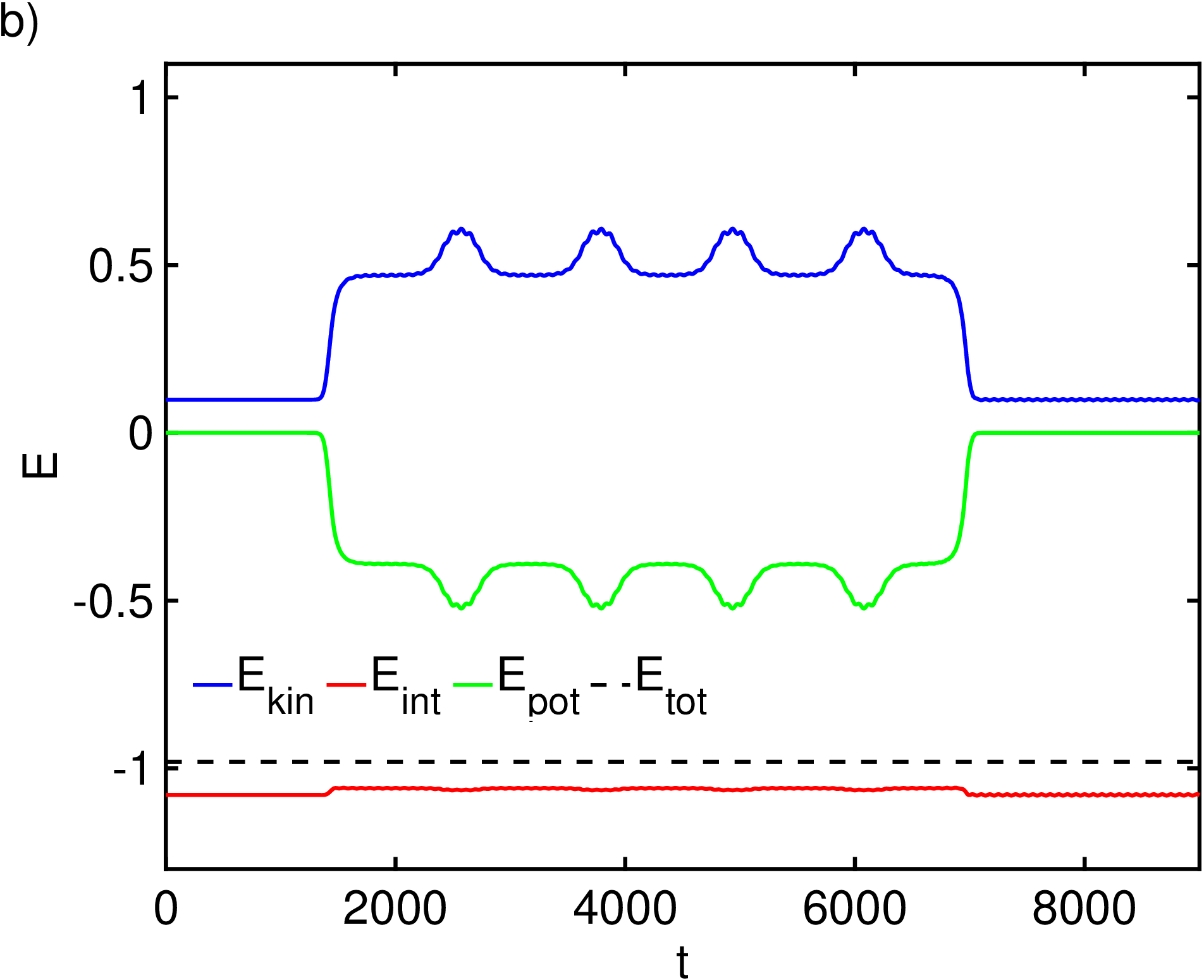}
\end{center}
\caption{Time evolution of the kinetic energy $E_{\mathrm{kin}}$, interaction energy $E_{\mathrm{int}}$, potential energy $E_{\mathrm{pot}}$ and total energy $E_{\mathrm{tot}}$ during the scattering of QDs. Panel (a) shows a small droplet $N=1$ with initial speed equal to the critical speed (corresponding to Fig.~\ref{fig:dynSmallQDandCoefC}(b)), while panel (b) shows a large droplet $N=10$ with initial speed equal to the critical speed (corresponding to Fig.~\ref{fig:dynLargeQDandCoefC}(b)). All other parameters are the same as in Fig.~\ref{fig:dynSmallQDandCoefC}.}
\label{fig:Energy}
\end{figure}
The energy in a QD comes from three parts: movement energy (kinetic), outside force (external potential), and how the parts of the droplet interact (interaction). To show how scattering works, we look at how these energies change as the droplet hits the special potential well. Figures~\ref{fig:Energy}(a) and~\ref{fig:Energy}(b) present the corresponding energy dynamics for a small ($N=1$) and a large ($N=10$) droplet, respectively. For incident speeds very close to the relevant critical speeds, these cases correspond to the density evolutions shown in Figs.~\ref{fig:dynSmallQDandCoefC}(b) and~\ref{fig:dynLargeQDandCoefC}(b).

When the droplet is far from the potential well, its overlap with $V(x)$ is negligible, so $E_{\mathrm{pot}} \approx 0$ and the droplet's motion changes little. The self-bound shape arises from a balance between quantum pressure and nonlinear interactions. On entering the well, a key transition occurs: potential energy decreases and, by energy conservation, is redistributed among other forms of energy. Early in the encounter, our simulations show kinetic (gradient) energy rises, while interaction energy varies little. The increased kinetic energy produces stronger spatial gradients in $\psi(x,t)$ as the droplet deforms and partially spreads into the trapped-mode shape imposed by the well. Consequently, more energy is shifted to the fine features of the wavefunction, such as sharper gradients and higher quantum pressure. This redistribution arises from wave-mechanical effects-interference and spreading of the matter wave-rather than simple classical centre-of-mass motion.

This redistribution matters most near the critical speed. When $v \simeq v_{\mathrm{cr}}$, a drop in external potential energy mostly becomes gradient energy. This lets the droplet efficiently couple to the trapped mode and strongly suppresses its centre-of-mass motion. As a result, kinetic energy usually settles at a long-lived, nearly constant level from internal gradient (quantum pressure) effects, not from whole-droplet motion. At this stage, small, regular peaks can appear on this plateau as the droplet moves through, where the potential affects the wavefunction. The droplet shifts to one side of the well, bounces back, and compresses and expands the wavefunction through self-interference. This changes the gradient energy. Oscillations in the plateau are a clear sign of the critical trapped state: the droplet stays near the well with little movement, while the remaining `kinetic' energy is distributed across internal gradients and collective movements.

The critical trapped state is usually unstable, poised at the verge of a tenuous balance. Even slight imbalances or small disturbances may push the droplet to one side. As a result, another key transition follows: the droplet can't remain trapped forever. After a period in the localised state, it leaves and is ejected. When it exits the well and resumes motion, it loses kinetic energy and gains potential energy.
Figure~\ref{fig:Energy} shows that after the scattering event, the interaction energy starts to oscillate. These oscillations signal excited internal collective modes, like breathing or shape oscillations, produced during the collision. This shows that some incident energy has shifted from the centre-of-mass motion to the droplet's internal motion.

\begin{figure}[t]
\centerline{\includegraphics[width=4.5cm]{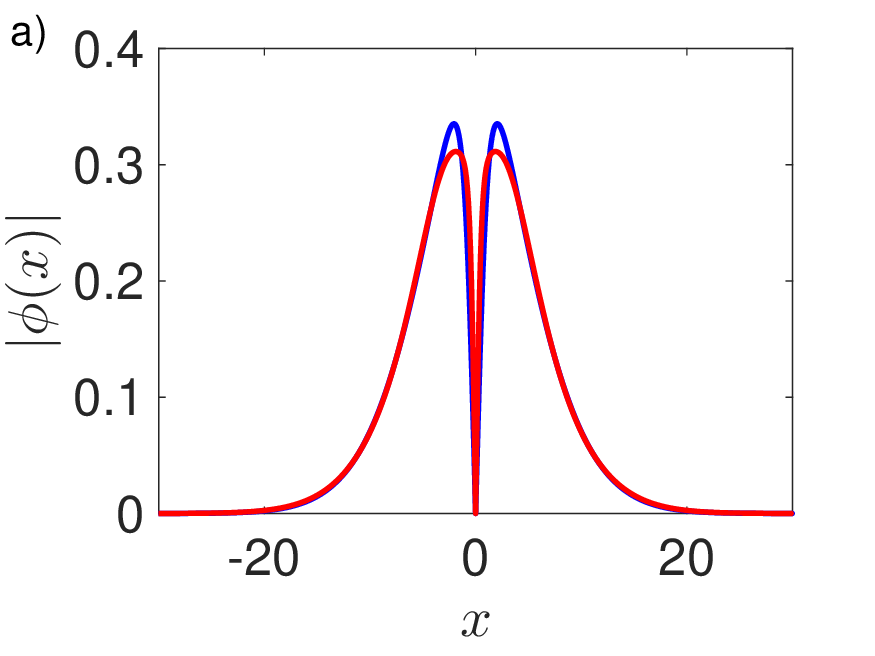} \includegraphics[width=4.5cm]{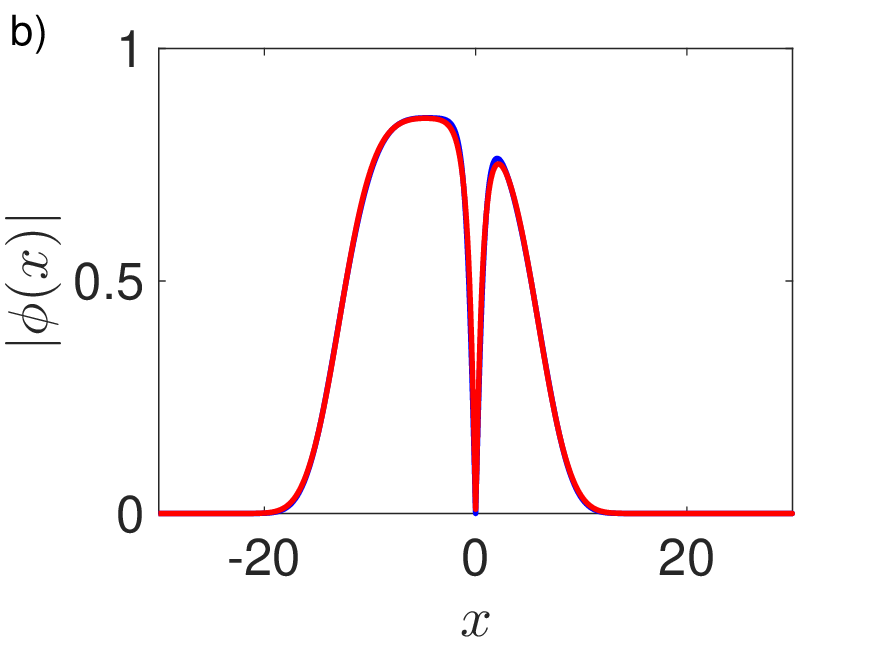}}
\centerline{\includegraphics[width=4.5cm]{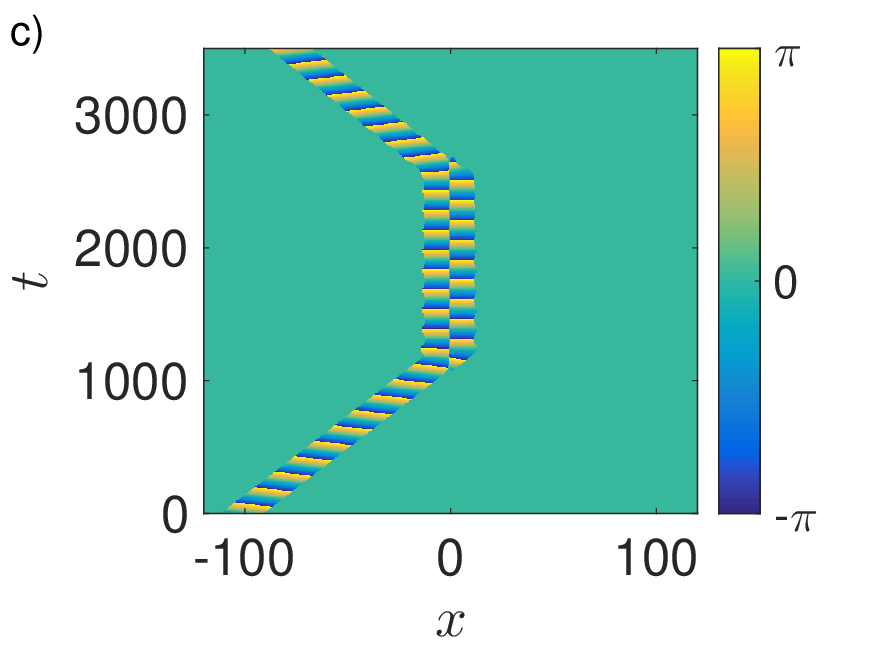} \includegraphics[width=4.5cm]{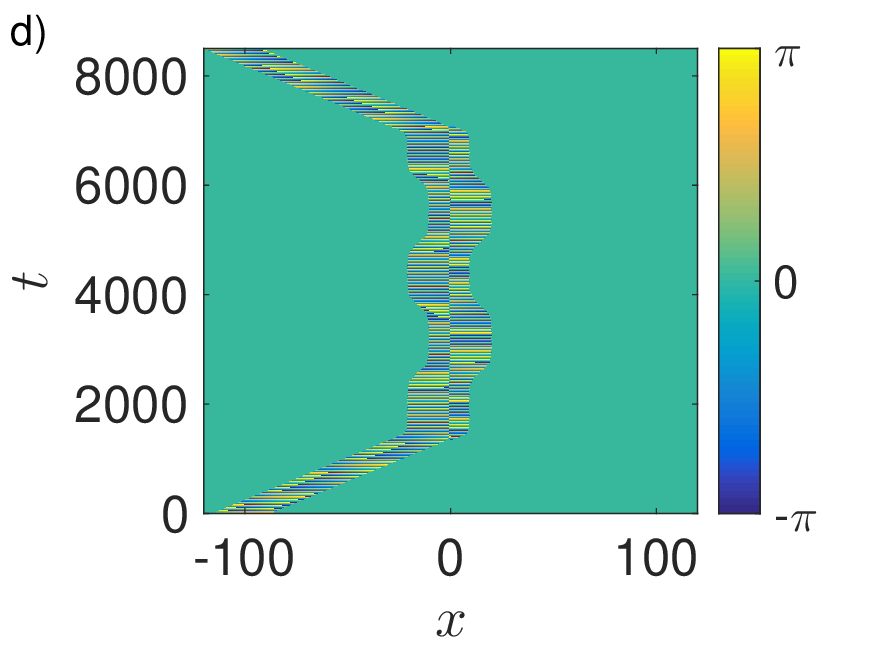}}
\caption{Comparison of trapped-mode profiles for (a) a small droplet with $N=1$ and $x_0=0$, and (b) a large droplet with $N=10$ whose density maximum is shifted to $x_0=-3.5$ relative to the centre of the potential well. The dynamical evolution corresponding to panel (b) is shown in Fig.~\ref{fig:TrappedMode}(d). In both panels, the red dashed curve represents the trapped mode obtained by imaginary-time propagation, while the blue solid curve shows the corresponding VA. Panels (c) and (d) display the phase evolution during the scattering process for the small and large droplets, respectively. The colorbar represents the phase, $\arg[\psi(x,t)]$. The corresponding density dynamics are the same as those shown in Figs.~\ref{fig:dynSmallQDandCoefC}(b) and \ref{fig:dynLargeQDandCoefC}(b). The other parameters are $U_0=4$ and $q=g=1$.}
\label{fig:TrapModeProf}
\end{figure}

Validation of the trapped-mode profiles predicted by the VA is achieved by computing stationary pinned states of the governing equation, Eq.~(\ref{eq:gpe}), and comparing them with the VA profiles derived from the trial function in Eq.~(\ref{eq:TrapModeTrialFunc}). The numerical stationary state is obtained by imaginary-time propagation, using the corresponding variational profile as the initial seed. Figures~\ref{fig:TrapModeProf}(a) and \ref{fig:TrapModeProf}(b) compare the results for small and large droplets, respectively. For the small droplet in Fig.~\ref{fig:TrapModeProf}(a), the droplet centre coincides with the centre of the PT well, $x_0=0$. The VA profile closely matches the stationary state obtained by imaginary-time propagation, confirming that the trial function accurately represents the pinned configuration in the compressible droplet regime.

For a larger droplet, a symmetric pinned state is also obtained when the droplet centre is placed exactly at the well centre. However, the dynamically relevant configurations formed near the critical scattering regime are typically off-centred. This behaviour is evident in the density evolution in Fig.~\ref{fig:dynLargeQDandCoefC}(b) and in the corresponding density cross sections in Figs.~\ref{fig-QD10CrossSec_abcdef}(c)--(e), where the trapped density repeatedly shifts between the left and right sides of the potential region. To reproduce this type of state, the imaginary-time calculation is initialised with an off-centred VA profile. The resulting pinned state, shown in Fig.~\ref{fig:TrapModeProf}(b), has its density maximum shifted to one side of the potential centre. The opposite-shifted state is obtained by applying the parity transformation $x\to -x$. Thus, the parity symmetry of Eq.~(\ref{eq:gpe}) is not explicitly broken, rather, the symmetric energy landscape supports two equivalent off-centred branches, and the incident direction or initial seed selects one of them dynamically.

The emergence of these asymmetric pinned profiles is a consequence of the extended flat-top structure of large droplets. When a large droplet is displaced slightly to the left or right of the localised well, one edge overlaps the defect more strongly than the other, producing an imbalanced trapped configuration. In the variational description, this mechanism is reflected in the two off-centre maxima of the minimised energy $E_z(x_0)$. These maxima correspond to critical configurations that are unstable under shifts of the coordinate $x_0$, which explains the repeated left-right motion observed in real-time simulations near the critical velocity. By contrast, the lower-energy symmetric pinned state centred at the well can remain dynamically trapped under real-time propagation. Small droplets, being compact and compressible, do not show a stable asymmetric pinned profile under small left- or rightward displacements, because their overlap with the well is dominated by the central part of the density. The comparison in Fig.~\ref{fig:TrapModeProf} therefore confirms that the VA captures both the symmetric trapped mode of small droplets and the off-centred pinned configurations characteristic of large flat-top droplets.

Figures~\ref{fig:TrapModeProf}(c) and (d) show the phase evolution near the critical velocity. Panel~(c) shows a small droplet, and panel~(d) shows a large droplet. Their density dynamics match those in Figs.~\ref{fig:dynSmallQDandCoefC}(b) and \ref{fig:dynLargeQDandCoefC}(b). In both cases, a clear $\pi$-phase jump across $x=0$ appears. This behaviour matches the nodal structure set by the variational ansatz~(\ref{eq:TrapModeTrialFunc}), where the $\tanh(x)$ factor adds a node at the well's centre and creates the phase discontinuity.

\begin{figure}[t]
  \centerline{\includegraphics[width=4.45cm]{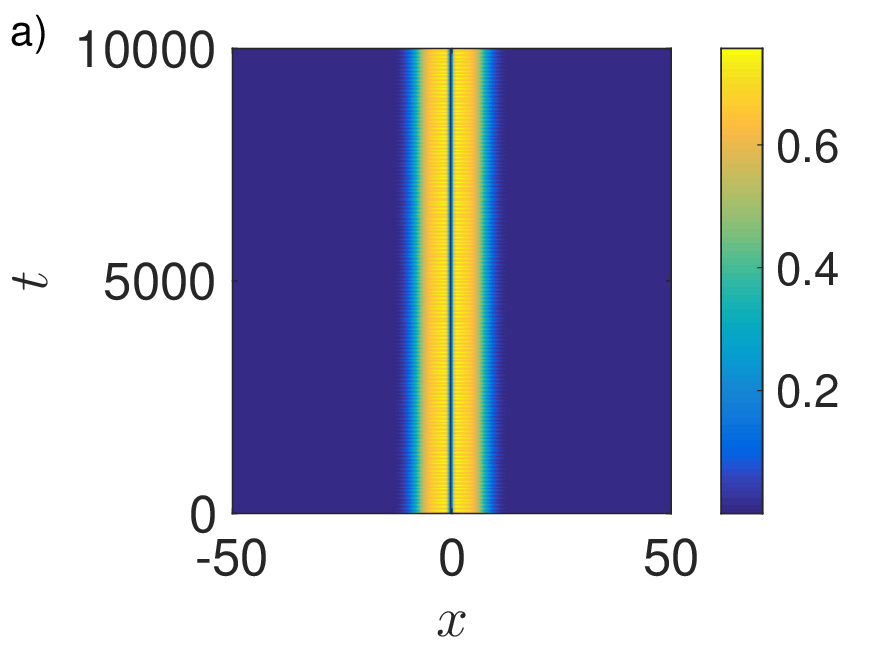} \includegraphics[width=4.45cm]{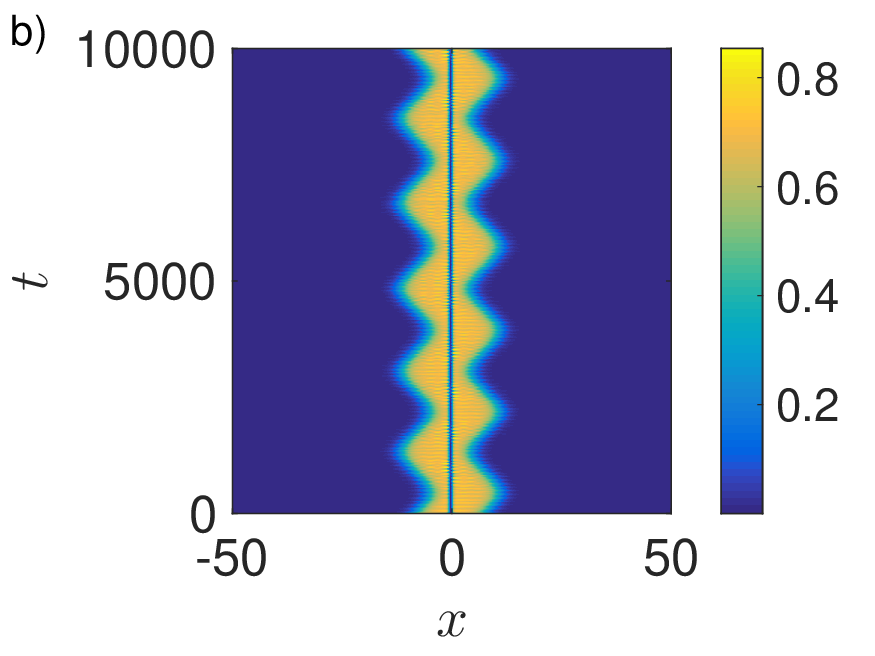}}
  \centerline{\includegraphics[width=4.45cm]{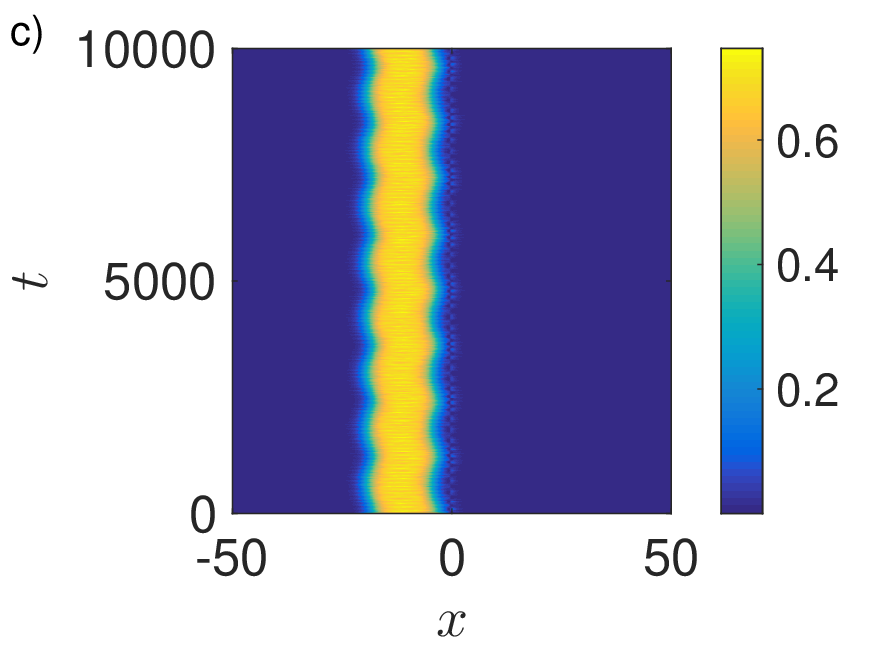} \includegraphics[width=4.45cm]{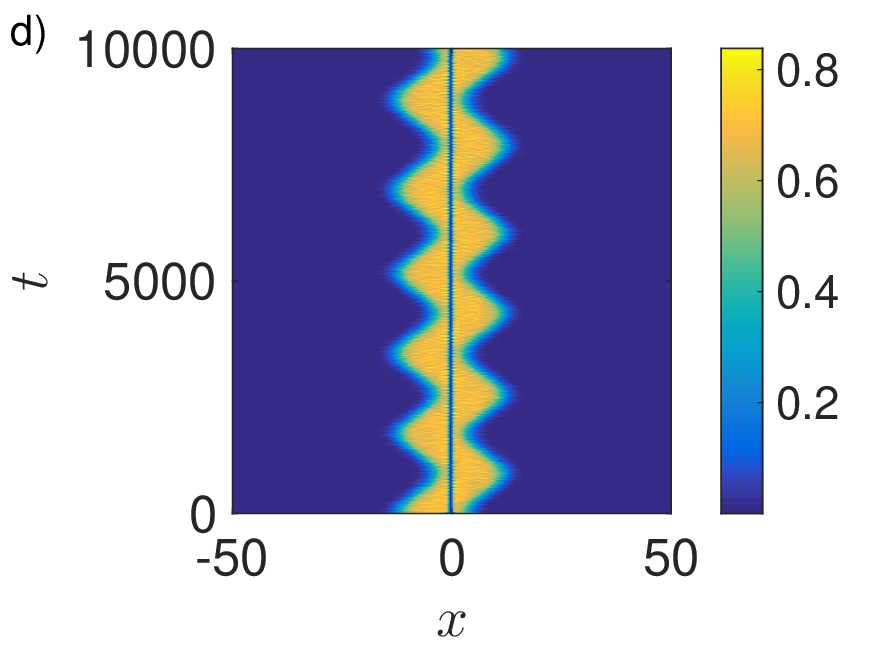}}
  \caption{Density evolution of trapped states for a large, flat-top QD with $N=10$. Panel (a) shows the time propagation of a QD initially centred at the well, $x_0=0$, confirming its time-invariant (pinned) character. Panel (b) shows the evolution of the same state after imprinting a small incident velocity, $v=0.01$, the droplet remains confined. It performs bounded oscillations about the reflectionless PT well, demonstrating trapping rather than reflection or transmission. Panel (c) displays the evolution of a low-energy zero-speed state predicted by the variational analysis, initialised at $x_0=-6$. Panel (d) shows that when the droplet is initially placed between the two energy maxima as in Fig.~\ref{fig:EzVSx0}(b), similar bounded oscillations can occur even without an initial velocity. The colorbar represents the density $|\psi|^2$. Other parameters are $U_0=4$ and $q=g=1$.}
\label{fig:TrappedMode}
\end{figure}
For the balanced trapped mode, the variational and numerical profiles nearly coincide. This demonstrates that the ansatz captures the pinned configuration centred on the well. The numerically obtained stationary state remains unchanged under real-time propagation. This confirms it is a stationary nonlinear mode of the system. A representative example appears in Fig.~\ref{fig:TrappedMode}(a). To assess dynamical robustness, we imprint a small centre-of-mass velocity on this state. As shown in Fig.~\ref{fig:TrappedMode}(b), the evolution exhibits neither transmission nor reflection. Instead, the droplet stays localised and performs bounded oscillations around the equilibrium. This sequence confirms that the mode is genuinely trapped by the PT well and behaves as a strong pinned state.

In addition, Fig.~\ref{fig:TrappedMode}(c) further supports the existence of extra low-energy pinned configurations suggested by the VA. When the droplet is initialised in a lower-energy state, its interaction with the well excites its internal degrees of freedom, leading to weak radiation emission. Here, the wave packet remains confined near the defect but does not remain in a single stationary configuration. Instead, it periodically reshapes and transitions between nearby low-energy pinned states, producing local oscillations around the well centre. Figure~\ref{fig:TrappedMode}(d) similarly shows that such bounded oscillations can also happen without initial velocity, when the droplet is placed between the two energy maxima as in Fig.~\ref{fig:EzVSx0}(b), i.e., within the effective trapping region defined by the VA.

\subsection{Collision-Induced Dynamics in P\"oschl-Teller well}
\label{subsec:Collision}

Collisions between interacting QDs can have different outcomes. They may merge into a single droplet, separate due to effective repulsion, or partially break up via radiation. These outcomes depend sensitively on their relative phase. Thus, the collision dynamics is governed not only by the droplet velocities and sizes. The phase difference and the phase gradients generated during the interaction also play key roles. Importantly, scattering from a reflectionless PT potential well can imprint an additional phase shift on each droplet. This phase shift is expected to strongly influence the subsequent collision outcome when the droplets meet near the well.
To isolate and quantify this effect, we perform direct numerical simulations of the governing GPE. We initialise the system with two droplets placed symmetrically about the potential well. The droplets are launched with opposite velocities toward $x=0$, so that they collide at the location of the reflectionless PT potential.
The initial condition was taken as a superposition of two QDs,
\begin{equation}
\Psi(x,t=0)=\Psi_0(x+x_0) \mathrm{e}^{i v x}+\Psi_0(x-x_0) \mathrm{e}^{-i v x +i \theta}\, .
\label{eq:superpozition}
\end{equation}
Here, $\Psi_0(x+x_0)$ and $\Psi_0(x-x_0)$ are identical stationary QD profiles, described by the super-Gaussian ansatz in Eq.~(\ref{gaussian}), and are initially centered at $x=-x_0$ and $x=+x_0$. The phase factors $e^{\pm ivx}$ give the droplets equal and opposite velocities, and $\theta$ sets their initial relative phase.

\begin{figure}[t]
\centerline{\includegraphics[width=4.45cm]{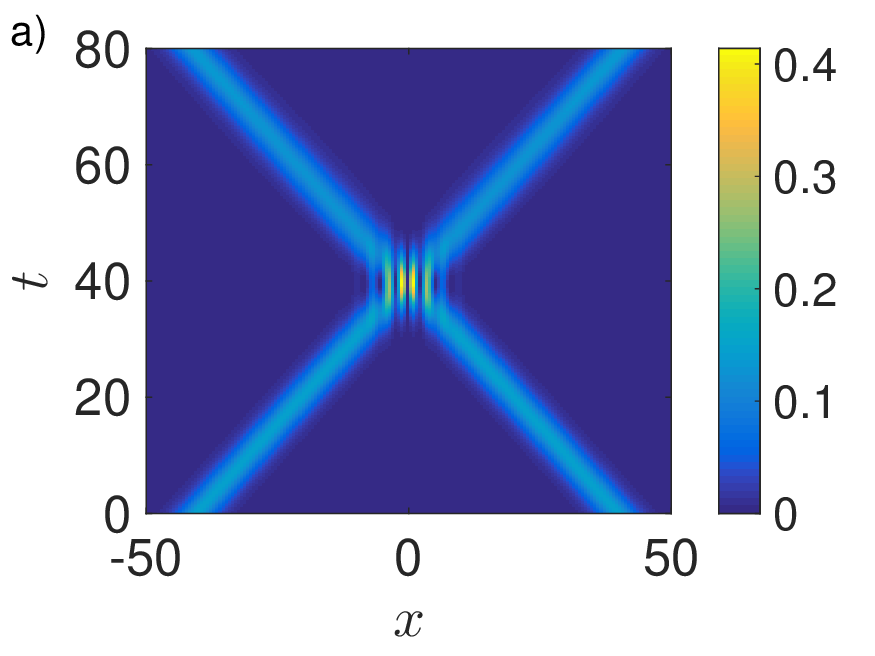} \includegraphics[width=4.45cm]{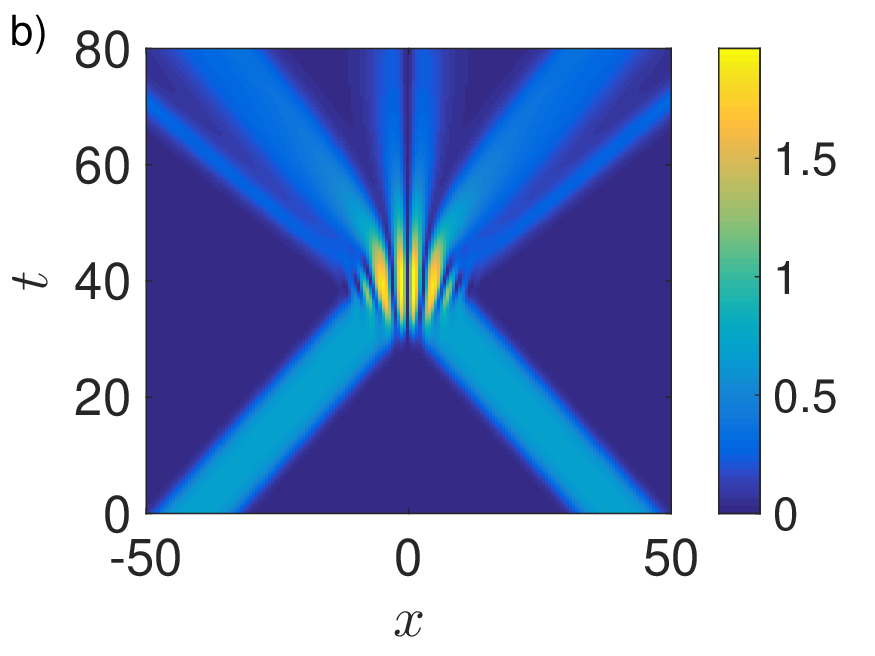}}
\centerline{\includegraphics[width=4.45cm]{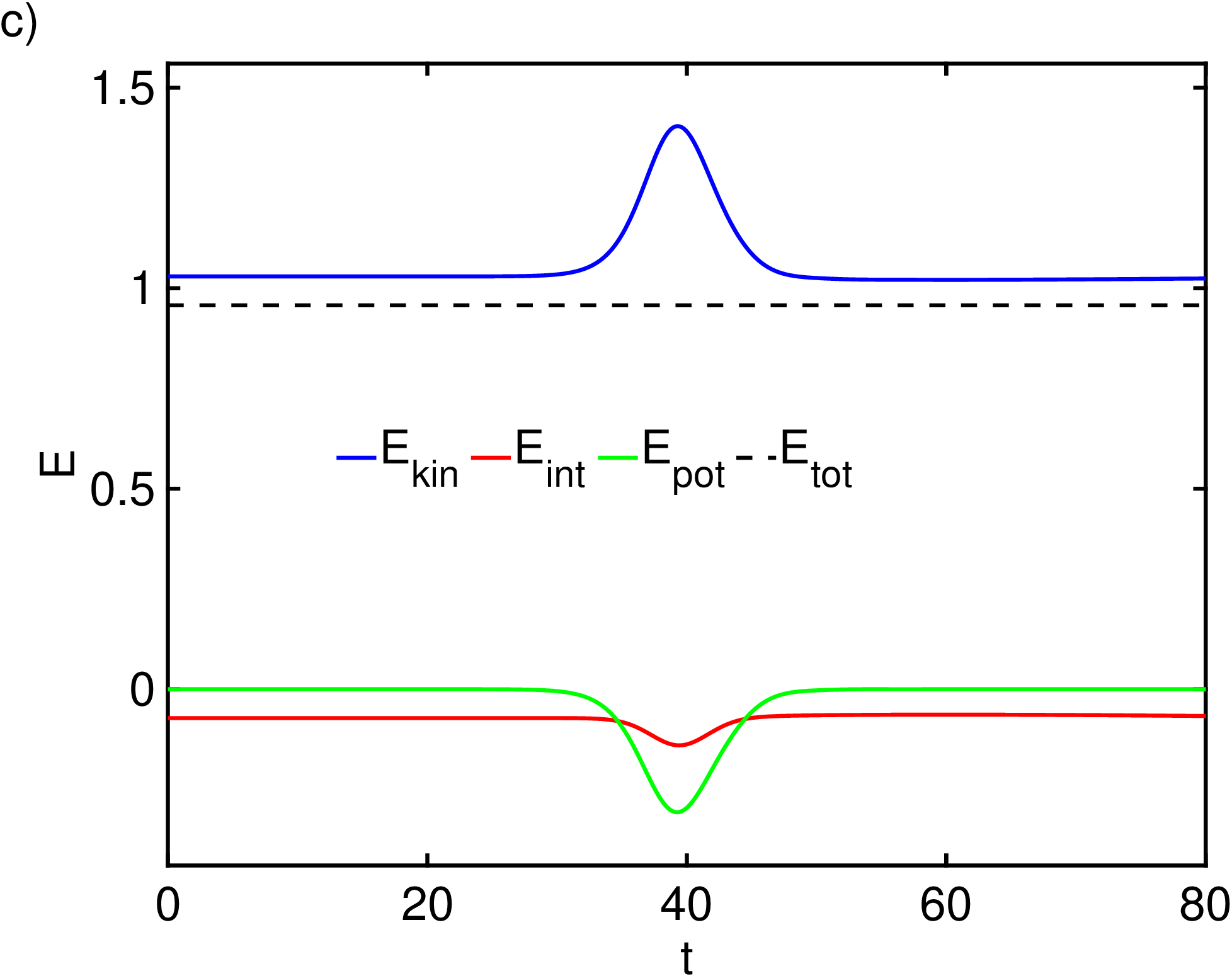} \includegraphics[width=4.45cm]{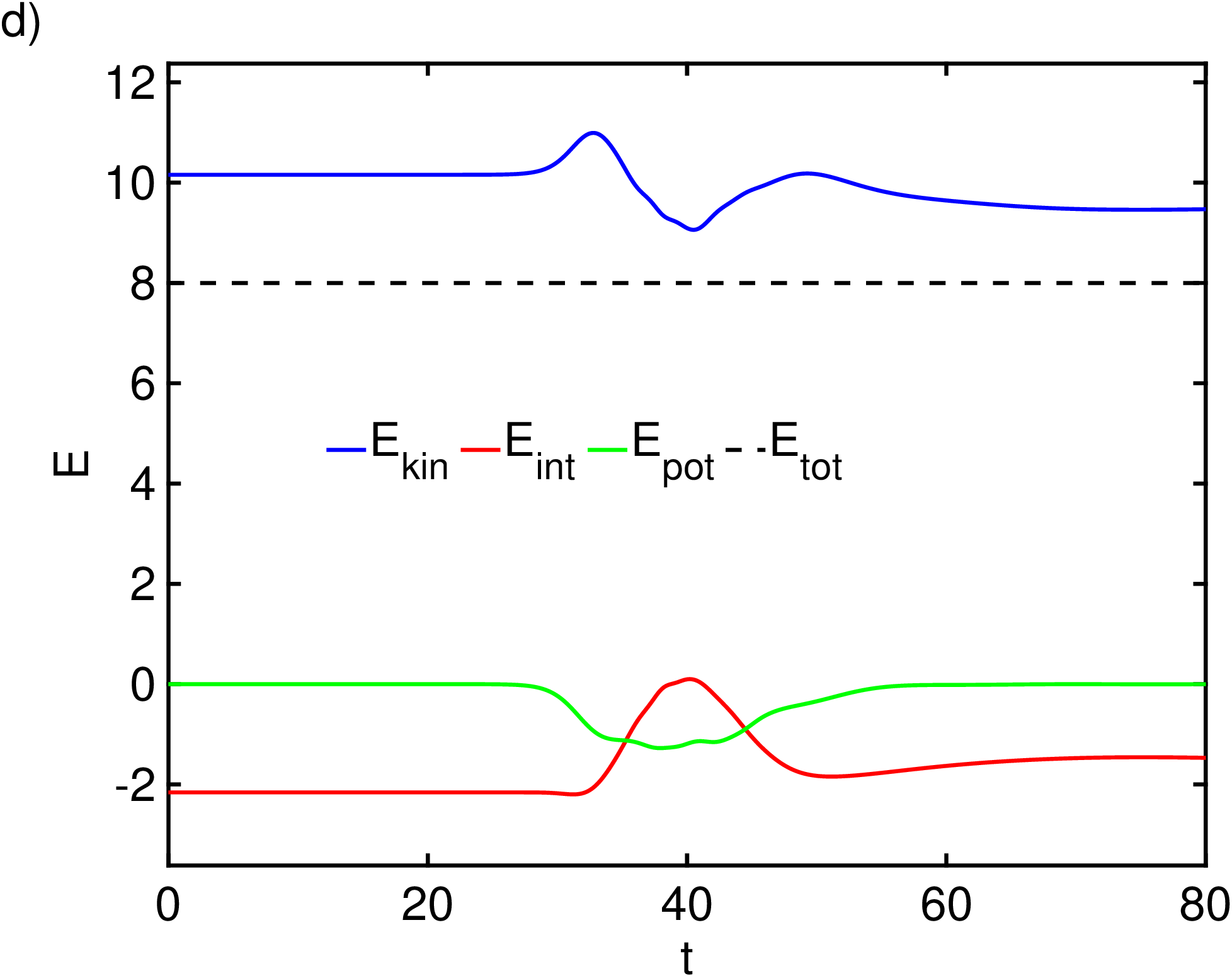}}
\caption{The space-time density plots illustrate the interference pattern generated when two identical, counter-propagating QDs collide with relative phase $\theta$. Each droplet is launched with the same incident speed, $v=1$, so that they meet near the potential region and overlap for a finite time, during which phase-sensitive interference modulates the combined density profile. Panel (a) corresponds to two small, bell-shaped droplets with $N=1$, while panel (b) shows two larger, flat-top droplets with $N=10$. In both cases, the initial relative phase is set to $\theta=\pi$. The colorbar represents the density $|\Psi|^2$. Panels (c) and (d) show the time evolution of the kinetic energy (solid blue), interaction energy (solid red), and potential energy (solid green) associated with the collision dynamics presented in panels (a) and (b), respectively. Although the collision redistributes energy among these components, the total energy, shown by the dashed black curve, remains conserved in both cases. The remaining parameters are $U_0=1$, $q=g=1$, and the initial droplet positions are $\pm x_0$ with $x_0=40$.}
\label{fig-InterferPattern}
\end{figure}

We now consider high-speed collisions of two QDs prepared with a relative phase difference $\pi$ in the presence of a reflectionless PT well. Figure~\ref{fig-InterferPattern}(a) shows the collision of fast, bell-shaped droplets with a small norm, whereas Fig.~\ref{fig-InterferPattern}(b) corresponds to fast collisions of larger flat-top droplets. In both cases, the collision is accompanied by a transient interference pattern at the overlap region, reflecting coherent superposition of the two matter-wave packets. Similar high-speed interference fringes were reported without an external potential well in the in-phase ($0$-phase) configuration for the same model~\cite{Otajonov2024}. Here, the reflectionless PT potential provides an additional phase-imprinting mechanism, thereby modifying the overlap dynamics near the collision point.

For small bell-shaped droplets (e.g., $N=1$, with a central peak and smooth edges) launched with $v=1$, the droplets behave in an essentially elastic manner (retaining energy and shape after colliding). After briefly overlapping, their density profiles largely recover, and the outgoing droplets retain their shapes. In contrast, for large flat-top droplets ($N=10$, meaning a uniform density across the central region), the collision is strongly inelastic (significant loss of initial structure, energy redistributes). The extended overlap region and stronger nonlinear interaction excite internal modes (collective oscillations within each droplet) and generate radiative emission (energy as small-amplitude waves). The initial flat-top profiles do not return after collision. Instead, the interaction produces a train of low-amplitude outgoing droplets, plus nearly stationary (quiescent) fragments near the collision region. This shows a partial breakup of the original droplets into smaller self-bound pieces.

Figures~\ref{fig-InterferPattern}(c) and (d) depict the time evolution of kinetic, interaction, potential, and total energy during the collision of two counter-propagating QDs within the PT well. These panels correspond to the dynamical evolutions illustrated in Figs.~\ref{fig-InterferPattern}(a) and (b), respectively.

Figure~\ref{fig-InterferPattern}(c) shows the energy exchange for two small droplets. In this case, the collision is nearly elastic. After passing through the interaction region, the droplets recover their initial density profiles and propagate away with only weak residual deformation. Consequently, the energy curves are almost symmetric with respect to the collision time. As the droplets approach the well and begin to overlap, their kinetic energy increases, while their potential energy becomes more negative due to stronger overlap with the attractive well. The interaction energy also changes during the overlap, reflecting the temporary increase of the local density and the interference between the two coherent matter-wave packets. These variations reach their extrema near the collision centre, around $t\approx 40$. After the droplets separate, the kinetic, potential, and interaction energies return close to their pre-collision values. This confirms the essentially elastic character of the small-droplet collision.

Figure~\ref{fig-InterferPattern}(d) shows the corresponding energy dynamics for larger flat-top droplets. In contrast to the small-droplet case, the energy exchange is no longer symmetric in time. This asymmetry indicates that the collision is strongly inelastic and excites internal degrees of freedom of the droplets. During the initial approach, the droplets' kinetic energy changes as they accelerate toward the attractive potential and begin to overlap. Near the collision centre, the potential energy reaches its minimum value, reflecting the maximum overlap with the well. At the same time, the interaction energy changes substantially because the flat-top droplets undergo stronger density deformation and norm redistribution during the collision. After the collision, the outgoing droplets do not fully recover their original flat-top profiles. Instead, part of the incident energy remains stored in internal shape oscillations and density redistribution. As a result, the interaction energy approaches a value different from its initial one, while the kinetic energy settles to a new nearly constant level. In both cases, the total energy is conserved throughout the evolution, confirming the numerical accuracy of the simulations.

\begin{figure}[t]
\centerline{\includegraphics[width=4.4cm]{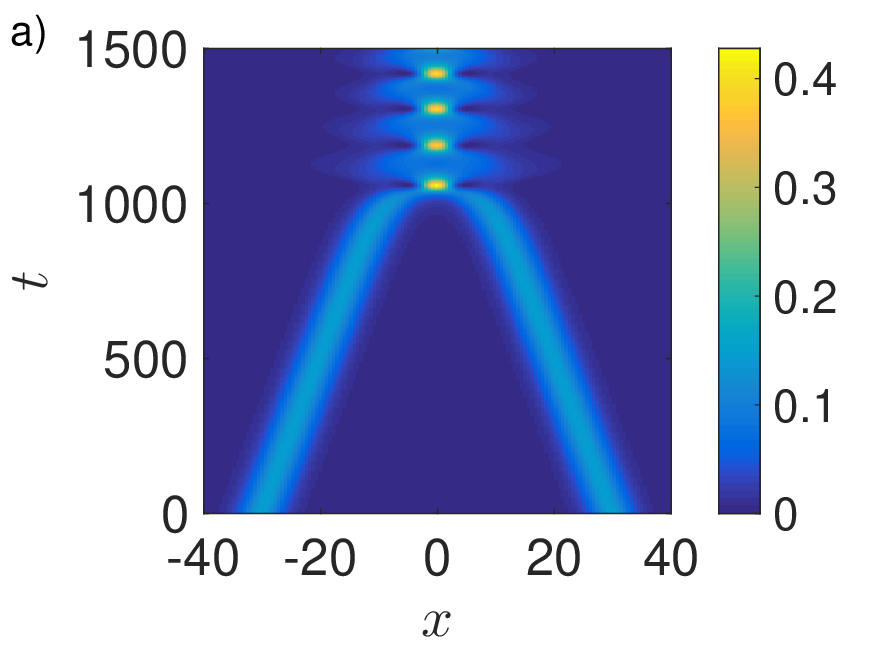} \hskip-0.1cm \includegraphics[width=4.4cm]{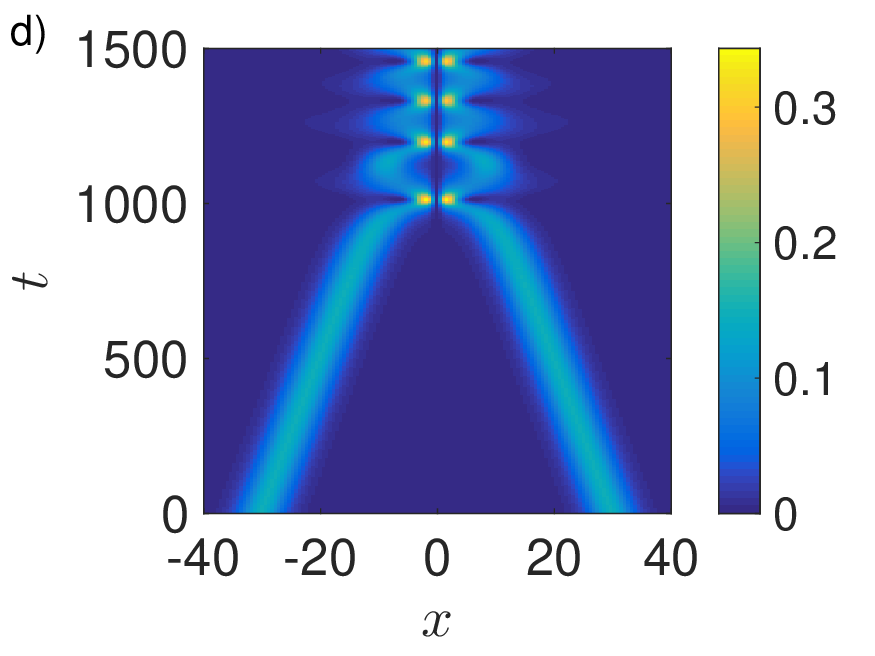}}
\centerline{\includegraphics[width=4.4cm]{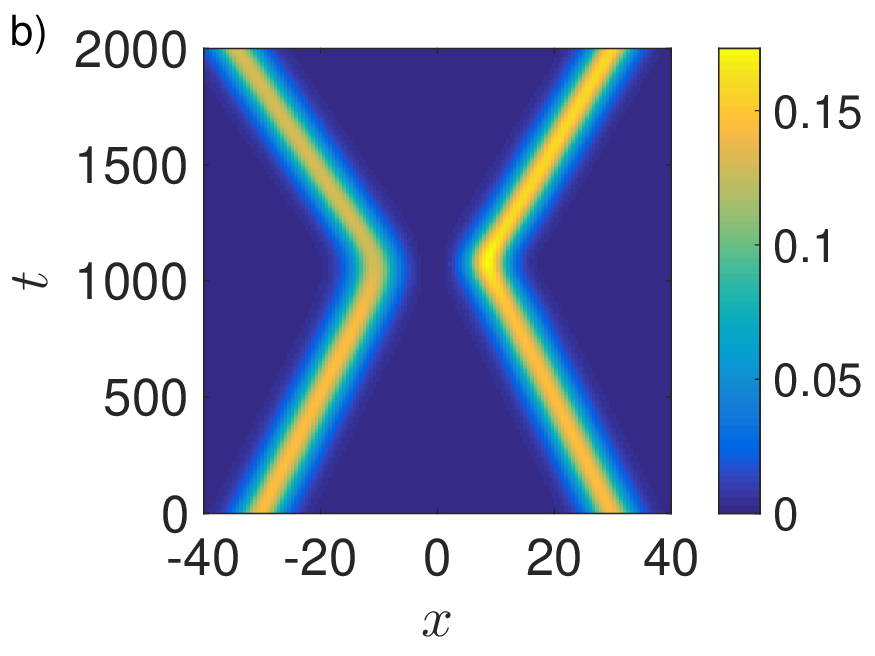} \hskip-0.1cm \includegraphics[width=4.4cm]{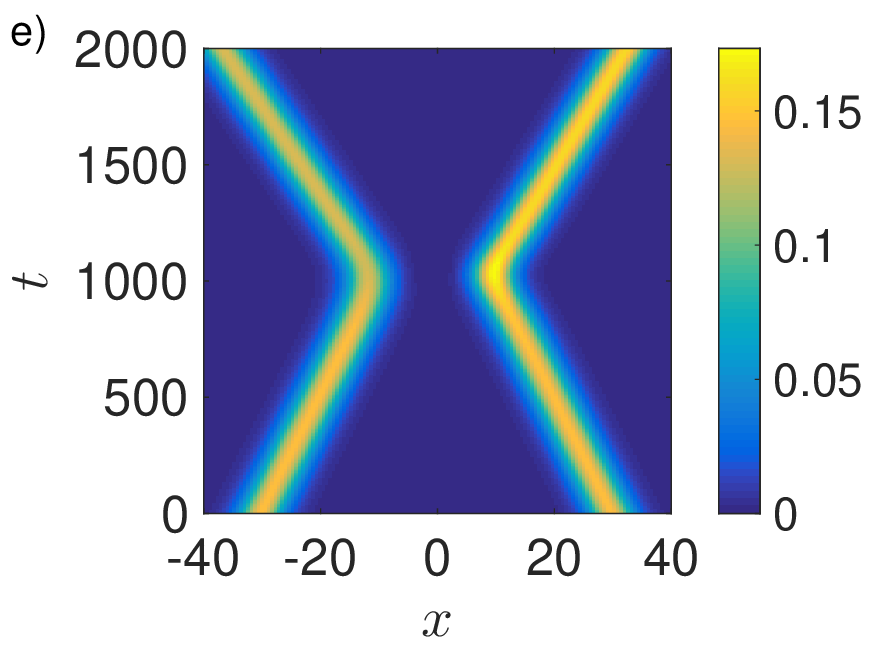}}
\centerline{\includegraphics[width=4.4cm]{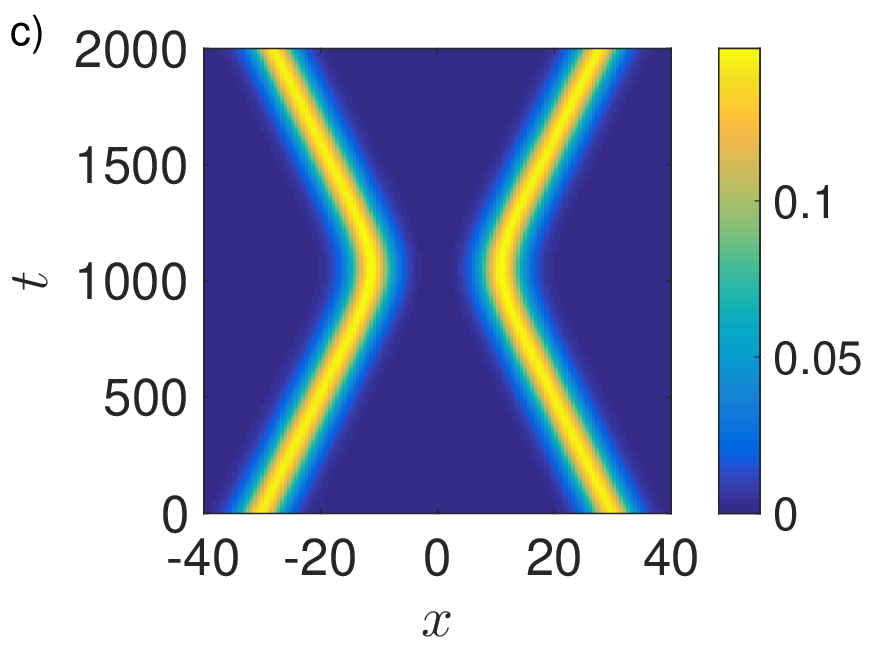} \hskip-0.1cm \includegraphics[width=4.4cm]{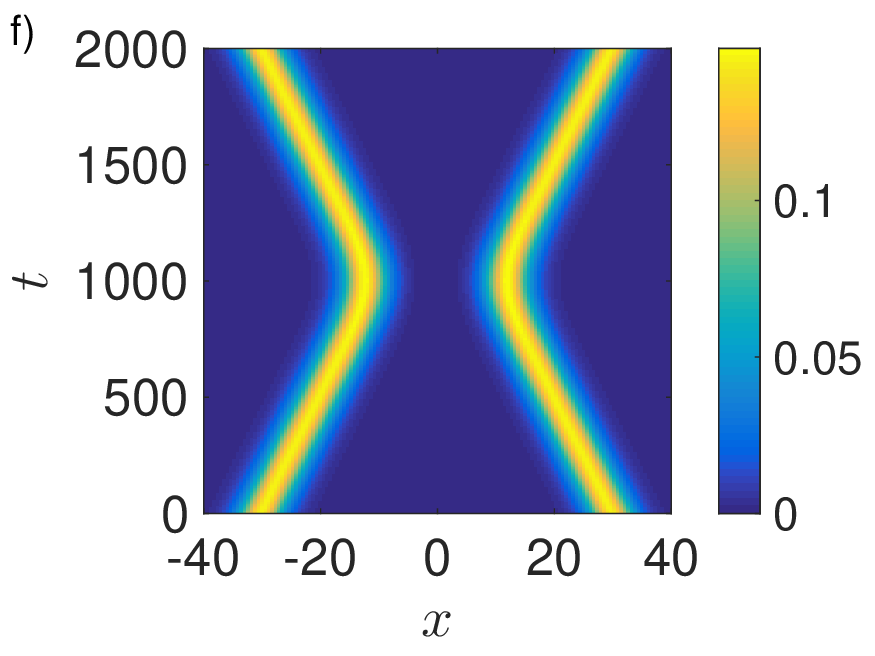}}
\caption{Scattering dynamics of QDs with a small number of particles ($N=1$) for different relative phases $\theta$, in free space and in the presence of a PT potential well. The left column shows the collision dynamics in free space ($U_0=0$): panel (a) corresponds to $\theta=0$, panel (b) to $\theta=\pi/2$, and panel (c) to $\theta=\pi$. The right column shows the collision dynamics in the presence of an attractive potential well of depth $U_0=1$, centred at $x=0$: panel (d) corresponds to $\theta=\pi$, panel (e) to $\theta=3\pi/2$, and panel (f) to $\theta=0$. The colorbar represents the density $|\psi|^2$. The comparison shows that, apart from a $\pi$-phase shift induced by the potential well, the QD dynamics remain essentially unchanged. In all panels, the initial velocity is fixed as $v=0.02$.}
\label{fig-CollisionsN1}
\end{figure}
Figure~\ref{fig-CollisionsN1} shows collisions of two identical, slowly moving small QDs with equal norms ($N=1$) and equal and opposite initial speed ($\pm v$, with $v=0.02$). The left column corresponds to free-space collisions (no external potential), while the right column shows the same setup in the presence of a reflectionless PT well centred at $x=0$.
In free space (left column), the results depend sensitively on the incident velocity and the relative phase $\theta$. For sufficiently small velocity, $v<v_{\mathrm{cr}}$, the droplets do not simply pass through one another, instead, they form a long-lived bound state (the ``weak-merging'' regime). As illustrated in Fig.~\ref{fig-CollisionsN1}(a), the initial impact produces a composite localised structure, but complete fusion into a single stationary droplet does not occur, the two density peaks re-emerge and undergo repeated collisions, resembling a breathing droplet bound state~\cite{Boudjema}. For larger velocity, $v>v_{\mathrm{cr}}$, two in-phase droplets ($\theta=0$) collide essentially elastically and pass through each other. For an intermediate phase difference, $\theta=\pi/2$, the collision becomes inelastic and is accompanied by noticeable particle-number transfer at the overlap, followed by separation due to effective repulsion, see Fig.~\ref{fig-CollisionsN1}(b). For out-of-phase collisions ($\theta=\pi$), the interaction is predominantly repulsive, and the droplets rebound, producing near mirror-like trajectories, see Fig.~\ref{fig-CollisionsN1}(c).
The right column panels, Figs.~\ref{fig-CollisionsN1}(d)--(f), show the same initial droplet pair colliding in the presence of the reflectionless PT well. The well generates negligible reflection but introduces an additional transmission phase. As a result, the relative phase at the moment of overlap shifts from that in free space. This shift can qualitatively change whether the droplets merge into a bound state, pass through, or rebound, even when all other initial parameters are unchanged.

\begin{figure}[t]
\centerline{\includegraphics[width=4.4cm]{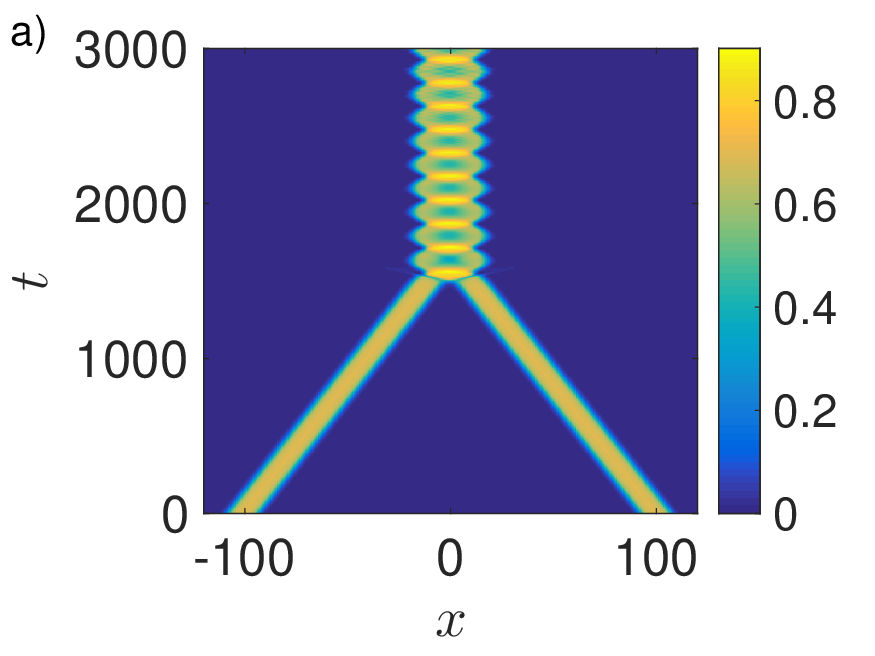} \hskip-0.1cm \includegraphics[width=4.4cm]{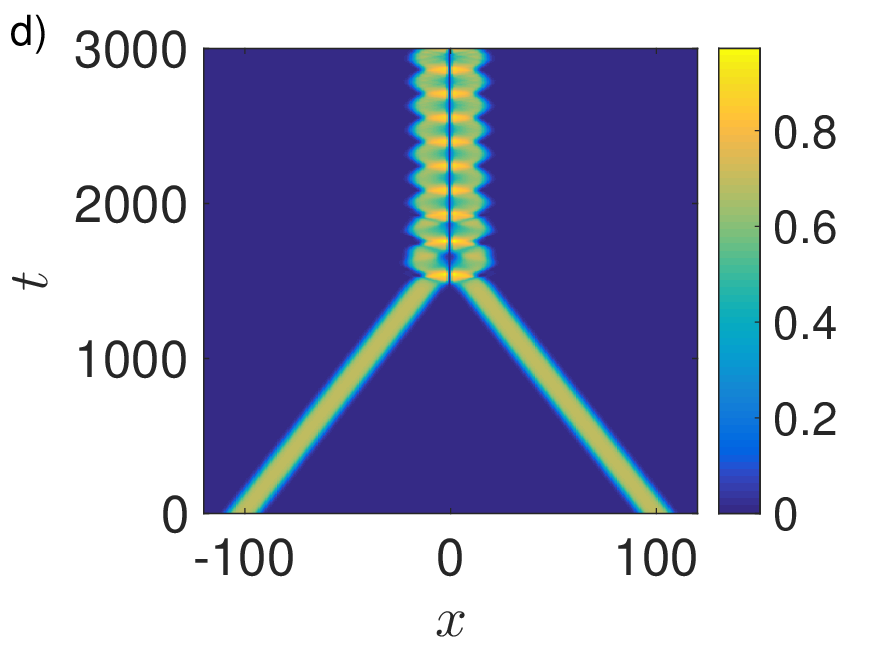}}
\centerline{\includegraphics[width=4.4cm]{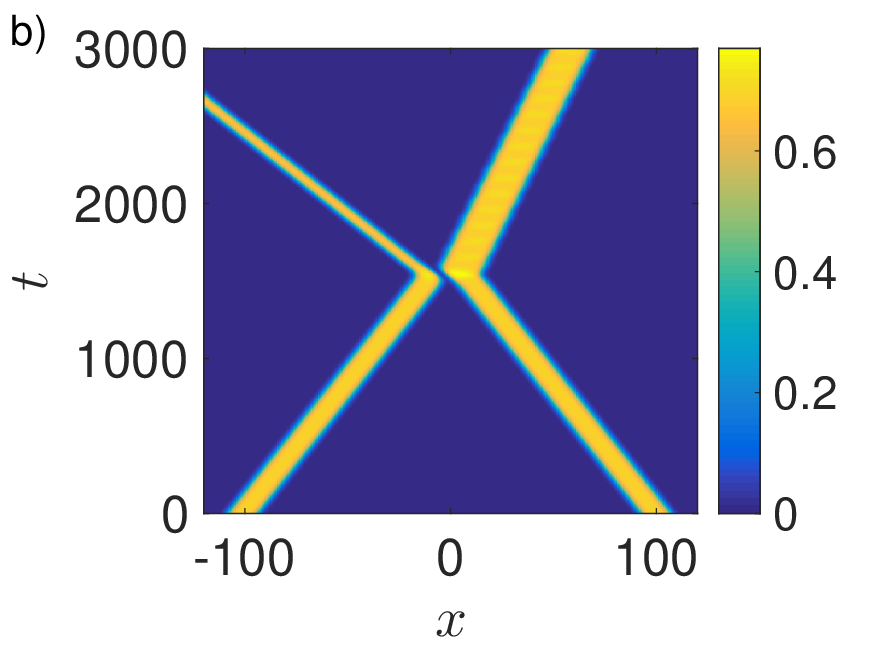} \hskip-0.1cm \includegraphics[width=4.4cm]{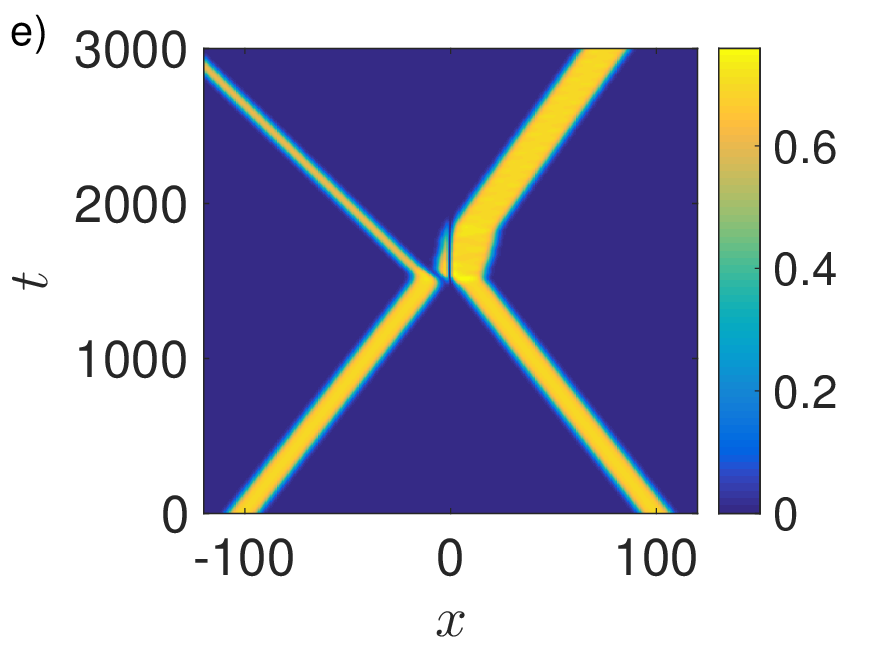}}
\centerline{\includegraphics[width=4.4cm]{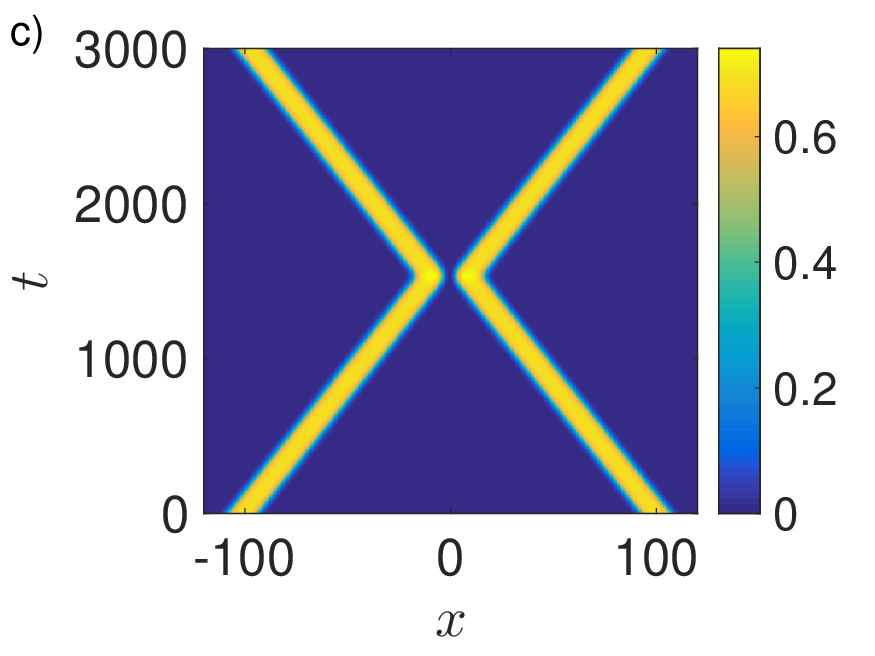} \hskip-0.1cm \includegraphics[width=4.4cm]{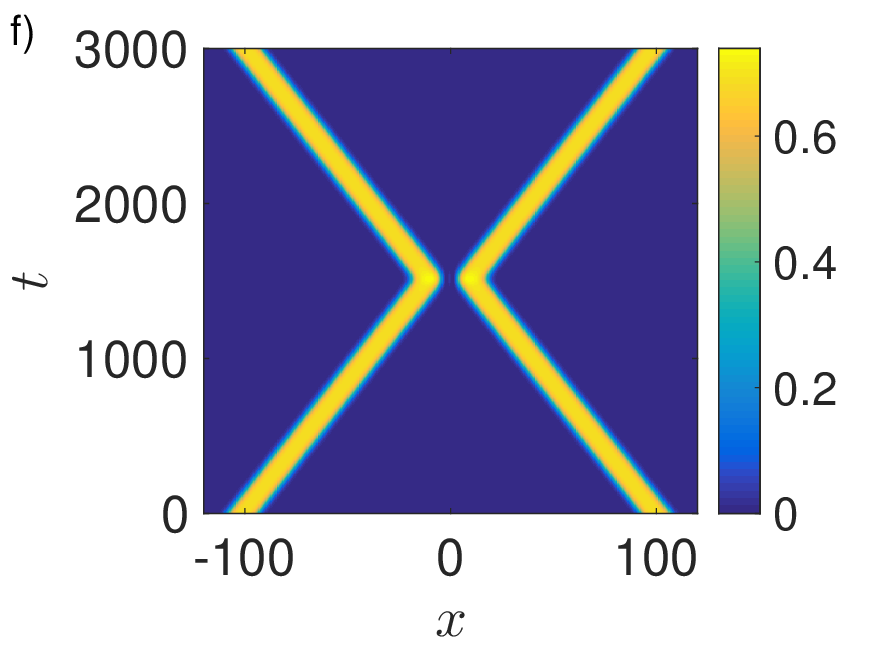}}
\caption{Scattering of larger flat-top QDs for $N=10$ and different relative phases $\theta$, in free space and in the presence of a PT potential well. The left column shows the collision dynamics for $U_0=0$: (a) $\theta=0$, (b) $\theta=\pi/2$, and (c) $\theta=\pi$. The right column shows the corresponding dynamics in an attractive potential well of depth $U_0=1$, centred at $x=0$: (d) $\theta=\pi$, (e) $\theta=3\pi/2$, and (f) $\theta=0$. The colorbar represents the density $|\psi|^2$. In all panels, the initial velocity is fixed at $v=0.06$.}
\label{fig-CollisionsN10}
\end{figure}
We also consider collisions between two large, flat-top droplets with equal norms, $N=10$. The corresponding free-space dynamics was analysed in detail in Ref.~\cite{Otajonov2024}. We compare collisions without the well (Figs.~\ref{fig-CollisionsN10}(a)--(c)) to those in the presence of the reflectionless PT potential (Figs.~\ref{fig-CollisionsN10}(d)--(f)). As with small droplets, the main effect of the well is phase imprinting rather than reflection. In this parameter regime, the scattering-induced phase shift is close to $\pi$, yielding an effective collision phase near $\pi$ and, consequently, a pronounced change in the droplets' interference pattern after the collision.
$$
\theta_{\mathrm{eff}} \approx \theta + \pi \quad (\mathrm{mod}\ 2\pi),
$$
which exchanges the roles of in-phase and out-of-phase collision scenarios (and maps $\pi/2$ to $3\pi/2$).

\begin{figure}[t]
\centerline{\includegraphics[width=4.3cm]{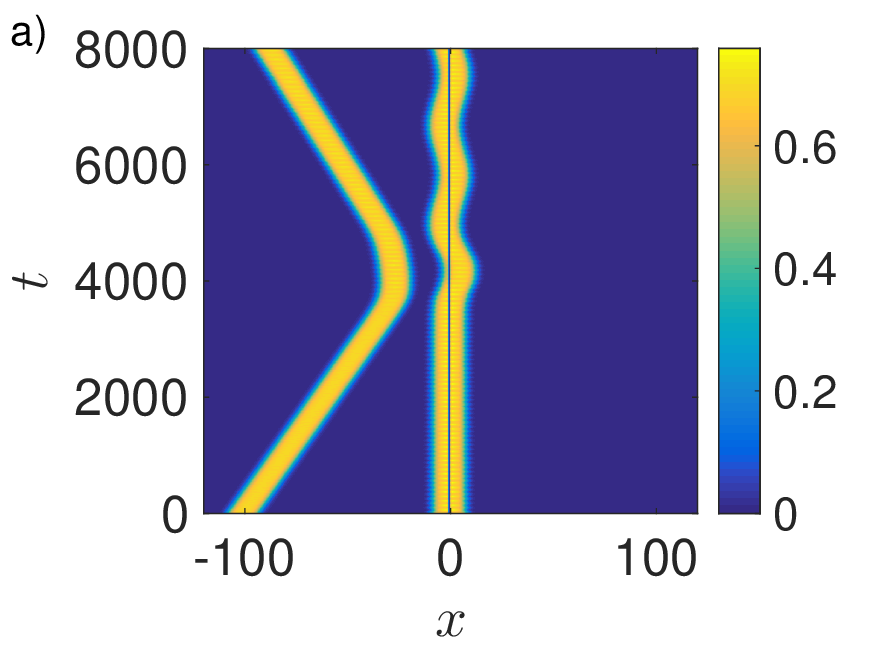} \includegraphics[width=4.3cm]{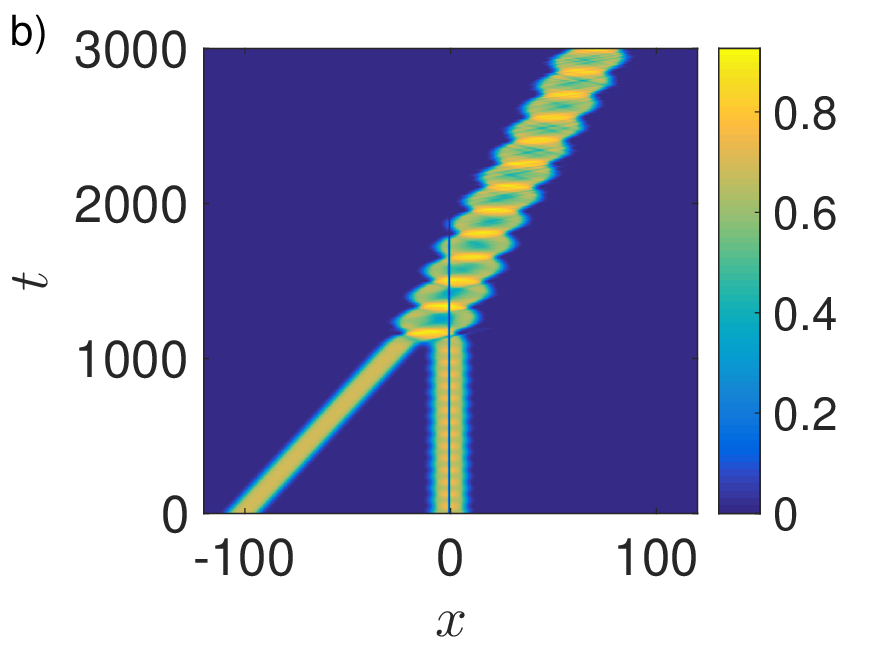}}
\centerline{\includegraphics[width=4.3cm]{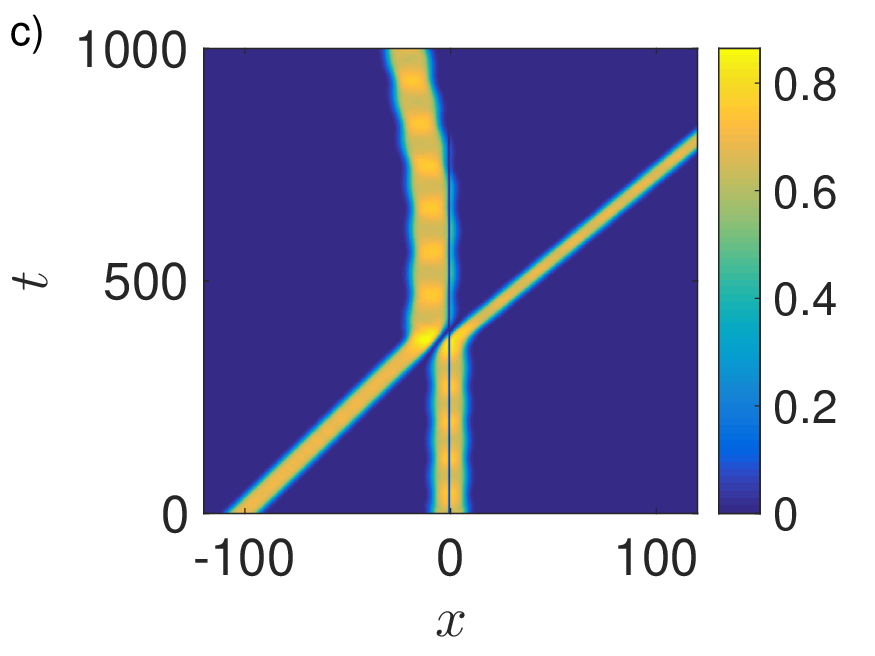}}
\caption{Collision dynamics of flat-top QDs with large particle numbers. The colorbar shows the density $|\psi|^2$. A droplet with $N=10$ is launched from $x_0=-100$ toward the potential with initial velocity $v$, and its collision with a second droplet of particle number $N=10$, initially trapped in the potential well centre, is shown for different values of $v$: (a) $v=0.02$, (b) $v=0.065$, and (c) $v=0.235$. All other parameters are the same as in Fig.~\ref{fig:TrappedMode}.}
\label{fig:TrappedModeScattering}
\end{figure}

The following analysis examines the collision dynamics between a QD initially confined in the PT well and a second droplet approaching from the left, with the goal of clarifying the collision setup. The initial condition is defined as
$$
\Psi(x,t=0)=\Psi_0(x+x_0) e^{i v x}+\phi_0(x).
$$
Here, $\Psi_0(x+x_0)$ denotes the stationary profile of the incident QD, initially centred at $x = -x_0$. This profile is characterised by the super-Gaussian ansatz in Eq.~(\ref{gaussian}). The factor $\exp(i v x)$ imparts the initial velocity $v$ to the incident droplet. It enables propagation toward the potential well at $x = 0$. The function $\phi_0(x)$ represents the trapped mode localised at the PT well, described by the modified trial function in Eq.~(\ref{eq:TrapModeTrialFunc}). Because the incident droplet is initially positioned far from the potential centre, its overlap with the trapped mode is negligible at $t = 0$. Therefore, the above superposition constitutes an appropriate initial state for investigating the subsequent collision-induced dynamics.
Figure~\ref{fig:TrappedModeScattering}(a) illustrates the collision dynamics of QDs for a small initial speed, $v=0.02$. In this case, the incident droplet reaches the potential well and interacts weakly with the droplet trapped at its centre. As a result, only a small amount of momentum is transferred to the trapped droplet, which remains localised in the well and performs small oscillations about the equilibrium position. The incoming droplet is then almost completely reflected.
Figure~\ref{fig:TrappedModeScattering}(b) shows the collision dynamics for a larger initial speed, $v=0.065$. In this regime, the interaction between the incident and trapped droplets is strong enough to form a bound state. This composite state subsequently escapes from the potential well, indicating a transmission process assisted by the droplet-droplet interaction.
Figure~\ref{fig:TrappedModeScattering}(c) presents the case of a relatively large initial speed, $v=0.235$. After the collision, part of the initially trapped droplet tunnels through the potential well and propagates to the right. The remaining part merges with the incident droplet inside the well and stays temporarily localised on the left side of the potential centre. After a finite time, this merged structure is reflected and eventually returns from the well. These results demonstrate that the collision outcome strongly depends on the incident speed, with reflection, bound-state-assisted transmission, and partial splitting occurring in different dynamical regimes.

\begin{figure}[t]
\centerline{\includegraphics[width=4.2cm]{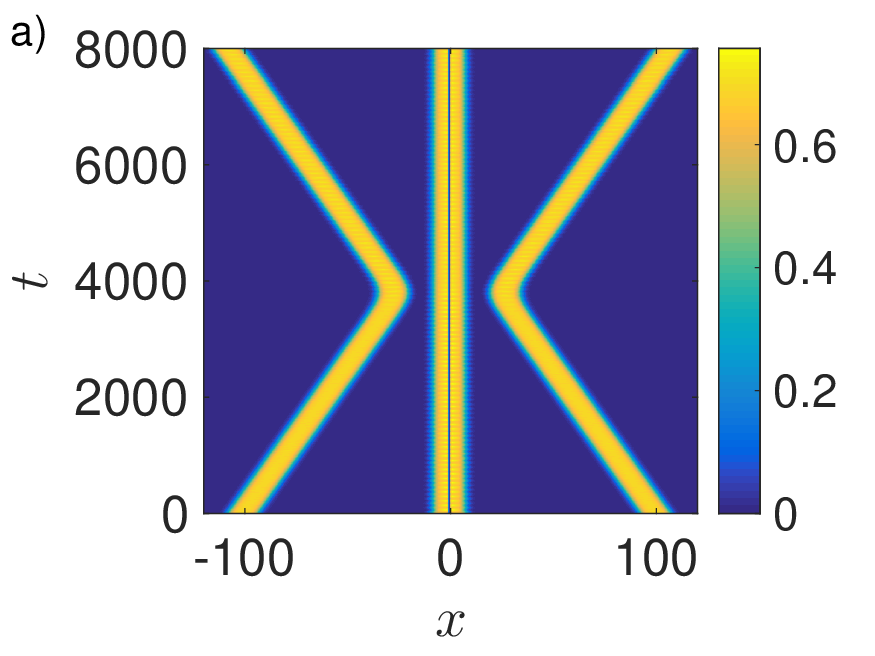} \hskip-0.1cm \includegraphics[width=4.2cm]{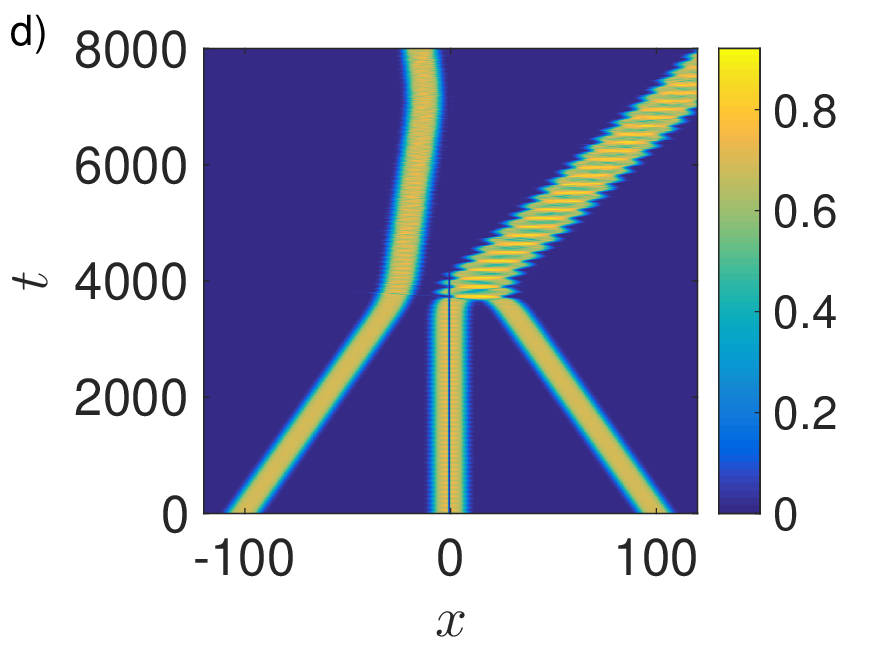}}
\centerline{\includegraphics[width=4.2cm]{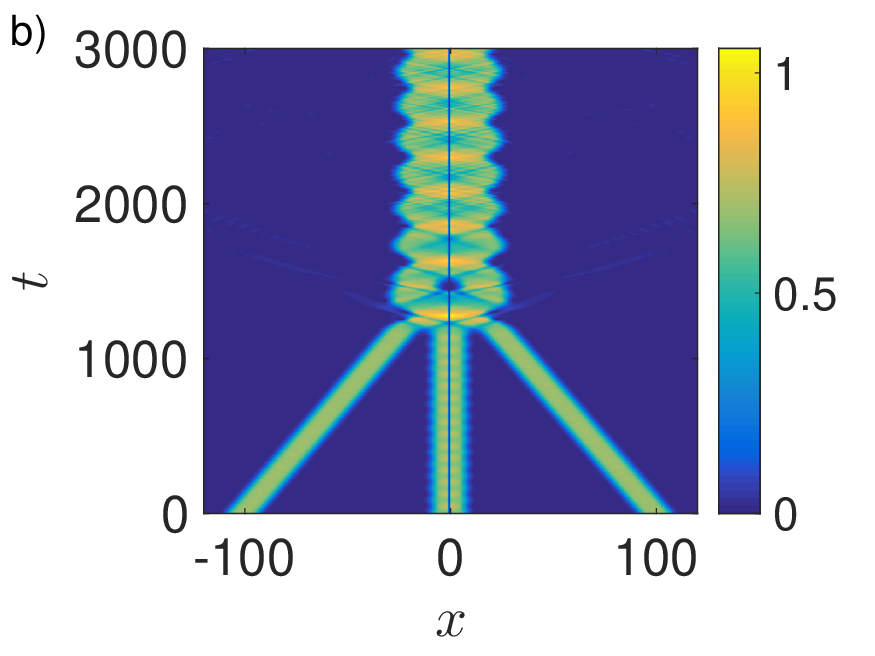} \hskip-0.1cm \includegraphics[width=4.2cm]{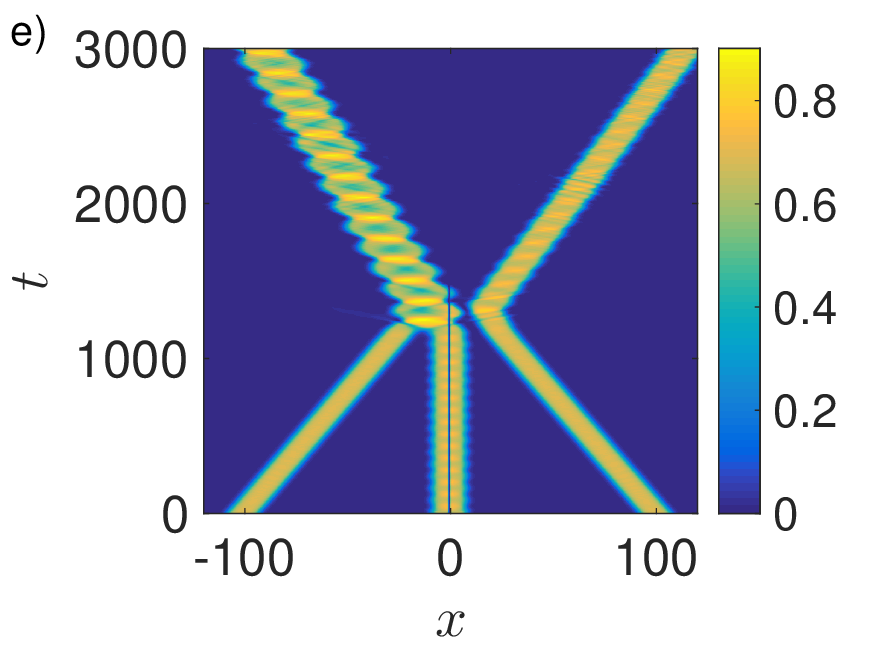}}
\centerline{\includegraphics[width=4.2cm]{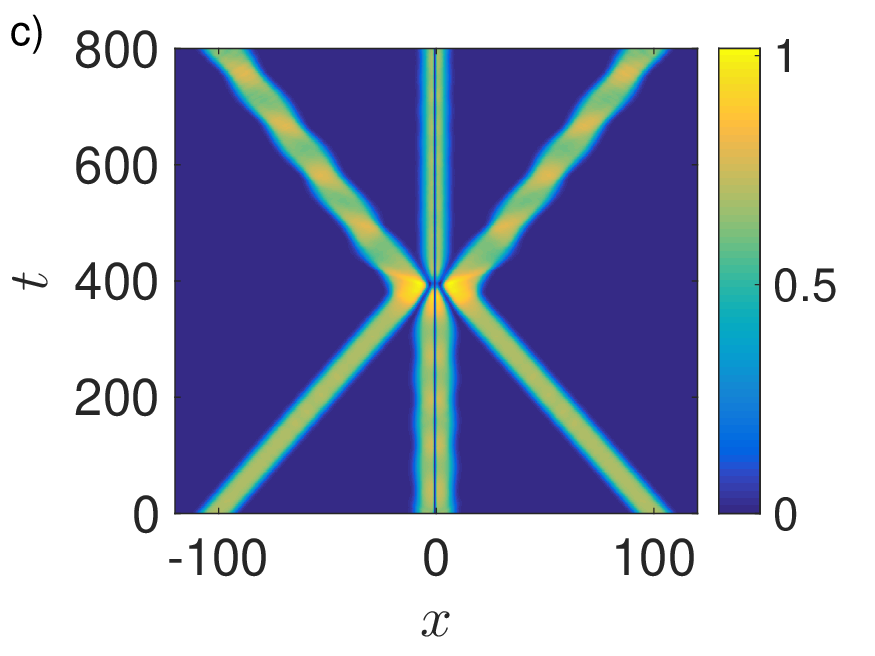} \hskip-0.1cm \includegraphics[width=4.2cm]{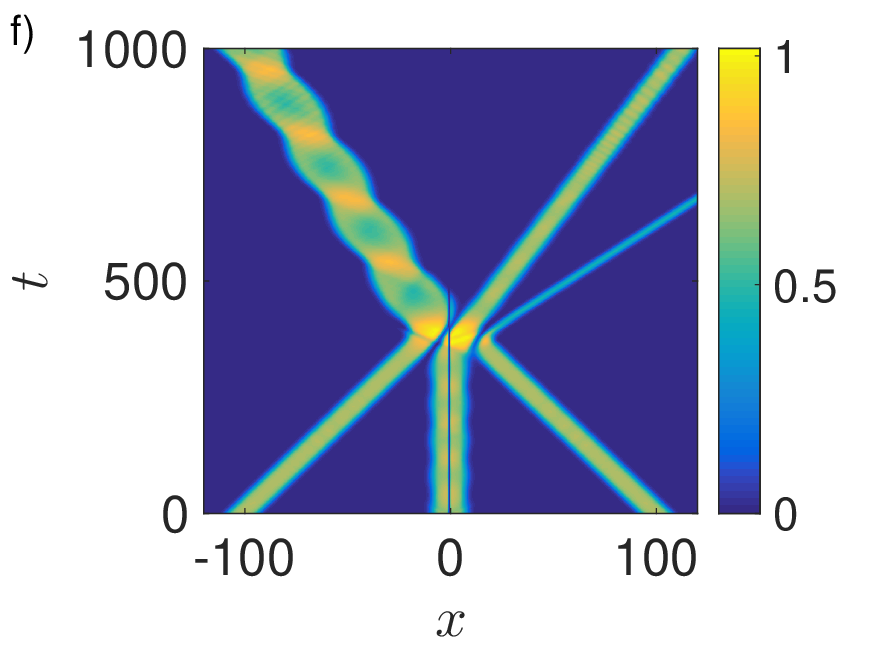}}
\caption{Collision dynamics of three flat-top QDs with equal particle numbers $N=10$ in an attractive PT potential. One droplet is trapped at the potential centre, $x=0$, while the other two approach symmetrically from opposite sides with the same initial speed $v$ and collide with it. The colorbar shows the density $|\psi|^2$. In the left column, the incident droplets differ in phase by $\pi$, in the right column, they are in phase. Panels (a,d), (b,e), and (c,f) correspond to $v=0.02$, $0.065$, and $0.235$, respectively. All other parameters are the same as in Fig.~\ref{fig:TrappedMode}.}
\label{fig:TrappedModeScattering2}
\end{figure}

Figures~\ref{fig:TrappedModeScattering2} show collision dynamics involving a QD initially confined within the PT well and two incident droplets launched symmetrically from opposite sides of the potential. The initial condition is specified as follows:
$$
\Psi(x,t=0)=\Psi_0(x+x_0)e^{i v x}+\phi_0(x)+\Psi_0(x-x_0)e^{-i v x+i\theta}.
\label{eq:superposition}
$$
Here, $\Psi_0(x\pm x_0)$ denotes the stationary super-Gaussian profiles of the incident droplets initially centred at $x=\mp x_0$, while the phase factors $\exp(\pm i v x)$ assign opposite velocities so that the droplets move toward the well at $x=0$. The parameter $\theta$ is their relative phase, and $\phi_0(x)$ represents the trapped mode localised in the PT well, described by Eq.~(\ref{eq:TrapModeTrialFunc}). Since the incident droplets are initially far from the defect, their overlap with each other and with the trapped mode is negligible. This initial condition therefore allows us to examine phase-sensitive collisions between two counter-propagating droplets and a pinned droplet state. The left column shows results for incident droplets with a relative phase difference of $\pi$ (i.e., the two incoming droplets are out of phase by $\pi$). The right column shows the case where the incident droplets are in phase. Each row shows a different initial speed: panels (a,d), (b,e), and (c,f) correspond to $v=0.02$, $0.065$, and $0.235$, respectively.

In the left column, the incident droplets differ in phase by $\pi$. The trapped mode remains robust after the collision. In Fig.~\ref{fig:TrappedModeScattering2}(a), for the smallest speed $v=0.02$, the two incident droplets are reflected nearly symmetrically after colliding with the trapped droplet. They largely preserve their initial density profiles. The central trapped droplet also remains localised and retains its shape. In Fig.~\ref{fig:TrappedModeScattering2}(b), for intermediate speed $v=0.065$, the collision leads to a composite bound state involving the incident droplets and the trapped droplet. This structure stays confined near the potential's centre during later evolution. In Fig.~\ref{fig:TrappedModeScattering2}(c), for the largest speed $v=0.235$, the density of the trapped droplet is redistributed: one part stays localised at the centre of the well, while the rest is transferred symmetrically back into the outgoing droplets. In all three cases with a relative phase difference of $\pi$, a trapped remnant persists in the well after the collision.

In the right column, the two incident droplets are in phase. The post-collision dynamics are qualitatively different. In Fig.~\ref{fig:TrappedModeScattering2}(d), for $v=0.02$, the droplet approaching from one side merges with the trapped droplet to form a bound state that escapes the well. The droplet incident from the opposite side remains trapped for some time before being reflected. In Fig.~\ref{fig:TrappedModeScattering2}(e), for $v=0.065$, a similar asymmetric outcome is observed, although the roles of the left- and right-incident droplets are interchanged. One droplet forms a bound state with the central mode, while the other continues its motion after the collision.
Figure~\ref{fig:TrappedModeScattering2}(f) corresponds to the largest incident velocity $v=0.235$ and demonstrates a significant redistribution of the norm among the outgoing QDs. Following the collision, most particles are transferred to droplets that form a bound structure in the left-return region of the potential well. The droplet initially trapped at rest near the potential minimum is displaced toward the transmission side of the well. Despite the reduction in particle number, the density profile remains flat-topped. In contrast, the droplet incident from the right does not retain its initial flat-top shape after the collision. It returns with a narrower, Gaussian-like profile, indicating that part of its norm has been transferred during the interaction.

These results show that the relative phase among incident droplets strongly affects the collision outcome. When the incident droplets are out of phase by $\pi$, the trapped mode survives and stays localised within the well. In contrast, when the two are in phase, the trapped mode is destabilised. It does not persist as a stationary localised state after the collision. This illustrates the strong phase responsiveness of multi-droplet interactions beneath an external trapping potential.

\section{Scattering of Quantum Droplets in P\"{o}schl-Teller potential barrier}
\label{sec:PTPotBar}

To investigate the scattering of QDs by a PT barrier, we combine the variational approximation, constructed from the super-Gaussian ansatz in Eq.~(\ref{gaussian}), with direct numerical simulations of the full dynamics. We launch variationally predicted droplet profiles toward the barrier at finite velocity. As velocity increases, we observe three regimes: complete reflection at low velocities, partial transmission with partial reflection at intermediate velocities, and complete transmission at high velocities. These regimes reflect competition among droplet kinetic energy, nonlinear self-binding, and the external potential.
Unlike the attractive $\ell=1$ PT well, the corresponding repulsive PT barrier is not reflectionless in the linear problem. Its exact linear transmission and reflection coefficients are given in~\ref{subsec:linear_repulsive_PT}. The same Appendix also shows that the first Born approximation predicts equal reflection probabilities for weak potentials $+U(x)$ and $-U(x)$, although the exact well and barrier scattering properties differ nonperturbatively. The barrier results below therefore contain both ordinary linear scattering and nonlinear effects associated with droplet self-binding, deformation, and fragmentation.

\begin{figure}[t]
\centerline{\includegraphics[width=4.3cm]{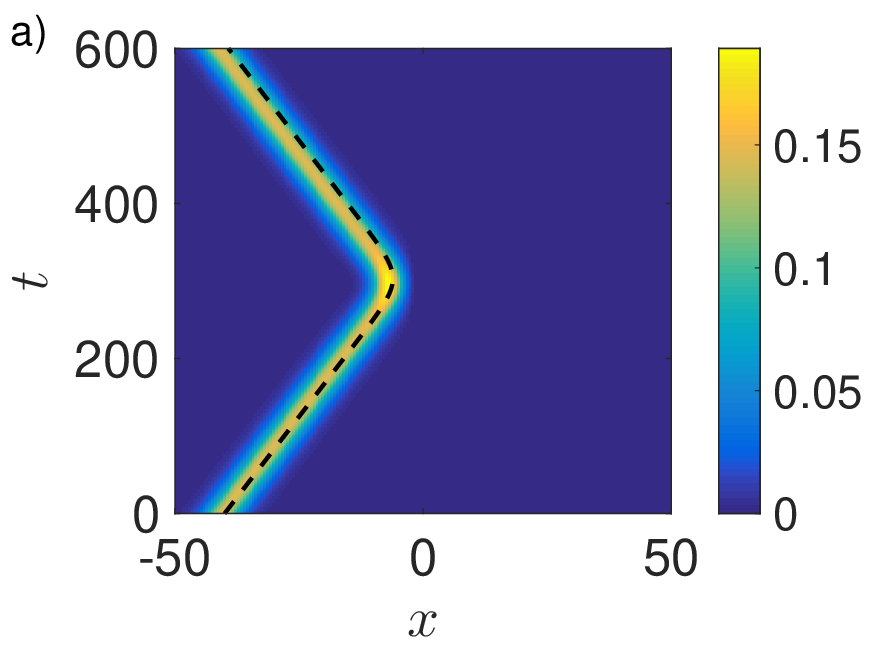} \hskip-0.1cm \includegraphics[width=4.3cm]{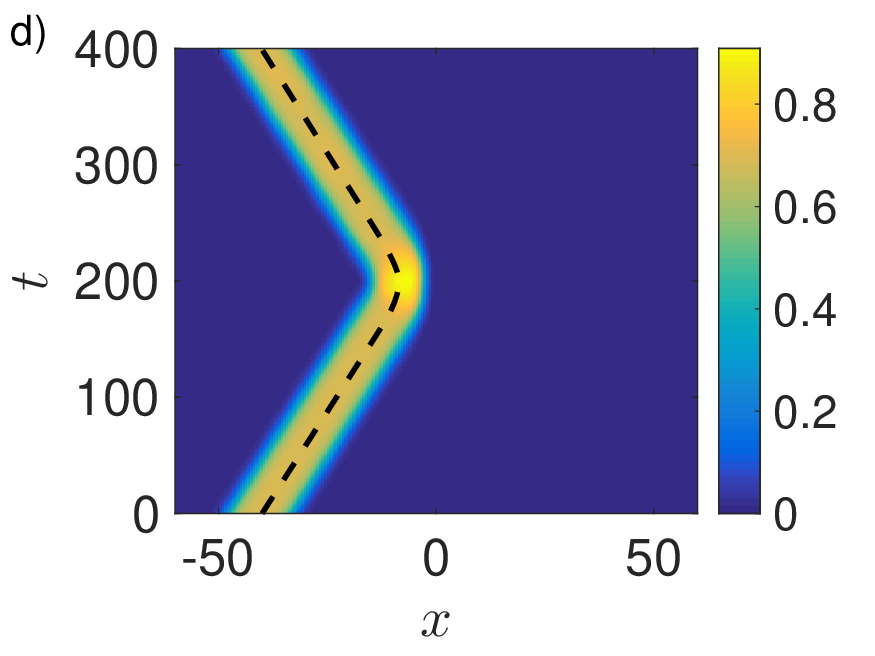}}
\centerline{\includegraphics[width=4.3cm]{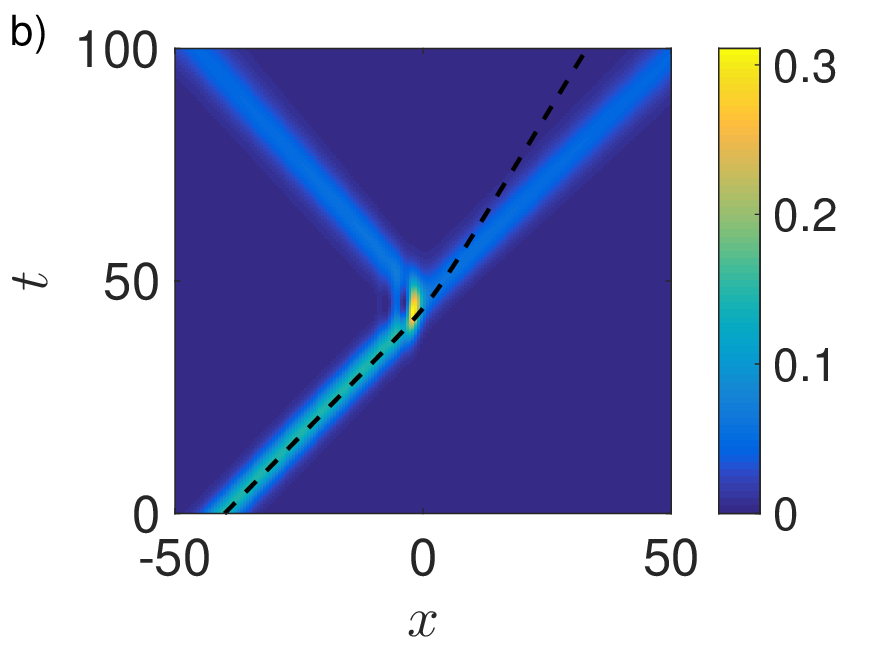} \hskip-0.1cm \includegraphics[width=4.3cm]{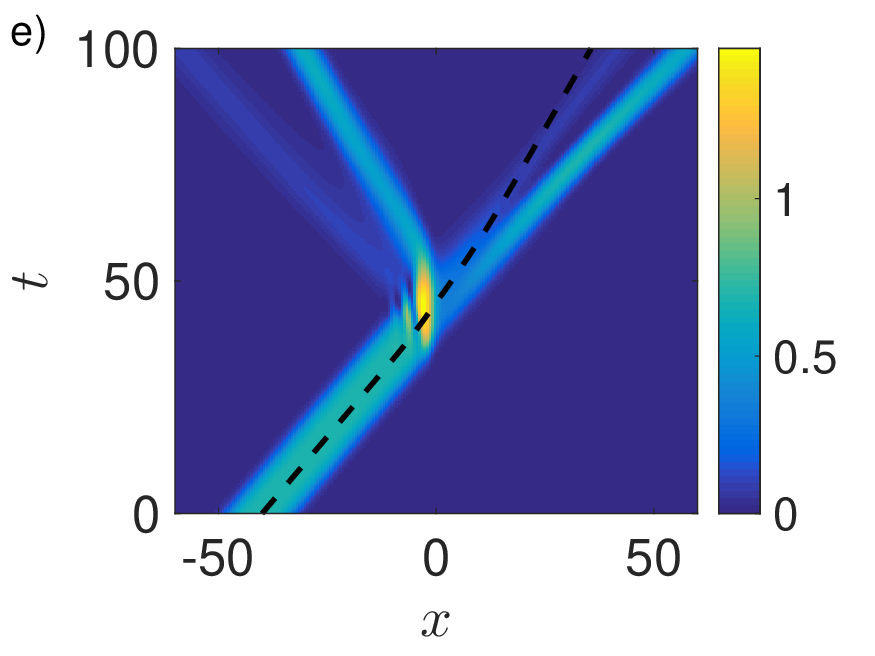}}
\centerline{\includegraphics[width=4.3cm]{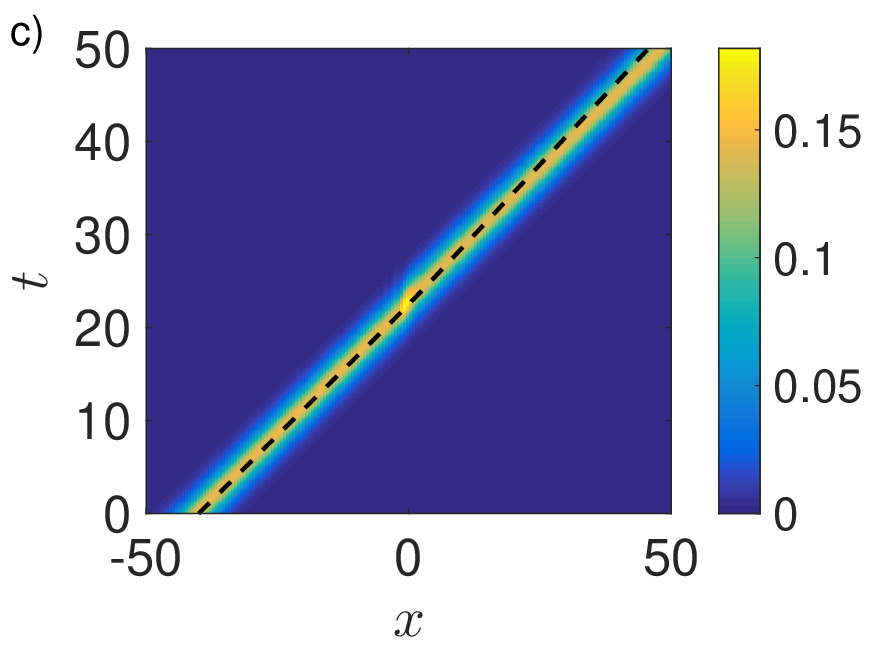} \hskip-0.1cm \includegraphics[width=4.3cm]{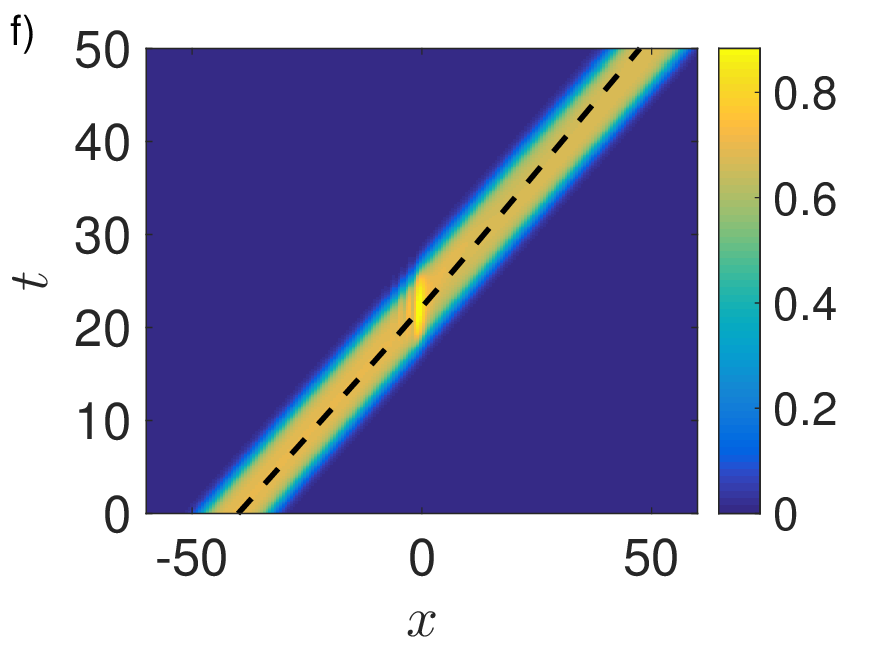}}
\caption{Scattering responses of QDs incident on a potential barrier at $x=0$ are shown for three representative incident velocities. In both columns, the three rows correspond to the main scattering regimes: total reflection, partial transmission with partial reflection, and total transmission. The left column presents the time evolution of the density profile, for a small compressible droplet with $N=1$: (a) $v=0.12$, (b) $v=0.918$, and (c) $v=1.783$. The right column shows the corresponding dynamics for a large incompressible flat-top droplet with $N=10$: (d) $v=0.171$, (e) $v=0.907$, and (f) $v=1.813$. In all panels, the dashed line represents the centre-of-mass dynamics of the VA-predicted results. The colorbar represents the density $|\psi|^2$. The remaining parameters are set at $U_0=-0.5$, $q=g=1$, and $x_0=-40$.}
\label{fig:dynSmallQDandCoefCBarrier}
\end{figure}

Figure~\ref{fig:dynSmallQDandCoefCBarrier} illustrates the time evolution of the density profile for two representative cases. The left column shows a small compressible droplet with $N=1$. The right column displays the dynamics of a larger, incompressible flat-top droplet with $N=10$. For each column, the three rows represent distinct scattering outcomes as the initial velocity varies. These outcomes are complete reflection, partial transmission with partial reflection, and complete transmission.

For sufficiently small incident velocities, when the droplet's initial kinetic energy is less than the effective barrier energy, it is fully reflected. The droplet preserves its overall shape after the collision. At sufficiently high velocities, when the droplet's kinetic energy exceeds the barrier energy, it is transmitted through the barrier and continues its motion without significant distortion. In both limits, the droplet behaves as an effective particle. The variational approximation provides an accurate description of the centre-of-mass dynamics. This agreement is demonstrated by the close correspondence between the direct numerical results and the dashed trajectories obtained from the variational equations.

A qualitatively different behaviour emerges in the intermediate regime. Here, the kinetic energy of the incident droplet is comparable to the barrier energy. In this case, the wave nature of the droplet becomes essential. Rather than behaving as a rigid particle, the droplet undergoes substantial deformation during the collision and splits into reflected and transmitted fragments. As a result, two secondary droplets emerge and propagate in opposite directions. This regime cannot be captured with the same accuracy by the reduced particle-like variational description. Internal excitations, shape deformation, and droplet splitting play a central role in the post-collision dynamics.

\begin{figure}[t]
\centerline{\includegraphics[width=4.45cm]{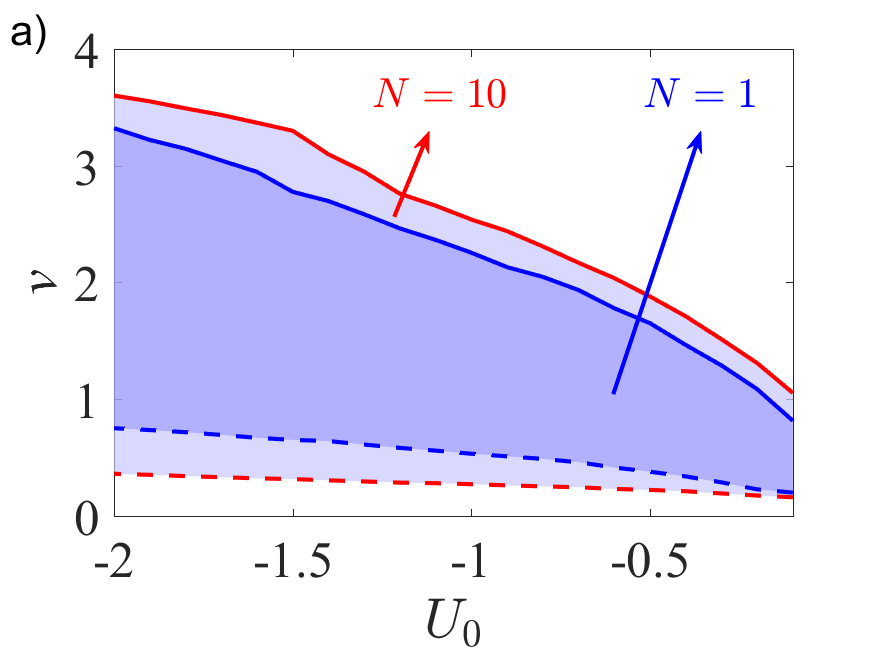} \includegraphics[width=4.45cm]{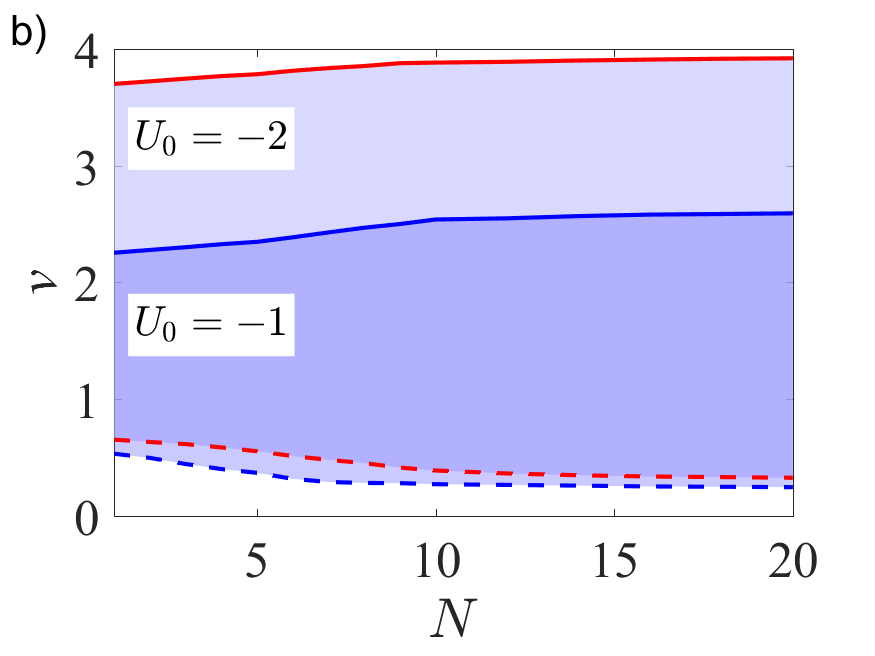}}
\caption{Scattering regimes of QDs colliding with a potential barrier are shown. Panel (a) displays the regimes in the $(U_0,v)$ plane for two particle numbers: $N=1$ (compressible) and $N=10$ (incompressible). Panel (b) shows the regimes in the $(N,v)$ plane for two barrier amplitudes: $U_0=-2$ and $U_0=-1$. In both panels, three regimes are identified: complete reflection, partial transmission with partial reflection, and complete transmission. Shaded areas mark the intermediate splitting regime. Below the lower boundary is total reflection. Above the upper boundary is total transmission. In panel (a), blue solid and dashed curves show thresholds for the compressible droplet ($N=1$), red solid and dashed curves show thresholds for the incompressible flat-top droplet ($N=10$). In panel (b), blue curves represent the weaker barrier ($U_0=-1$), and red curves indicate the stronger barrier ($U_0=-2$). Other parameters are $q = g = 1$, $x_0 = -100$.}
\label{fig-UandNBarrier}
\end{figure}
Figure~\ref {fig-UandNBarrier}(a) shows how the critical incident velocity depends on barrier strength for two representative droplets: $N=1$ (compressible) and $N=10$ (incompressible). In each case, three distinct scattering regimes are defined: complete reflection, mixed transmission and reflection, and complete transmission. For the incompressible flat-top droplet ($N=10$), the intermediate regime is indicated by the light-blue shaded area, bounded by the upper red solid and lower red dashed curves. Below the dashed lower threshold, the droplet undergoes total reflection. Above the solid upper threshold, it achieves full transmission across the barrier. The corresponding thresholds for the compressible droplet ($N=1$) are denoted by the blue solid and dashed curves. As $N$ increases, the intermediate regime broadens. Thus, higher $N$ droplets feature an expanded parameter window for droplet splitting.

The variational approximation based on the super-Gaussian ansatz accurately models droplet dynamics in the regimes of total reflection and transmission. Numerical simulations confirm this agreement. When fragmentation and distortion are minimal, the reduced variational approach reproduces the core dynamics. By contrast, accuracy decreases in intermediate regimes with partial transmission and reflection. This shortcoming is particularly pronounced post-collision, when subsequent dynamics are increasingly dominated by deformation and internal excitation phenomena that the reduced model cannot fully resolve. In this regime, following the approach of Ref.~\cite{Holmer2007}, the splitting of the droplet by the repulsive barrier admits an analytical explanation.

Figure~\ref{fig-UandNBarrier}(b) presents the scattering dynamics in the $(N,v)$ parameter space for two barrier amplitudes: $U_0=-2$ and $U_0=-1$. For the stronger barrier ($U_0 = -2$), the light-blue shaded zone between the upper red solid and lower red dashed contours denotes partial transmission and partial reflection. This zone is the crossover between total transmission (above the upper curve) and total reflection (below the lower curve). The boundaries for the weaker barrier, $U_0=-1$, are indicated by the blue solid and dashed lines. As expected, increasing the barrier amplitude extends both the total-reflection region and the intermediate scattering zone. A higher incident velocity is then needed for complete transmission through a stronger barrier.

\section{Numerical Methods and Experimental Estimates for the Model Parameters}
\label{sec:estim}

To validate the variational approach, predictions are compared with direct numerical simulations of the governing equation~(\ref{eq:gpe}). Stationary QD profiles and pinned trapped states are obtained via imaginary-time propagation. To accelerate convergence and minimise numerical transients, variationally predicted profiles serve as initial guesses. For droplets located outside the interaction region of the external potential, the super-Gaussian ansatz in Eq.~(\ref{gaussian}), with optimised variational parameters, is used as the initial condition. For trapped modes localised near the PT well, the trial function in Eq.~(\ref{eq:TrapModeTrialFunc}) is used as the initial condition. In this case, the additional tanh factor is essential because it introduces the appropriate nodal structure and phase change across the well centre, enabling the imaginary-time method to converge to the relevant pinned state.

After convergence, the resulting stationary profiles are propagated in real time to assess dynamical stability and to compare variational predictions with full time-dependent simulations. Real-time evolution is performed using a fourth-order split-step Fourier method with periodic boundary conditions. The computational domain is set as $L = 600$ and discretised using $N_x = 2048$ Fourier modes, with a time step of $dt = 10^{-4}$. This domain size is sufficient to prevent scattered droplets from interacting with their periodic images during the simulated time interval. In the scattering simulations, the corresponding VA-predicted QD profiles are used as initial conditions. For simulations involving trapped modes, the VA-predicted pinned profiles are used unless otherwise specified.

The numerical implementation follows a modified split-step Fourier scheme based on the algorithm described in Ref.~\cite{JYang}. Further details concerning the method, convergence criteria, and stability properties are given in that reference. In the present simulations, numerical convergence was verified by increasing the spatial resolution and decreasing the time step. The resulting reflection, trapping, and transmission coefficients, together with the critical velocities and density-evolution patterns, remained unchanged within numerical accuracy. Conservation of the norm during real-time propagation was also monitored as an additional consistency check.

To estimate the physical scales corresponding to the dimensionless model parameters, we consider the same symmetric $^{39}\mathrm{K}$ binary mixture used in the three-dimensional validation. The atomic mass is taken as $m_0=6.49\times10^{-26}~\mathrm{kg}$, the intraspecies scattering lengths are $a_{11}=a_{22}=a=50a_0$, and the residual mean-field interaction is chosen as $\delta a=-2a_0$, so that $a_{12}=-52a_0$. For the transverse confinement $\omega_\perp=2\pi\times500~\mathrm{Hz}$, the oscillator length is $l_\perp=\sqrt{\hbar/(m_0\omega_\perp)}\simeq0.720~\mu\mathrm{m}$, which remains much larger than the scattering length $a$ and is therefore consistent with an elongated quasi-one-dimensional reduction. For this parameter set, the physical-to-dimensionless conversion factors obtained from the effective one-dimensional cubic-quartic model are
$\psi_s=2.75\times10^4~\mathrm{m}^{-1/2}$, $t_s=1.98\times10^{-3}~\mathrm{s}$, and $x_s=1.80~\mu\mathrm{m}$.
Thus, one dimensionless time unit corresponds to $1.98~\mathrm{ms}$, and one dimensionless length unit corresponds to $1.80~\mu\mathrm{m}$. A dimensionless computational domain $L=600$ therefore corresponds to
$L_{\mathrm{phys}}=Lx_s\simeq1.08~\mathrm{mm}$.
The atom-number conversion is
$N_{\mathrm{atoms}}=\psi_s^2x_sN$.
Consequently, for a dimensionless norm $N=3$, one obtains
$N_{\mathrm{atoms}}\simeq4.1\times10^3$,
placing the simulated droplet in a realistic few-thousand-atom regime. For comparison, the validated state with $N_{\mathrm{atoms}}=1.5\times10^4$ corresponds to $N\simeq11.01$ in the same dimensionless units.

For the critical dimensionless velocity $v_{\mathrm{cr}}=0.114$, the corresponding physical speed is
$v_{\mathrm{cr}}^{\mathrm{phys}}=v_{\mathrm{cr}}x_s/t_s \simeq 0.104~\mathrm{mm\,s^{-1}}$.
A droplet launched from $x_0=-100$ reaches a defect at $x=0$ after
$t_{\mathrm{hit}}=|x_0|t_s/v_{\mathrm{cr}}\simeq1.74~\mathrm{s}$.
This relatively long time can be reduced either by using a smaller initial displacement or by launching the droplet at a higher velocity, provided the intended scattering regime is preserved.
The PT potential is written as
$V(x)=-U_0\,\mathrm{sech}^2(\alpha x)$.
For $U_0=\alpha=1$, the physical depth is set by the energy scale $\hbar/t_s$, giving
$U_{0,\mathrm{phys}}/k_B\simeq3.9~\mathrm{nK}$,
or equivalently $U_{0,\mathrm{phys}}/h\simeq80.3~\mathrm{Hz}$.
The full width at half maximum is
$\mathrm{FWHM}=2\,\mathrm{arcosh}\left(\sqrt{2}\right)\,x_s\simeq3.17~\mu\mathrm{m}$,
and the corresponding crossing time is
$t_{\mathrm{cross}}\sim \mathrm{FWHM}/v_{\mathrm{cr}}^{\mathrm{phys}}\simeq30.7~\mathrm{ms}$. These estimates agree with the experimental values reported in Ref.~\cite{Marchant2}. The corresponding physical scales are readily tunable in experiments: by adjusting the s-wave scattering lengths and the transverse confinement frequency, one can control the characteristic length and time scales.

\section{Conclusions}
\label{sec:conc}

 We study the scattering of elongated one-dimensional droplets from localised PT defects using the extended Gross-Pitaevskii model with a beyond-mean-field correction. We developed a two-stage variational method: first, applying a super-Gaussian profile to the stationary droplet, second, using a position-dependent ansatz to determine turning points and trapped states near the defect.
In the attractive reflectionless PT well, scattering interactions show a sharp transition at a critical velocity between quantum reflection and transmission. Near this threshold, the droplet forms long-lived metastable configurations, whose structure depends on particle number. Small droplets act like solitons and create symmetric trapped states at the threshold. Large flat-top droplets form related asymmetric trapped states away from the well centre, revealing their incompressible, liquid-like nature. The turning-point energy landscape and effective potential explain the prolonged droplet residence and the emergence of near-critical states.

Our central result is the nonmonotonic dependence of critical velocity on atom number. In small, compressible droplets, critical velocity rises with atom number, peaks at the flat-top crossover, then falls in the incompressible regime. This pattern shows a structural change: small droplets localise and densify as they grow, while large ones keep peak density and simply grow in size. Thus, scattering marks the crossover from soliton-like to liquid-like behaviour.
We showed that the reflectionless PT well acts as a phase-imprinting defect, imparting an approximate $\pi$-phase shift that alters droplet collisions relative to free space. This effect is clearest with a pinned droplet. Whether a localised remnant remains depends strongly on relative phase: out-of-phase collisions preserve the trapped mode, while in-phase collisions destabilise it. Collision outcomes also depend on droplet size: small droplets remain robust and nearly elastic, while large, flat-top droplets exhibit greater inelasticity, internal excitation.

For the repulsive PT barrier, dynamics exhibit three velocity regimes: (i) full reflection at low velocities, where kinetic energy is insufficient to cross, (ii) partial reflection and transmission at intermediate velocities, with the droplet splitting and deforming, (iii) full transmission at high velocities. Regime boundaries depend on velocity, barrier height, and particle number. The variational particle-like model captures full reflection and transmission but is less accurate in the intermediate regime, where wave effects and excitations dominate.

Overall, quantum droplet scattering by localised defects depends on nonlinear self-binding, internal structure, collective motion, phase imprinting, and external confinement. This study expands previous work on droplet scattering to the experimentally more relevant case of elongated traps and demonstrates that localised defects can control reflection, transmission, trapping, and phase-sensitive collisions in nonlinear matter-wave systems.  The developed variational approach may also be useful for investigating defect-induced dynamics in other self-bound nonlinear states for which no exact analytical profiles are available.

For future investigations will be interesting to consider the processes of the generation of solitonic systems due to the scattering process \cite{Arazo2021} and the extension to the scattering of the higher dimensional quantum droplets (see \cite{Chen2024,Deekshita2025,Sanjay2025}) on the different defects.

\appendix
\section{Linear scattering by the P\"oschl-Teller potential}
\label{sec:linear_PT_limit}

Prior to analysing nonlinear QD scattering, it is instructive to recall the corresponding linear scattering problem~\cite{Flugge,Landau3,Lekner2007}. The linear reference case serves three purposes. First, it clarifies the precise meaning of the term reflectionless, which applies strictly to linear waves. Second, it explains why the choice $\alpha^2=U_0$ selects a specific member of the PT family. Third, it provides a clear comparison between attractive wells and repulsive barriers, including their distinct phase-imprinting properties.

Upon omission of the nonlinear terms in the governing equation, the resulting stationary linear problem is formulated as follows.

$$
\left[-\frac{1}{2}\frac{d^2}{dx^2}+V(x)\right]\phi=E\phi,
\qquad E=\frac{k^2}{2},
$$

where $k$ is the incident wave number. For a unified notation, we write the PT profile as

$$
V_s(x)=s\,U_0\operatorname{sech}^2(\alpha x),
\qquad U_0>0,
\qquad s=\pm1 .
$$

The choice $s=-1$ gives the attractive well, while $s=+1$ gives the repulsive barrier. With $y=\alpha x$, $K=k/\alpha$, and $\Lambda=2U_0/\alpha^2$, the linear equation becomes

$$
\frac{d^2\phi}{dy^2}
+
\left[
K^2-s\Lambda\operatorname{sech}^2y
\right]\phi=0 .
$$

The attractive and repulsive PT potentials exhibit identical spatial profiles, however, they enter the scaled scattering problem with opposite signs. While this sign difference is not significant at the level of the Born approximation, it becomes critical in the exact scattering problem.

\subsection{Attractive reflectionless well}
\label{subsec:linear_attractive_PT}

For the attractive well $V_{\rm w}(x)=-U_0\operatorname{sech}^2(\alpha x)$, the scaled equation becomes

$$
\frac{d^2\phi}{dy^2}
+
\left[
K^2+\Lambda\operatorname{sech}^2y
\right]\phi=0 .
$$

The standard reflectionless PT family is obtained when $\Lambda=\ell(\ell+1)$, where $\ell=1,2,3,\ldots$. Equivalently,

$$
\frac{2U_0}{\alpha^2}=\ell(\ell+1),
\qquad
\alpha^2=\frac{2U_0}{\ell(\ell+1)} .
$$

The reflectionless $\operatorname{sech}^2$ potential family, as discussed in Ref.~\cite{Lekner2007}, exhibits vanishing linear reflection coefficients for all incident wave numbers when $\ell$ is an integer. Consequently, $\mathcal R_{\rm w}(k)=0$ and $\mathcal T_{\rm w}(k)=1$.

For a general attractive PT well, $\lambda$ can be defined such that $\lambda(\lambda+1)=\Lambda$, or equivalently, $\lambda=(-1+\sqrt{1+4\Lambda})/2$. The exact reflection probability is then given by

$$
\mathcal R_{\rm w}(k)
=
\frac{\sin^2(\pi\lambda)}
{\sinh^2(\pi K)+\sin^2(\pi\lambda)} .
$$

This expression demonstrates that the attractive PT well is reflectionless when $\lambda=\ell$ is an integer.

In this study, $\alpha^2$ is set equal to $U_0$, resulting in $\Lambda=2U_0/\alpha^2=2$. This choice corresponds to the $\ell=1$ member of the reflectionless PT family. Consequently, the linear reference problem is the one-bound-state reflectionless PT well. Its normalizable bound state is

$$
\phi_0(x)=C_0\operatorname{sech}(\alpha x),
\qquad
E_0=-\frac{\alpha^2}{2}=-\frac{U_0}{2},
$$

where $C_0$ is a normalization constant.

The same linear problem also admits the zero-energy threshold solution $\phi_{\rm th}(x)\propto\tanh(\alpha x)$. Specifically,
$$
\left[
-\frac{1}{2}\frac{d^2}{dx^2}
-U_0\operatorname{sech}^2(\alpha x)
\right]\tanh(\alpha x)=0
$$
when $\alpha^2=U_0$, the threshold state is not square-integrable and thus does not constitute a second bound state. However, this state is physically significant because it changes sign across the center of the well and is associated with the low-energy phase jump of the reflectionless transmission channel.

For the $\ell=1$ attractive well, the transmission amplitude can be expressed, up to a convention-dependent overall phase, as follows:

$$
t_{\rm w}(k)=\frac{k+i\alpha}{k-i\alpha},
\qquad |t_{\rm w}(k)|^2=1 .
$$

The transmitted linear wave therefore acquires the phase

$$
\Delta\theta_{\rm w}(k)
=
2\arctan\left(\frac{\alpha}{k}\right),
$$
which approaches $\pi$ as $k\rightarrow0$. This phase-imprinting property is particularly valuable in the current nonlinear context, as it links the linear reflectionless channel to the emergence of a nodal nonlinear trapped mode near the reflection and transmission threshold.

This analysis also clarifies why increasing $U_0$ does not generate an additional linear bound state when the constraint $\alpha^2=U_0$ is maintained. Although increasing $U_0$ deepens the well, it simultaneously narrows it. The ratio $2U_0/\alpha^2=2$ remains constant, so the scaled linear problem stays within the $\ell=1$ reflectionless class. Achieving a two-bound-state reflectionless PT well would require $\ell=2$, that is, $2U_0/\alpha^2=6$, or $\alpha^2=U_0/3$. Therefore, along the trajectory $\alpha^2=U_0$, there is no particular value, such as $U_0\simeq2$, at which an additional normalizable linear bound state emerges. If the variational approximation loses accuracy for larger $U_0$, this discrepancy should be attributed to nonlinear droplet deformation, enhanced localization by the defect, excitation of internal modes, and the limited adaptability of the trial function. This interpretation aligns with the observation of Ref.~\cite{Khawaja2021}: multinode trapped modes require relaxing the one-bound-state reflectionless condition and employing a broader potential that can accommodate additional nodes.

The same threshold-state structure also provides justification for the nodal factor employed later in the variational ansatz,
$$
\phi(x)=A\psi_0[\gamma(x-x_0)]\tanh(\beta x).
$$

The factor $\tanh(\beta x)$ does not serve to introduce an additional normalizable bound state. Instead, it enforces the odd threshold-channel structure and the corresponding $\pi$-phase jump. The parameter $\beta$ enables renormalization of the nodal length scale by accounting for both the finite droplet width and nonlinear self-interaction.

\subsection{Repulsive P\"oschl-Teller barrier}
\label{subsec:linear_repulsive_PT}

For the repulsive barrier $V_{\rm b}(x)=U_0\operatorname{sech}^2(\alpha x)$, the scaled linear equation is

$$
\frac{d^2\phi}{dy^2}
+
\left[
K^2-\Lambda\operatorname{sech}^2y
\right]\phi=0 .
$$

This scenario does not correspond to the attractive reflectionless PT problem. The barrier lacks a normalizable bound state and generally exhibits a nonzero reflection coefficient. The exact transmission probability can be expressed as

$$
\mathcal T_{\rm b}(k)
=
\frac{\sinh^2(\pi K)}
{\sinh^2(\pi K)+
\cos^2\!\left[\pi\sqrt{\frac{1}{4}-\Lambda}\right]} .
$$

For $\Lambda>1/4$, this expression is understood by analytic continuation, so that

$$
\mathcal T_{\rm b}(k)
=
\frac{\sinh^2(\pi K)}
{\sinh^2(\pi K)+
\cosh^2\!\left[\pi\sqrt{\Lambda-\frac{1}{4}}\right]} .
$$

The reflection probability is $\mathcal R_{\rm b}(k)=1-\mathcal T_{\rm b}(k)$.

For the same scaling used in the attractive case, $\alpha^2=U_0$, one has $\Lambda=2$. Therefore,

$$
\mathcal T_{\rm b}(k)
=
\frac{\sinh^2(\pi k/\alpha)}
{\sinh^2(\pi k/\alpha)+
\cosh^2\!\left(\frac{\pi\sqrt{7}}{2}\right)} .
$$

These results indicate that the repulsive PT barrier does not exhibit reflectionless behavior in the linear limit. The barrier becomes nearly transparent only in the high-energy limit $k/\alpha\gg1$.

Although the repulsive barrier is not reflectionless, it still generates scattering phases. When the transmission amplitude is expressed as $t_{\rm b}(k)=|t_{\rm b}(k)|e^{i\delta_{\rm b}(k)}$, $\delta_{\rm b}(k)$ represents the transmitted phase shift. In contrast to the attractive reflectionless well, the barrier typically produces both transmitted and reflected waves. Therefore, it does not impart a single, well-defined global phase to the entire incident wave packet, except approximately in the high-energy regime where transmission is nearly complete.

This distinction is also evident in the first Born approximation. For the linear equation

$$
-\frac{1}{2}\phi_{xx}+V(x)\phi=\frac{k^2}{2}\phi,
$$

the first Born reflection amplitude is

$$
r^{(1)}(k)
=
-\frac{i}{k}
\int_{-\infty}^{+\infty}V(x)e^{2ikx}\,dx .
$$

Let $U_{\rm PT}(x)=U_0\operatorname{sech}^2(\alpha x)$, and compare $V_+(x)=+U_{\rm PT}(x)$ with $V_-(x)=-U_{\rm PT}(x)$. The first Born amplitudes satisfy $r_-^{(1)}(k)=-r_+^{(1)}(k)$, and hence $\mathcal R_-^{(1)}(k)=\mathcal R_+^{(1)}(k)$. Thus, in the first Born approximation, a weak PT well and the corresponding weak PT barrier have the same reflection probability.

For the PT profile, the Fourier transform

$$
\int_{-\infty}^{+\infty}
\operatorname{sech}^2(\alpha x)e^{iqx}\,dx
=
\frac{\pi q/\alpha^2}
{\sinh\!\left(\pi q/2\alpha\right)}
$$

is used with its $q\rightarrow0$ limit understood by continuity. Setting $q=2k$, one obtains

$$
\mathcal R_\pm^{(1)}(k)
=
\left[
\frac{2\pi U_0}
{\alpha^2\sinh(\pi k/\alpha)}
\right]^2 .
$$

The equivalence between the reflection probabilities for the well and barrier is valid only at leading order in perturbation theory. The Born approximation becomes invalid outside the weak-scattering or high-energy regime and should not be used when its value approaches or exceeds unity. For the exact attractive $\ell=1$ well, $\mathcal R_{\rm w}=0$ for all $k$, whereas the repulsive barrier generally results in $\mathcal R_{\rm b}\neq0$. This difference is due to nonperturbative effects that are not accounted for at first Born order.

Within the same approximation, the transmitted phase remains unaffected. The first Born transmission amplitude is

$$
t^{(1)}(k)
\simeq
1-\frac{i}{k}\int_{-\infty}^{+\infty}V(x)\,dx,
$$

so that the leading transmitted phase shift is

$$
\delta_T^{(1)}(k)
\simeq
-\frac{1}{k}\int_{-\infty}^{+\infty}V(x)\,dx .
$$

For the repulsive PT barrier this gives

$$
\delta_{T,{\rm b}}^{(1)}(k)
\simeq
-\frac{2U_0}{\alpha k},
$$

while for the corresponding attractive well the sign is reversed,
$$
\delta_{T,{\rm w}}^{(1)}(k)
\simeq
+\frac{2U_0}{\alpha k}.
$$
Thus, the barrier imparts a phase to the transmitted component, however, this phase depends on velocity and does not remain fixed at $\pi$. In contrast, the attractive reflectionless well in the $\ell=1$ case exhibits an exact low-energy phase shift that approaches $\pi$.

It is important to emphasize both the connection and the limitations of this linear analysis when applied to the nonlinear Gross-Pitaevskii equation. In the nonlinear model, the external potential contributes to the local dynamical phase through a factor of the form
$$
\exp\!\left[-i\int V(x(t))\,dt\right],
$$
however, the cubic attraction and quartic LHY repulsion also result in density-dependent nonlinear phase accumulation. For rapidly moving, weakly deformed transmitted droplets, the barrier-induced phase may be estimated semiclassically as
$$
\Delta\theta_{\rm b}\simeq-\frac{2U_0}{\alpha v},
$$
apart from nonlinear self-phase corrections, the outgoing field in both the reflection regime and the intermediate splitting regime does not constitute a single coherent transmitted object. Reflected and transmitted fragments may possess distinct phases and internal excitations, rendering a single global phase imprint ill-defined.

The PT potential is employed not solely for mathematical convenience. The attractive reflectionless well serves as a benchmark that eliminates linear backscattering, thereby isolating the nonlinear mechanisms of droplet reflection, trapping, transmission, and phase-sensitive collisions. By contrast, a generic smooth well or barrier alters the numerical thresholds and may introduce ordinary linear reflection. However, the qualitative mechanisms discussed below, including the formation of near-critical zero-speed states, the separatrix between reflection and transmission, the significance of phase imprinting, and the distinction between compressible and flat-top droplets, do not depend on the exact solvability of the PT potential.

\section*{Acknowledgements}
This work has been supported by the State Budget of the Republic of Uzbekistan (Grant No. 2026 year award).



\begin{thebibliography}{99}
%
\bibitem{Frid}
H. Friedrich, A. Jurisch, Quantum reflection times for attractive potential tails, Phys. Rev. Lett. {\bf 92}, 103202 (2004).
%
\bibitem{Druzhinina}
V. Druzhinina, M. DeKieviet, Experimental observation of Quantum reflection far from threshold, Phys. Rev. Lett. {\bf 91}, 193202 (2003).
%
\bibitem{Cote}
R. C\^{o}t\'{e}, B. Segev, Quantum reflection engineering: The bichromatic evanescent-wave mirror, Phys. Rev. A {\bf 67}, 041604(R) (2003).
%
\bibitem{Shimizu}
F. Shimizu, Specular Reflection of very slow metastable neon atoms from a solid surface, Phys. Rev. Lett. {\bf 86}, 987 (2001).
%
\bibitem{Pasquini}
T. A. Pasquini, Y. Shin, C. Sanner, M. Saba, A. Schirotzek, D. E. Pritchard, W. Ketterle, Quantum reflection from a solid surface at normal incidence, Phys. Rev. Lett. {\bf 93}, 223201 (2004).
%
\bibitem{atom}
L. Amico, M. Boshier, G. Birkl, A. Minguzzi et al., Roadmap on Atomtronics: State of the art and perspective, AVS Quantum Sci. {\bf 3}, 039201 (2021).
%
\bibitem{QI}
B. Gertjerenken, T. P. Wiles, C. Weiss, Progress towards quantum-enhanced interferometry with harmonically trapped quantum matter-wave bright solitons, Phys. Rev. A {\bf 94}, 053638 (2016).
%
\bibitem{QI1}
A. D. Martin, J. Ruostekoski, Quantum dynamics of atomic bright solitons under splitting and recollision, and implications for interferometry, New J. Phys. {\bf 14}, 043040 (2012).
%
\bibitem{Goodman}
R. H. Goodman, P. J. Holmes, M. I. Weinstein, Strong NLS soliton-defect interactions, Physica D {\bf 192}, 215 (2004).
%
\bibitem{Lee}
C. Lee, J. Brand, Enhanced quantum reflection of matter-wave solitons, EuroPhys. Lett. {\bf 73}, 321 (2006).
%
\bibitem{Cao}
X. D. Cao, B. A. Malomed, Soliton-defect collisions in the nonlinear Schr\"{o}dinger equation, Phys. Rev. Lett. A {\bf 206}, 177 (1995).
%
\bibitem{Fornish}
K. Forinash, M. Peyrard, B. Malomed, Interaction of discrete breathers with impurity modes, Phys.Rev. E {\bf 49}, 3400 (1994).
%
\bibitem{Marchant}
A. L. Marchant, T. P. Billam, T. P. Wiles, M. M. H. Yu, S. A. Gardiner, S. L. Cornish, Controlled formation and reflection of a bright solitary matter-wave, Nat. Commun. {\bf 4}, 1865 (2013).
%
\bibitem{Sakaguchi}
H. Sakaguchi, M. Tamura, Scattering and trapping of nonlinear Schr\"{o}dinger solitons in external potentials, J. Phys. Soc. Jpn. {\bf 73}, 503 (2004).
%
\bibitem{Sakkaf}
L. Al Sakkaf, U. Al Khawaja, Reflectionless potentials and resonant scattering of flat-top and thin-top solitons, Phys. Rev. E {\bf 107}, 014202 (2023).
%
\bibitem{Zeng}
L. Zeng, J. Zeng, Gaussian-like and flat-top solitons of atoms with spatially modulated repulsive interactions, JOSA B {\bf 36}, 2278 (2019).
%
\bibitem{Umarov}
B. A. Umarov, N. A. B. Aklan, M. R. Rosly, T. H. Hassan, Flat top solitons on linear gaussian potential, J. Phys.: Conf. Ser. {\bf 890} (1), 012071 (2017).
%
\bibitem{Kagan}
P. O. Fedichev, Yu. Kagan, G. V. Shlyapnikov, J. T. M. Walraven, Influence of nearly resonant light on the scattering length in low-temperature atomic gases, Phys. Rev. Lett. {\bf 77}, 2913 (1996).
%
\bibitem{Jap1}
R. Yamazaki, S. Taie, S. Sugawa, Y. Takahashi, Submicron spatial modulation of an interatomic interaction in a Bose-Einstein condensate, Phys. Rev. Lett. {\bf 105}, 050405 (2010).
%
\bibitem{Kev}
G. Theocharis, P. Schmelcher, P. G. Kevrekidis, D. J. Frantzeskakis, Dynamical trapping and transmission of matter-wave solitons in a collisionally inhomogeneous environment, Phys. Rev. A {\bf 74}, 053614 (2006).
%
\bibitem{Abd1}
J. Garnier, F. Kh. Abdullaev, Transmission of matter-wave solitons through nonlinear traps and barriers, Phys. Rev. A {\bf 74}, 013604 (2006).
%
\bibitem{Abd2}
F. Kh. Abdullaev, A. Gammal, L. Tomio, Dynamics of bright matter-wave solitons in a Bose-Einstein condensate with inhomogeneous scattering length, J. Phys. B: At. Mol. Opt. Phys. {\bf 37} (3), 635 (2004).
%
\bibitem{AlKhawadja}
U. Al Khawaja, Stability and dynamics of two-soliton molecules, Phys. Rev. E {\bf 81}, 056603 (2010).
%
\bibitem{Hulet2}
O. V. Marchukov, B. A. Malomed, V. A. Yurovsky, M. Olshanii, V. Dunjko, R. G. Hulet, Splitting of nonlinear-Schr\"{o}dinger equation breathers by linear and nonlinear localized potentials, Phys. Rev. A  {\bf 99}, 063623 (2019).
%
\bibitem{Xu2026} S. Xu, Y. Wang, Y. Zhao, Z. Fan, S. Chen, Steering photonic breathers via position-modulated Kerr nonlinearity in a segmented optomechanical array, Phys. Rev. A {\bf 113}, 033509 (2026).
%
\bibitem{Zhao2025} Y. Zhao, Y. Hou, Y. Wang, Z. Fan, W. Peng, Q. Zhou, S. Xu, S. Chen, Breather dynamics in an optomechanical array with staggered coupling, Phys. Rev. A {\bf 111}, 033502 (2025).
%
\bibitem{Hulet}
J. Cuevas, P. G. Kevrekidis, B. A. Malomed, P. Dyke, R. G. Hulet, Interactions of solitons with a Gaussian barrier: splitting and recombination in quasi-one-dimensional and three-dimensional settings, New J. Phys. {\bf 15}, 063006 (2013).
%
\bibitem{Petr15}
D. S. Petrov, Quantum mechanical stabilization of a collapsing Bose-Bose mixture, Phys. Rev. Lett. {\bf 115}, 155302 (2015).
%
\bibitem{Petr16}
D. S. Petrov, G. E. Astrakharchik, Ultradilute low-dimensional liquids, Phys. Rev. Lett. {\bf 117}, 100401 (2016).
%
\bibitem{LHY}
T. D. Lee, K. Huang, C. N. Yang, Eigenvalues and eigenfunctions of a Bose system of hard spheres and its low-temperature properties, Phys. Rev. {\bf 106}, 1135 (1957).
%
\bibitem{Cabr18}
C. R. Cabrera, L. Tanzi, J. Sanz, B. Naylor, P. Thomas, P.Cheiney, L. Tarruell, Quantum liquid droplets in a mixture of Bose-Einstein condensates, Science {\bf 359}, 301 (2018).
%
\bibitem{Seme18}
G. Semeghini, G. Ferioli, L. Masi, C. Mazzinghi, L. Wolswijk, F. Minardi, M. Modugno, G. Modugno, M. Inguscio, M. Fattori, Self-bound quantum droplets of atomic mixtures in free space, Phys. Rev. Lett. {\bf 120}, 235301 (2018).
%
\bibitem{Ferr16}
I. Ferrier-Barbut, H. Kadau, M. Schmitt, M. Wenzel, T. Pfau, Observation of quantum droplets in a strongly dipolar Bose gas, Phys. Rev. Lett. {\bf 116}, 215301 (2016).
%
\bibitem{Hu}
X. Hu, Z. Li, Y. Guo, Y. Chen, X. Luo, Scattering of one-dimensional quantum droplets by a reflectionless potential well, Phys. Rev. A {\bf 108}, 053306 (2023).
%
\bibitem{Debnath}
A. Debnath, A. Khan, B. Malomed, Interaction of one-dimensional quantum droplets with potential wells and barriers, Commun. Nonlinear Sci. Numer. Simul. {\bf 126}, 107457 (2023).
%
\bibitem{Abdullaev}
F. Kh. Abdullaev, R. M. Galimzyanov, Bosonic impurity in a one-dimensional quantum droplet in the Bose-Bose mixture, J. Phys. B: At. Mol. Opt. Phys. {\bf 53 }, 165301 (2020).
%
\bibitem{Bighin}
G. Bighin, A. Burchianti, F. Minardi, T. Macr\`{i}, Impurity in a heteronuclear two-component Bose mixture, Phys.Rev. A {\bf 106}, 023301 (2022).
%
\bibitem{Sinha}
S. Sinha, S. Biswas, L. Santos, S. Sinha, Impurities in quasi-one-dimensional droplets of binary Bose mixtures, Phys. Rev. A {\bf 108}, 023311 (2023).
%
\bibitem{Bristy}
F. Bristy, G. A. Bougas, G. C. Katsimiga, S. I. Mistakidis, Localization and splitting of a quantum droplet with a potential defect, Chaos, Solitons, and Fractals {\bf 201}, 117383 (2025).
%
\bibitem{Zin}
P. Zin, M. Pylak, T. Wasak, K. Jachymski, Z. Idziaszek, Quantum droplets in a dipolar Bose gas at a dimensional crossover, J. Phys. B: At. Mol. Opt. Phys. {\bf 54}, 165302 (2021).
%
\bibitem{Debnaz}
A. Debnath, A. Khan, Investigation of quantum droplets: an analytical approach, Ann. der Phys. {\bf 533}, 2000549 (2021).
%
\bibitem{Otajonov2019}
Sh. R. Otajonov, E. N. Tsoy, F. Kh. Abdullaev, Stationary and dynamical properties of one-dimensional quantum droplets, Phys. Lett. A {\bf 383}, 125980 (2019).
%
\bibitem{Otajonov2024}
Sh. R. Otajonov, B. A. Umarov, F. Kh. Abdullaev, Dynamics of quasi-one-dimensional quantum droplets in Bose-Bose mixtures, Chaos, Solitons, and Fractals {\bf 186}, 115212 (2024).
%
\bibitem{Otajonov2025}
Sh. R. Otajonov, B. A. Umarov, F. Kh. Abdullaev, Modulational instability and discrete quantum droplets in a deep quasi-one-dimensional optical lattice, Phys. Rev. E {\bf 111}, 054206 (2025).
%
\bibitem{Khawaja2021}
U. Al Khawaja, Critical soliton speed for quantum reflection by a reflectionless potential well, Phys. Rev. E {\bf 103}, 062202 (2021).
%
\bibitem{Ilg2018} T. Ilg, J. Kumlin, L. Santos, D. S. Petrov, H. P. B\"{u}chler, Dimensional crossover for the beyond-mean-field correction in Bose gases, Phys. Rev. A {\bf 98}, 051604(R) (2018).
%
\bibitem{Zin2018} P. Zin, M. Pylak, T. Wasak, M. Gajda, Z. Idziaszek, Quantum Bose-Bose droplets at a dimensional crossover, Phys. Rev. A {\bf 98}, 051603(R) (2018).
%
\bibitem{Lavoine2021} L. Lavoine, T. Bourdel, Beyond-mean-field crossover from one dimension to three dimensions in quantum droplets of binary mixtures, Phys. Rev. A {\bf 103}, 033312 (2021).
%
\bibitem{Debnath2021} A. Debnath and A. Khan, Investigation of quantum droplets: An analytical approach, Ann. Phys. (Berlin) {\bf 533}, 2000549 (2021).
%
\bibitem{Debnath2022} A. Debnath, A. Khan, and S. Basu, Dropleton-soliton crossover mediated via trap modulation, Phys. Lett. A {\bf 439}, 128137 (2022).
%
\bibitem{Shamriz2020} E. Shamriz, Z. Chen, and B. A. Malomed, Suppression of the quasi-two-dimensional quantum collapse in the attraction field by the Lee-Huang-Yang effect, Phys. Rev. A {\bf 101}, 063628 (2020).
%
\bibitem{Liu2023} B. Liu, X. Cai, X. Qin, X. Jiang, J. Xie, B. A. Malomed, and Y. Li, Ring-shaped quantum droplets with hidden vorticity in a radially periodic potential, Phys. Rev. E {\bf 108}, 044210 (2023).
%
\bibitem{AlMarzoug2026} S. M. Al-Marzoug, Coherent two-level dynamics of a single impurity in a self-bound quantum droplet, Eur. Phys. J. Plus {\bf 141}, 44 (2026).
%
\bibitem{Huang2026} W. Huang, Y. Liu, and X. Ruan, Ground states and droplet regimes of the extended Gross-Pitaevskii equation with Lee-Huang-Yang correction, arXiv:2604.04460 (2026).
%
\bibitem{Arora2021} A. K. Arora, C. Sparber, Self-Bound vortex states in nonlinear Schr\"{o}dinger equations with LHY correction, arXiv:2112.09213
%
\bibitem{Chen2024} G.-H. Chen, H.-C. Wang, H.-M. Deng, and B. A. Malomed, Vortex quantum droplets under competing nonlinearities, Chin. Phys. Lett. {\bf 41}, 020501 (2024).
%
\bibitem{Pelayo2025} J. C. Pelayo, G. Bougas, T. Fogarty, T. Busch, S. I. Mistakidis, Phases and dynamics of quantum droplets in the crossover to two-dimensions, SciPost Phys. {\bf 18}, 129 (2025).
%
\bibitem{JYang} J. Yang, Nonlinear Waves in Integrable and Nonintegrable Systems. SIAM, Philadelphia, 2010.
%
\bibitem{Otajonov2020}
Sh. R. Otajonov, E. N. Tsoy, F. Kh. Abdullaev, Variational approximation for two-dimensional quantum droplets, Phys. Rev. E {\bf 102}, 062217 (2020).
%
\bibitem{Otajonov2022}
Sh. R. Otajonov, Quantum droplets in three-dimensional Bose-Einstein condensates, J. Phys. B: At. Mol. Opt. Phys. {\bf 55}, 085001 (2022).
%
\bibitem{Brand2010} T. Ernst, J. Brand, Resonant trapping in the transport of a matter-wave soliton through a quantum well, Phys. Rev. A {\bf 81}, 033614 (2010).
%
\bibitem{Hansen} S. D. Hansen, N. Nygaard, K. M{\o}lmer, Scattering of matter wave solitons on localized potentials, Appl. Sci. {\bf 11}, 2294 (2021).
%
\bibitem{Boudjema}
Kh. M. Elhadj, L. Al Sakkaf, A. Boudjem\^{a}a, U. Al Khawaja, Quantum droplet molecules in Bose-Bose mixtures, Phys. Lett. A {\bf 494}, 129274 (2024).
%
\bibitem{Holmer2007} J. Holmer, J. Marzuola, M. Zworski, Soliton splitting by external delta potentials, J. Nonlinear Sci. Vol. {\bf 17}, 349-367 (2007).
%
\bibitem{Marchant2}
A. L. Marchant, T. P. Billam, M. M. H. Yu, A. Rakonjac, J. L. Helm, J. Polo, C. Weiss, S. A. Gardiner, S. L. Cornish, Quantum reflection of bright solitary matter waves from a narrow attractive potential, Phys. Rev. A {\bf 93}, 021604(R) (2016).
%
\bibitem{Arazo2021} M. Arazo, M. Guilleumas, R. Mayol, M. Modugno, Dynamical generation of dark-bright solitons through the domain wall of two immiscible Bose-Einstein condensates, Phys. Rev. A {\bf 104}, 043312 (2021).
%
\bibitem{Deekshita2025} S. Deekshita, S. Sanjay, S. S. Veni, C. B. Tabi, T. C. Kofan\"e, Synergistic effects of spin-orbit coupling and intercomponent interactions in two-component (2+1)D photonic fields, Chaos, Solitons \& Fractals, {\bf 199}, 116806 (2025).
%
\bibitem{Sanjay2025} S. Sanjay, S. S. Veni, B. A. Malomed, Vortex droplets and lattice patterns in two-dimensional traps: A photonic spin-orbit-coupling perspective, Chaos, Solitons \& Fractals, {\bf 197}, 116441 (2025).
%
\bibitem{Flugge} S. Fl\"{u}gge, Practical Quantum Mechanics, Second Printing, Springer, Berlin, 1994.
%
\bibitem{Landau3} L. D. Landau, E. M. Lifshitz, Course of Theoretical Physics, Vol. 3: Quantum Mechanics--Non-relativistic Theory, 3rd ed., Pergamon Press, Oxford, 1977.
%
\bibitem{Lekner2007} J. Lekner, Reflectionless eigenstates of the $\mathrm{sech}^2$ potential, Am. J. Phys. {\bf 75}, 1151-1157 (2007).
%

\end{thebibliography}
\end{document}